\documentclass[lettersize,journal]{IEEEtran}
\usepackage{amsmath,amsfonts}
\usepackage{algorithmic}
\usepackage{algorithm}
\usepackage{array}
\usepackage[caption=false,font=normalsize,labelfont=sf,textfont=sf]{subfig}
\usepackage{textcomp}
\usepackage{stfloats}
\usepackage{url}
\usepackage{verbatim}
\usepackage{graphicx}
\usepackage{cite}

\usepackage{hyperref}

\graphicspath{{figs/}{figures/}{pictures/}{images/}{./}} 
\usepackage{tabu}                      
\usepackage{booktabs}                  
\usepackage{lipsum}                    
\usepackage{mwe}                       

\usepackage{mathptmx}                  

\usepackage{afterpage}
\usepackage{amstext}
\usepackage{amsmath}
\usepackage{amssymb}
\usepackage{empheq}
\usepackage{xcolor}
\usepackage{amsthm}

\usepackage{tikz}
\usetikzlibrary{calc}

\DeclareMathAlphabet{\mathcal}{OMS}{cmsy}{m}{n}

\begin{document}

\title{Region-Aware Wasserstein Distances of \\ Persistence Diagrams and Merge Trees}

\author{Mathieu Pont and Christoph Garth
\thanks{Mathieu Pont and Christoph Garth are with RPTU Kaiserslautern-Landau. E-mails: \{mathieu.pont,garth\}@rptu.de}
}

\markboth{
}%
{Shell \MakeLowercase{\textit{et al.}}: A Sample Article Using IEEEtran.cls for IEEE Journals}


\newcommand{\domain}{M}
\newcommand{\complex}{\mathcal{M}}
\newcommand{\complexVertexSet}{\mathcal{V}}
\newcommand{\range}{\mathbb{R}}
\newcommand{\sublevelset}[1]{#1^{-1}_{-\infty}}
\newcommand{\superlevelset}[1]{#1^{-1}_{+\infty}}
\newcommand{\Star}{St}
\newcommand{\Link}{Lk}
\newcommand{\simplex}{\sigma}
\newcommand{\face}{\tau}
\newcommand{\lowerlink}{\Link^{-}}
\newcommand{\upperlink}{\Link^{+}}
\newcommand{\Index}{\mathcal{I}}
\newcommand{\offset}{o}
\newcommand{\Natural}{\mathbb{N}}
\newcommand{\criticalSet}{\mathcal{C}}
\newcommand{\diagram}{\mathcal{D}}
\newcommand{\wasserstein}[1]{W^{\diagram}_#1}
\newcommand{\projection}{\Delta}
\newcommand{\hierarchy}{\mathcal{H}}
\newcommand{\decimation}{D}
\newcommand{\xDimD}{L_x^\decimation}
\newcommand{\yDimD}{L_y^\decimation}
\newcommand{\zDimD}{L_z^\decimation}
\newcommand{\xDim}{L_x}
\newcommand{\yDim}{L_y}
\newcommand{\zDim}{L_z}
\newcommand{\Grid}{\mathcal{G}}
\newcommand{\GridD}{\mathcal{G}^\decimation}
\newcommand{\x}{\phantom{x}}
\newcommand{\Mod}{\;\mathrm{mod}\;}
\newcommand{\NN}{\mathbb{N}}
\newcommand{\forwardIntegralLine}{\mathcal{L}^+}
\newcommand{\backwardIntegralLine}{\mathcal{L}^-}
\newcommand{\triangulationOp}{\phi}
\newcommand{\decimationOp}{\Pi}
\newcommand{\isovalue}{w}
\newcommand{\persistence}{\mathcal{P}}
\newcommand{\pointMetric}{d}
\newcommand{\diagramSet}{\mathcal{S}_\mathcal{D}}
\newcommand{\diagramSpace}{\mathbb{D}}
\newcommand{\jointree}{\mathcal{T}^-}
\newcommand{\splittree}{\mathcal{T}^+}
\newcommand{\mergetree}{\mathcal{T}}
\newcommand{\mergetreeSet}{\mathcal{S}_\mathcal{T}}
\newcommand{\branchset}{\mathcal{S}_\mathcal{B}}
\newcommand{\branchspace}{\mathbb{B}}
\newcommand{\mergetreeSpace}{\mathbb{T}}
\newcommand{\editdistance}{D_E}
\newcommand{\wassersteinTree}[1][2]{W^{\mergetree}_{#1}}
\newcommand{\distanceSequence}{d_S}
\newcommand{\branchtree}{\mathcal{B}}
\newcommand{\branchtreeSet}{\mathcal{S}_\mathcal{B}}
\newcommand{\branchtreeSpace}{\mathbb{B}}
\newcommand{\forest}{\mathcal{F}}
\newcommand{\sequenceSpace}{\mathbb{S}}
\newcommand{\forestMatrix}{\mathbb{F}}
\newcommand{\treeMatrix}{\mathbb{T}}
\newcommand{\normalizedLocation}{\mathcal{N}}
\newcommand{\normalizedWasserstein}{W^{\normalizedLocation}_2}
\newcommand{\geodesictree}{\mathcal{G}}
\newcommand{\dummyVector}{\mathcal{V}}
\newcommand{\geodesictreeVec}{g}
\newcommand{\geodesicAxis}{\mathcal{A}}
\newcommand{\directionVector}{\mathcal{V}}
\newcommand{\geodesicdiagram}{\mathcal{G}^{\diagram}}
\newcommand{\reconstructionError}{E_{L_2}}
\newcommand{\pcaBasis}{B_{\mathbb{R}^d}}
\newcommand{\waepcaBasis}{B}
\newcommand{\origin}{o_b}
\newcommand{\sizeEncoding}{n_e}
\newcommand{\sizeDecoding}{n_d}
\newcommand{\linearTransformation}{\psi}
\newcommand{\unitTransformation}{\Psi}
\newcommand{\waeorigin}{o}
\newcommand{\bdtOrigin}{\mathcal{O}}
\newcommand{\activation}{\sigma}
\newcommand{\validBDT}{\gamma}
\newcommand{\mtPgaBasis}{B_{\branchtreeSpace}}
\newcommand{\mtPgaError}{E_{\wassersteinTree}}
\newcommand{\frechetEnergy}{E_F}
\newcommand{\geodesicExtremity}{\mathcal{E}}
\newcommand{\vectorNotation}[1]{\protect\vv{#1}}
\newcommand{\waevectorNotation}[1]{#1}
\newcommand{\axisNotation}[1]{\protect\overleftrightarrow{#1}}
\newcommand{\individualEnergy}{E}
\newcommand{\ensembleSize}{N}
\newcommand{\numberBranchinBarycenter}{N_1}
\newcommand{\numberGeodesicSamples}{N_2}
\newcommand{\planarGridX}{N_x}
\newcommand{\planarGridY}{N_y}
\newcommand{\regularGrid}{G}
\newcommand{\distanceMatrix}{\mathbb{D}}
\newcommand{\maxDimensions}{{d_{max}}}
\newcommand{\projectionOperator}{\mathcal{P}}
\newcommand{\reconstructed}[1]{\widehat{#1}}
\newcommand{\gt}{>}
\newcommand{\lt}{<}
\newcommand{\branch}{b}
\newcommand{\nonLinearFunction}{\sigma}
\newcommand{\batchSequence}{S}

\newcommand{\volume}{\mathcal{V}}
\newcommand{\volumeWasserstein}[1]{W^{\volume\diagram}_#1}
\newcommand{\volumeWassersteinTree}{W^{\volume\mergetree}_2}

\newcommand{\lifting}{\mathcal{L}}
\newcommand{\liftingWasserstein}[1]{W^{\lifting\diagram}_#1}
\newcommand{\liftingWassersteinTree}{W^{\lifting\mergetree}_2}

\newcommand{\continuousRegion}{R}
\newcommand{\region}{\mathcal{R}}
\newcommand{\regionNumber}[1]{N_{\region^#1}}
\newcommand{\regionWasserstein}[1]{W^{\region\diagram}_{#1}}
\newcommand{\regionWassersteinTree}[1][2]{W^{\region\mergetree}_{#1}}

\newcommand{\regionalDiscrepancy}{C}
\newcommand{\projectionRegion}{\delta}
\newcommand{\subsampleRegion}{\mathcal{S}}

\newcommand{\subsampleParameter}{\lambda}
\newcommand{\compressionParameter}{\tau}


\newcommand{\appendixMetricProof}{\mbox{Appendix A}}
\newcommand{\appendixMetricProofAny}{\mbox{Appendix B}}
\newcommand{\appendixStability}{\mbox{Appendix C}}
\newcommand{\appendixStabilitySameOrder}{\mbox{Appendix C-B}}
\newcommand{\appendixStabilityMultiSaddle}{\mbox{Appendix C-B (i)}}
\newcommand{\appendixStabilityEvaluation}{\mbox{Appendix D}}
\newcommand{\appendixDimRedExperiments}{\mbox{Appendix E}}
\newcommand{\appendixMemory}{\mbox{Appendix F}}
\newcommand{\appendixCompression}{\mbox{Appendix G}}
\newcommand{\appendixGeometric}{\mbox{Appendix H}}
\newcommand{\appendixDimRedIndicators}{\mbox{Appendix I}}

\newcommand{\equationGroundMetric}{\mbox{Eq. 1}}

\newcommand{\secIntroduction}{\mbox{Sec. I}}
\newcommand{\secRelatedWork}{\mbox{Sec. I-A}}
\newcommand{\secPreliminariesInput}{\mbox{Sec. II-A}}
\newcommand{\secPreliminariesPersistenceDiagrams}{\mbox{Sec. II-B}}
\newcommand{\secPreliminariesMergeTrees}{\mbox{Sec. II-C}}
\newcommand{\secPreliminariesWasserstein}{\mbox{Sec. II-D}}
\newcommand{\secDistance}{\mbox{Sec. III}}
\newcommand{\secDistanceMotivation}{\mbox{Sec. III-A}}
\newcommand{\secDistanceRABDT}{\mbox{Sec. III-B}}
\newcommand{\secDistanceDefinition}{\mbox{Sec. III-C}}
\newcommand{\secDistanceDefinitionGroundMetric}{\mbox{Sec. III-C1}}
\newcommand{\secDistanceDefinitionProjection}{\mbox{Sec. III-C2}}
\newcommand{\secDistanceDefinitionDistance}{\mbox{Sec. III-C3}}
\newcommand{\secDistanceProperties}{\mbox{Sec. III-D}}
\newcommand{\secDistancePropertiesGeneralization}{\mbox{Prop. 1}}
\newcommand{\secAlgorithm}{\mbox{Sec. IV}}
\newcommand{\secCompression}{\mbox{Sec. IV-B}}
\newcommand{\secResults}{\mbox{Sec. V}}
\newcommand{\secBaseline}{\mbox{Sec. V-1}}
\newcommand{\secTiming}{\mbox{Sec. V-A}}
\newcommand{\secFeatureTracking}{\mbox{Sec. V-B}}
\newcommand{\secDimRed}{\mbox{Sec. V-C}}
\newcommand{\secFrameworkQuality}{\mbox{Sec. V-D}}
\newcommand{\secLimitations}{\mbox{Sec. V-E}}
\newcommand{\secConclusion}{\mbox{Sec. VI}}

\newcommand{\figTeaser}{\mbox{Fig. 1}}
\newcommand{\figTA}{\mbox{Fig. 2}}
\newcommand{\figRegion}{\mbox{Fig. 6}}
\newcommand{\figRegionDistance}{\mbox{Fig. 7}}
\newcommand{\figDarkSky}{\mbox{Fig. 8}}
\newcommand{\figCloud}{\mbox{Fig. 9}}
\newcommand{\figHeatedCylinder}{\mbox{Fig. 10}}
\newcommand{\figStability}{\mbox{Fig. 12}}

\newcommand{\tableTime}{\mbox{Tab. 1}}

\newcommand{\figStabilityIsabel}{\mbox{Fig. 2}}
\newcommand{\figStabilityEnsemble}{\mbox{Fig. 3}}
\newcommand{\figEucComparison}{\mbox{Fig. 4}}


\newcommand{\julien}[1]{\textcolor{red}{#1}}
 \renewcommand{\julien}[1]{\textcolor{black}{#1}}
\newcommand{\mathieu}[1]{\textcolor{blue}{#1}}
\newcommand{\jules}[1]{\textcolor{orange}{#1}}
\renewcommand{\jules}[1]{\textcolor{black}{#1}}
\newcommand{\note}[1]{\textcolor{magenta}{#1}}
\newcommand{\cutout}[1]{\textcolor{blue}{#1}}
\renewcommand{\cutout}[1]{}
\newcommand{\journal}[1]{\textcolor{blue}{#1}}
\renewcommand{\journal}[1]{\textcolor{black}{#1}}

\newif\ifinsideRevision
\DeclareRobustCommand{\revision}[1]{%
  \begingroup
    \ifinsideRevision
      #1%
    \else
      \color{black}#1%
    \fi
  \endgroup
}

\DeclareRobustCommand{\revisionTVCG}[1]{%
  \begingroup
    \ifinsideRevision
      #1%
    \else
      \insideRevisiontrue
      \color{black}#1%
    \fi
  \endgroup
}

\DeclareRobustCommand{\revisionTVCGminor}[1]{%
  \begingroup
    \insideRevisiontrue
    \color{black}#1%
  \endgroup
}

\newcommand{\myTodo}[1]{\begingroup\color{red}#1\endgroup}

\newcommand{\new}[1]{{\leavevmode\color{black}#1}}
\newcommand{\newT}[1]{{\leavevmode\color{blue}#1}}

\renewcommand{\figureautorefname}{Fig.}
\renewcommand{\sectionautorefname}{Sec.}
\renewcommand{\subsectionautorefname}{Sec.}
\renewcommand{\subsubsectionautorefname}{Sec.}
\renewcommand{\equationautorefname}{Eq.}
\renewcommand{\tableautorefname}{Tab.}
\newcommand{\algorithmautorefname}{Alg.}
\newcommand{\lineautorefname}{Alg.}

\newcommand{\eqSpace}{-1.75ex}

\setlength{\fboxrule}{0.8pt} 
\renewcommand\fbox{\fcolorbox{sorblue}{white}}

\newcommand{\definitionText}[2]{\textit{(#1)} #2}

\newcommand{\definition}[2]{\definitionf{\definitionText{#1}{#2}}}

\newcommand{\definitionBox}[2]{
  \definitionf{
    \fbox{
      \begin{minipage}[t]{\textwidth}
      \definitionText{#1}{#2}
      \end{minipage}
    }
  }
}

\newlength{\figDouble}
\setlength{\figDouble}{0.85\paperwidth}

\newtheorem{theorem}{Theorem}
\newtheorem{proposition}{Proposition}

\providecommand*{\propositionautorefname}{Prop.}

\maketitle

\begin{abstract}
This paper presents 
\revision{a generalization of the Wasserstein distance for both \revisionTVCG{0th} persistence diagrams and merge trees \cite{edelsbrunner09, pont_vis21}}
that takes advantage of the regions of their topological features in the input domain.
Specifically, we redefine the comparison of topological features as a distance between the values of their extrema-aligned regions.
\revision{It results in a more discriminative metric than the classical Wasserstein distance} and generalizes it through an input parameter adjusting the impact of the region properties in the distance.
We present two strategies to control both computation time and memory storage of our method by respectively enabling the use of subsets of the regions in the computation, and by compressing the regions' properties to obtain low-memory representations.
Extensive experiments on openly available ensemble data demonstrate the efficiency of our method, with running times on the orders of minutes on average.
We show the utility of our contributions with two applications.
First, we use the assignments between topological features provided by our method to track their evolution in time-varying ensembles and propose the \emph{temporal persistence curves} to facilitate the understanding of how these features appear, disappear and change over time.
Second, our method allows to compute a distance matrix of an ensemble that can be used for dimensionality reduction purposes and visually represent in 2D all its members, we show that such distance matrices also allow to detect key phases in the ensemble.
Finally, we provide a C++ implementation that can be used to reproduce our results.
\end{abstract}

\begin{IEEEkeywords}
Topological data analysis, ensemble data, merge trees, persistence diagrams.
\end{IEEEkeywords}

\begin{figure*}
    \centering
    \includegraphics[width=\linewidth]{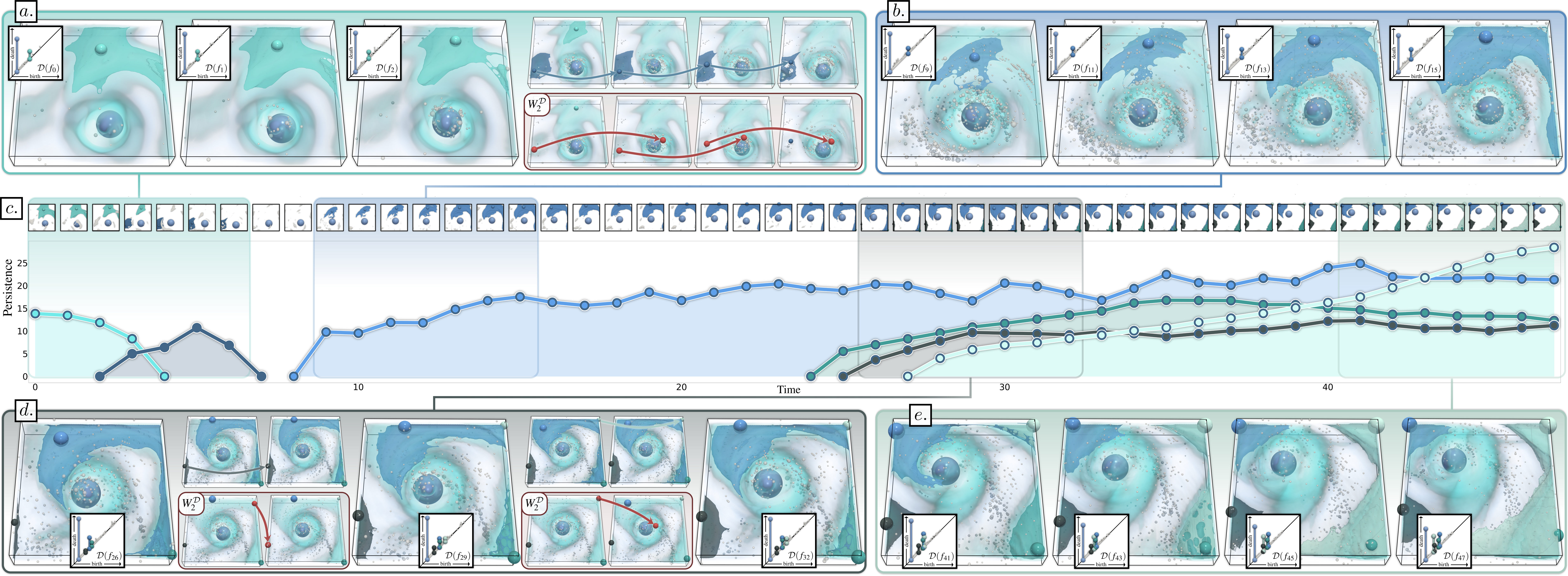}
    \caption{
\revisionTVCG{
Visual analysis of the 48 time steps of the \emph{Isabel} ensemble (wind velocity) with our novel region-aware Wasserstein distance.
Members at different points in time are shown in $(a)$, $(b)$, $(d)$ and $(e)$.
The persistence diagram of each member (black insets) allows to identify its features of interest, their maxima are visualized in the input domain with a sphere whose size is \revisionTVCGminor{proportional} to their persistence (a notion of importance) and the regions of the highest persistent features are shown with a surface.
Our method provides assignments between the features of consecutive members, allowing to track their evolution over time (with matched features having the same color).
In this example, it correctly matches the same features together over the whole ensemble while the classical Wasserstein distance $\wasserstein{2}$ \revisionTVCGminor{provides} some mismatches (red arrows in red insets of $(a)$ and $(d)$) preventing their correct tracking.
The \emph{temporal persistence curves} in $(c)$ represent the evolution over time of the persistence of the matched features with our method, only curves for the highest persistent features are shown with the same color \revisionTVCGminor{as} in the input domain.
This allows to visually convey the changes in the ensemble in one view and permits the user to identify and select important time steps to be visualized.
For instance, in the second time window $(b)$ the blue feature starts to appear and grow, while in the last time window $(e)$ the white feature \revisionTVCGminor{overtakes} the other ones.
}}
    \label{fig_teaser}
\end{figure*}

\section{Introduction}

\IEEEPARstart{A}{cquisition} devices \revisionTVCG{are becoming} more precise and computational power continues to grow, \revisionTVCG{making} modern datasets, acquired or simulated, increasingly detailed and complex.
\revisionTVCG{This}
poses a challenge for interactive exploration, making data analysis and interpretation more difficult for end users 
\revisionTVCG{and}
has motivated the development of intermediary representations which summarize the data in a concise and informative manner.
In that direction, Topological Data Analysis (TDA) methods \cite{edelsbrunner09, zomorodianBook} have been proven useful, such as in the visualization community \cite{heine16}, by providing representations that extract the features of interest within the data while having memory storage requirements orders of magnitude lower than the data themselves.
Such methods have been successfully applied to several domains \cite{heine16}\revision{, for example in} turbulent combustion \cite{laney_vis06, bremer_tvcg11, gyulassy_ev14}, material sciences \cite{gyulassy_vis07, gyulassy_vis15, favelier16}, nuclear energy \cite{beiNuclear16}, fluid dynamics \cite{kasten_tvcg11, nauleau_ldav22}, bioimaging \cite{carr04, topoAngler, beiBrain18}, quantum chemistry \cite{chemistry_vis14, harshChemistry, Malgorzata19} or astrophysics \cite{sousbie11, shivashankar2016felix}.
Popular instances in the visualization 
community of representations studied in TDA include the persistence diagram (\autoref{fig_tas}, top insets) \cite{dipha, edelsbrunner09, edelsbrunner02, guillou_tvcg23} and the merge tree (\autoref{fig_tas}, middle insets) \cite{carr00, carr04, bremer_tvcg11, topoAngler}. 

In modern applications, alongside the growing complexity of datasets, a significant challenge arises with the notion of \emph{ensemble datasets} 
\cite{scivis2015, scivis2004, favelier2018}.  
Instead of representing a phenomenon with a single dataset, these consist of multiple ones, each of them referred \revisionTVCGminor{to} as an \emph{ensemble member}. 
A common approach in TDA is to compute a topological abstraction, such as a persistence diagram or \revision{a} merge tree, for each ensemble member in order to encode its features of interest in a compact manner.
Then, a strategy consists in analyzing the given ensemble of datasets through the resulting ensemble of topological representations.
While this has several benefits, \revision{like} the direct processing of features of interest and a reduced memory footprint, it necessitates the development of analysis methods for these specific objects. 

Several works explored this direction, \revision{notably} with the computation of an average topological representation 
\cite{Turner2014, vidal_vis19, pont_vis21, wetzels23}
that is representative of an ensemble and can be used for clustering tasks, or with variability analysis methods for ensembles of topological representations \cite{pont_tvcg23, pont_tvcg24}.
The basic and main building block of all these methods is the notion of distances between topological representations.
While practically useful and well studied metrics have been developed for such objects, persistence diagrams and merge trees themselves can suffer from a lack of specificity, where geometrically different datasets have the same topological representation, \revision{preventing to discriminate them} (\autoref{fig_geom}).

By encoding features too compactly, from a topological point of view, they introduce a limitation that causes important geometric details to be lost.
This lack of geometric specificity impacts the subsequent data analysis pipeline after their computation and can prevent the identification of important feature differences in an ensemble\revision{.
This can lead} to wrong scientific conclusions where datasets seem to be \revision{identical} from a topological point of view but are not from a geometrical one.
\revision{To tackle this limitation, topological representations must be extended with geometric properties and for their analysis, a metric being computable in practice needs to be defined.}
\revision{Earlier works enrich topological descriptors with only coarse and aggregated geometric properties, particularly with size and position\revision{\mbox{\cite{soler_ldav18, lin_geometryAware23, mingzheTracking, SaikiaSW14_branch_decomposition_comparison}}}, therefore overlooking the fine-grained and vertex-level inner structure of each feature.}

This paper addresses this \revision{limitation} by first \revision{associating} to \revision{each topological feature encoded in} \revisionTVCG{0th} persistence diagrams \revision{or} merge trees \revision{a detailed characterization of the geometry of its inner structure}.
\revision{Specifically, \revisionTVCG{in contrast to previous \revisionTVCGminor{works} that use explicit geometric descriptors such as volume or position,} we 
\revisionTVCG{represent}
each topological feature \revisionTVCG{by} a \emph{vertex-level characterization}\revisionTVCG{, i.e. the scalar field sampled on} its region in the input domain. 
}
We then introduce a distance metric for both \revision{descriptors} that takes advantage of this geometric information \revisionTVCG{with a pointwise comparison of the restricted scalar fields of two features after aligning their regions at their extrema}.
Our approach \revision{is a natural generalization of} the Wasserstein distance between \revisionTVCG{0th} persistence diagrams and merge trees \cite{pont_vis21, edelsbrunner09}\revision{,} that we extend to incorporate such geometric properties.
\revision{We show that the proposed metric is more discriminative than the original Wasserstein distance that eventually corresponds to a special case of our method.}
We also propose and evaluate strategies, subject to input parameters, to control both the computation time and memory footprint of our method (\autoref{sec_algo}), by respectively allowing to use more or less the geometric information in the computation and by compressing the geometric characterizations to reduce memory storage.
We show the utility of our contributions with two visualization tasks.
First, we use the assignments between the features provided by our distance to track their evolution in time-varying datasets (\autoref{sec_featureTracking}) and propose a visualization tool to easily see how they evolve over time, allowing to detect phases and gain insights about the ensemble.
Second, the distance matrix of an ensemble can be used for dimensionality reduction (\autoref{sec_dimRed}) in order to visually represent in 2D all the members of the ensemble and how they relate to each other.

\subsection{Related work}
\label{sec_relatedWork}

The literature related to our work can be classified into two groups: \emph{(i)} topological representations and \emph{(ii)} \revision{their} metrics.

\begin{figure}
    \centering
    \includegraphics[width=\linewidth]{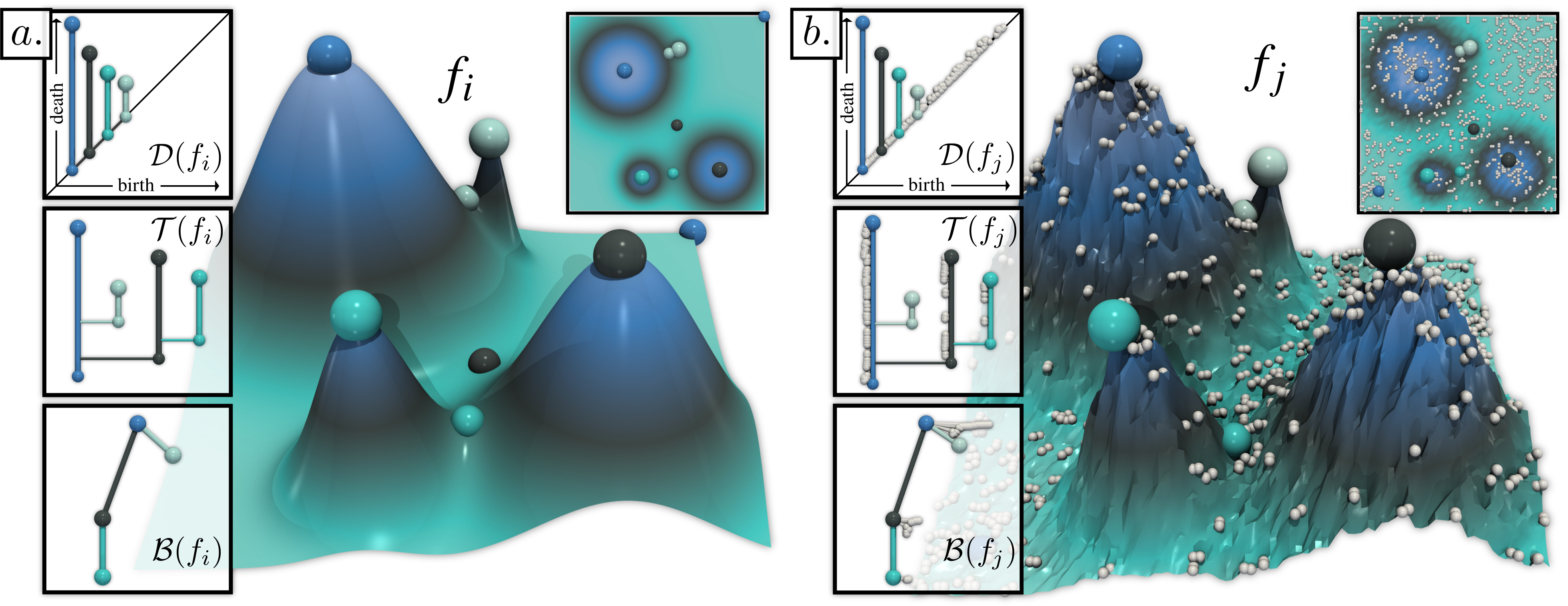}
    \caption{
\revisionTVCG{
Illustration of the topological descriptors studied in this work, computed on a 2D \revision{minimal example} (a) and a noisy version of it (b). 
For all descriptors, the color code represents the persistence of the corresponding saddle-maximum pair. 
Critical points are visualized with spheres, the larger ones indicating maxima. 
Persistence diagrams, merge trees, and branch decomposition trees (BDTs) are shown in the top, center, and bottom insets, respectively. 
In both datasets, the four primary features (the most prominent hills) are highlighted with salient pairs in the diagram and the merge tree. 
To maintain clarity in the visualization, pairs with low persistence (below 10\% of the function range) are displayed with small white arcs and spheres, while the larger and colored ones \revision{correspond to those with higher persistence}.
}}
    \label{fig_tas}
\end{figure}

\begin{figure}
    \centering
    \includegraphics[width=\linewidth]{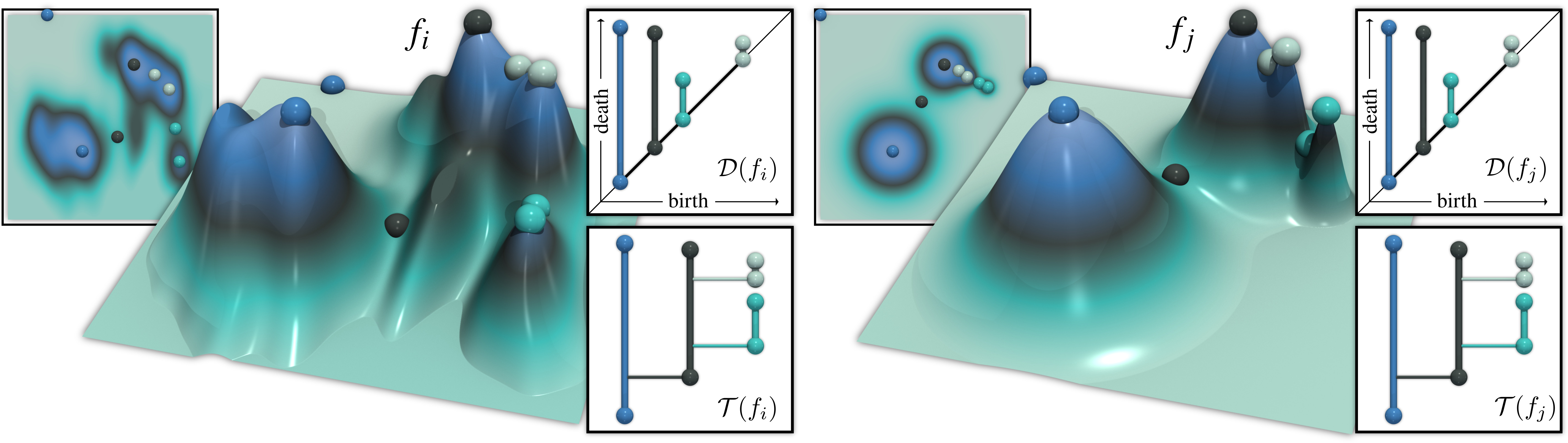}
    \caption{
      The persistence diagram, $\diagram(f_i)$, and the merge tree, $\mergetree(f_i)$, provide a visual summary of the number, data range, and salience of the features of interest present in the data. 
      However, they represent each feature without considering its geometric properties. 
      This results in a lack of specificity for these objects, which can yield identical data representations for significantly distinct datasets. 
      In this example, the four features of each dataset are identically encoded in the corresponding topological representation whereas their respective geometry differs.
      \revision{These datasets can therefore not be discriminated using the celebrated Wasserstein distance for topological representations.}
    }
    \label{fig_geom}
\end{figure}

\noindent
\textbf{\emph{(i)} Topological representations:}
Methods from computational topology \cite{edelsbrunner09} have been explored and extended by the visualization community for a diverse set of applications \cite{surveyComparison2021}.
The persistence diagram \cite{dipha, edelsbrunner09, edelsbrunner02, guillou_tvcg23} (\autoref{sec_pd} and \autoref{fig_tas} top insets) and the merge tree \cite{carr00, tarasov98} (\autoref{sec_mt} and \autoref{fig_tas} middle insets) intuitively 
segment the domain into regions (\autoref{fig_region}), each of them \revision{being} a feature \revision{of interest}.
It can be computed through matrix reduction \cite{dipha, edelsbrunner09} or using discrete morse theory \cite{guillou_tvcg23}.
The merge tree additionally encodes the connectivity of the features in a hierarchical manner and efficient algorithms for its computation (and its generalization, the contour tree) have been documented \cite{carr00, tarasov98, AcharyaN15, CarrWSA16, gueunet_tpds19, MaadasamyDN12, extreem}.
Other examples of topological representations include the Reeb graph \cite{biasotti08, GueunetFJT19, Parsa12, pascucci07, reeb1946points}
or the Morse-Smale complex \cite{BremerEHP04, Defl15, gyulassy_vis18, gyulassy_vis14, GyulassyNPBH06, ShivashankarN12}.

\noindent
\textbf{\emph{(ii)} Metrics for topological representations:}
Using concepts from optimal transport \cite{Kantorovich, monge81}, the Wasserstein distance between persistence diagrams \cite{edelsbrunner09} (\autoref{sec_wassersteinMetric}) is based on a bipartite assignment problem 
for which exact computations \cite{Munkres1957} or fast approximations \cite{Bertsekas81, Kerber2016} are available in open-source \cite{ttk17}.
It was proven to be stable under perturbations of the input data \cite{CohenSteinerEH05}.
However, persistence diagrams may lack \revision{in} specificity and \revision{can} represent significantly different datasets in the same way, failing to discriminate \revision{those} that differ in feature connectivity (a simple example can be found in \cite{pont_vis21} Figure 3).
This has motivated the development of metrics for more advanced descriptors, \revision{including} merge trees.

A stable distance between merge trees has been proposed \cite{morozov14} \revision{having} however 
\revisionTVCG{a runtime exponential in a structural parameter of the trees}, therefore
not tractable for real-life datasets.
A variant has been developed using pre-existing correspondence labels between merge trees nodes \cite{intrinsicMTdistance}, since manual labeling is impractical for real-life datasets, some heuristics need to be considered \cite{YanWMGW20}.
Apart from that, Beketayev et al. \cite{BeketayevYMWH14} focused on a dual representation, the \emph{branch decomposition tree} (BDT, see \autoref{sec_mt}), by minimizing a term over all possible branch decompositions and Wetzels et al. \cite{WetzelsLG22} provided a similar but faster approach based on edit distances.
They also proposed a constrained edit distance based on deformations of the trees \cite{WetzelsTopoInVis22} that provide meaningful matchings.
Both methods have however a quartic time complexity allowing to only process merge trees with a very small number of nodes.
An unconstrained version of the latter was shown to be stable but NP-complete with exponential time complexity \cite{wetzels_taming} for which some heuristics have been explored to choose a trade-off between the quartic and exponential time complexity \cite{wetzels_accelerating}.
Sridharamurthy et al. \cite{SridharamurthyM20} adapted an efficient algorithm for the computation of constrained edit distances between trees \cite{zhang96} to merge trees by defining edit costs related to these objects.
It results in an efficient distance with quadratic time complexity that can be easily computed for real-life datasets and whose stability can be controlled thanks to an input parameter.
Based on this, Pont et al. \cite{pont_vis21} generalized the Wasserstein distance between extremum persistence diagrams to merge trees, allowing to choose a trade-off between both representations given an input parameter.
One benefit of this framework, in addition to its efficiency, is that an analysis method developed using this distance can trivially process both persistence diagrams and merge trees without additional effort.

Some works have taken an initial step toward incorporating geometric information into distance computation by focusing on \revision{coarse and aggregated properties}.
It usually involves the addition of a term to the distance between two topological features, originally defined as the distance between their minimum and maximum values respectively (their \emph{birth} and their \emph{death}, see \autoref{sec_pd}).
For instance, the 
distance between the coordinates in the domain of the extrema of two topological features can be used\revision{\mbox{\cite{soler_ldav18, lin_geometryAware23, mingzheTracking}}}.
Note that while Yan et al. \cite{lin_geometryAware23} or \revision{\revisionTVCG{Li} et al. \cite{mingzheTracking}} propose a framework to incorporate geometric information into distance computation, they \revision{mainly} consider coordinates in the domain.
The size, i.e. the number of vertices, of the segmented region corresponding to a feature can also be considered \cite{SaikiaSW14_branch_decomposition_comparison}.
Inspired by \cite{ThomasN13}, the geodesic distance between the extrema of a feature in the hyperplane plot of the scalar function can provide some geometric insights about topological features.
\revision{
\emph{Decorated} approaches (for merge trees \cite{curryDecorated} and Reeb spaces \cite{curryDecoratedReeb}) attach auxiliary signals, for example with local persistence diagrams.
}
\revisionTVCG{Another approach compares point clouds by incorporating geometric proximity information \revisionTVCGminor{along with} persistence diagrams \cite{Zhang2025}.}

However these methods have several limitations, first they only add \revision{coarse} information about the geometric properties of the features, they are still missing detailed characterizations of the \revisionTVCGminor{features'} geometry related to their shape etc.
Second, they extend the comparison of the scalar values of topological features with a term of a different nature and scale (distances in the domain, differences in size etc.), challenging the correct weighting and mixture of both.
In contrast, our approach uses a more detailed characterization of the features.
It takes advantage of all their scalar values in order to precisely characterize their geometry and extends the usual distance between topological features to not only compare their minimum and maximum values, but up to all their values given an input parameter controlling the amount of information to use.


\subsection{Contributions}

This paper makes the following new contributions:

\begin{enumerate}
 \item \emph{A region-aware metric for \revisionTVCG{0th} persistence diagrams and merge trees}: We \revision{generalize} recent works on Wasserstein distances for \revisionTVCG{0th} persistence diagrams and merge trees \cite{pont_vis21, edelsbrunner09} 
 \revision{to} introduce a new metric \revision{that}
 incorporates richer and more detailed geometric characterizations \revision{of} the regions of the topological features.
 \revision{It makes}
 them special cases of our method, by allowing to control how much of this detailed characterization should be used depending on the applications.
 It can be computed efficiently and \revision{we prove theoretical properties such as it is \revisionTVCGminor{a} metric, stable under some conditions and more discriminative than the original metrics it is based on}.
 \item \emph{Time and memory optimization strategies:} We present and evaluate strategies to control both the computation time and the memory footprint of our method. 
 First, we propose a way to simplify the comparisons of the \revisionTVCGminor{features'} geometry inspired by subsampling.
 Second, we show how traditional compression methods can be used to compress the detailed characterization of the features to reduce memory storage and \revision{how it impacts} our metric.
 \item \emph{An application to feature tracking}: The matchings between topological features provided by our novel metric can be used to track their evolution in time-varying ensembles (\autoref{sec_featureTracking}). We also introduce the \emph{Temporal Persistence Curves} that visually convey how topological features evolve over time allowing to easily detect when they appear, disappear or change.
 \item \emph{An application to dimensionality reduction}: Our method provides distance matrices that can be used by traditional dimensionality reduction algorithms to embed each \revisionTVCGminor{topological} representation as a 2D point and visually see how they relate to each other (\autoref{sec_dimRed}). We \revision{also} show that such distance matrices allow to detect when important changes occur in an ensemble.
 \item \emph{Implementation}: We provide a C++ implementation of our algorithms for reproduction purposes.
\end{enumerate}

\section{Preliminaries}

This section introduces the various concepts upon which our work is based.
We refer to the reference books \cite{edelsbrunner09, zomorodianBook} for a detailed introduction to computational topology.

\subsection{Input data}

\revision{We consider}
an ensemble of $\ensembleSize$ piecewise linear (PL) scalar fields $f_i : \revisionTVCGminor{\domain} \rightarrow \range$, with $i \in \{1, \dots,  \ensembleSize\}$, 
defined on a 
PL \mbox{$d$-manifold} 
\revisionTVCGminor{$\domain$ triangulated by a finite}
simplicial complex 
\revisionTVCGminor{$\complex$}
with $d \leq 3$ in our applications (\revision{\revisionTVCGminor{such as}} triangulated 2D images or 3D volumes).
\revisionTVCGminor{For simplicity, we identify $\domain$ with the underlying space $|\complex|$.}
Each $f_i$ is \revisionTVCGminor{specified} on \revisionTVCGminor{the} vertices \revisionTVCGminor{of $\complex$} and \revisionTVCGminor{extended to} the simplices of higher dimension \revisionTVCGminor{using barycentric interpolation}.
A topological representation can be computed for each ensemble member $f_i$ to summarize its features of interest in a compact manner.
We focus in this work on the extremum \emph{Persistence Diagram} (PD), noted $\diagram(f_i)$, as well as a dual representation of the \emph{Merge Tree} (MT), called the \emph{Branch Decomposition Tree} (BDT)\revisionTVCGminor{\mbox{\cite{pascucci_mr04}}}, noted $\branchtree(f_i)$.

\subsection{Persistence diagrams}
\label{sec_pd}
\revisionTVCGminor{\mbox{We refer in this work to the 0th} topological features of a} PL scalar field $f : \revisionTVCGminor{\domain} \rightarrow \range$ as\revisionTVCGminor{, intuitively, the pits or peaks} of $\revisionTVCGminor{f}$.
\revisionTVCGminor{Their specific construction is described in the following paragraphs and their corresponding regions in $\domain$ are formalized in \autoref{sec_regionAwareBDTs}.}
\revisionTVCGminor{Each feature is characterized by a pair of critical points of $f$, a minimum $c$ and a saddle $c'$ (or a saddle $c$ and a maximum $c'$), and is} represented as a 2D point $\big(f(c), f(c')\big)$.
All \revisionTVCGminor{0th} topological features of $f$ \revisionTVCGminor{are} then summarized as a set of 2D points called the extremum persistence diagram $\diagram(f)$.

\revisionTVCGminor{More specifically,} the topological features involving \revisionTVCGminor{the} minima
of $f$ can be uniquely identified \revisionTVCGminor{by tracking the evolution of} its sub-level set \mbox{$\sublevelset{{f}}(\isovalue)=\{p \in \revisionTVCGminor{\domain} \mid f(p) \leq \isovalue\}$}, defined as \revisionTVCGminor{\mbox{$f^{-1}_{-\infty}(w) \coloneqq f^{-1}\big((-\infty,w]\big)$}}, the inverse of $(-\infty, \isovalue]$ by $f$, i.e. all the points of $\revisionTVCGminor{\domain}$ 
\revisionTVCGminor{at which}
$f$ \revisionTVCGminor{is} below or equal to $\isovalue$.
When $\isovalue$ is increased from the lowest values to the highest ones, topological changes, such as the appearance or the merge of connected components, occur when $\sublevelset{{f}}(\isovalue)$ reaches specific vertices of $\complex$ called the critical points of $f$ \cite{banchoff70}.
For instance, for a PL $d$-manifold $\revisionTVCGminor{\domain}$ with $d \gt 1$, connected components appear at minima and merge at saddles.
For such merges, the Elder rule \cite{edelsbrunner09} states that the \emph{younger} component (created last) \emph{dies} in favor of the \emph{older} one.
The dying component can then be associated with a unique pair of critical points, the minimum $c$ at which it appeared and the saddle $c'$ at which it merges with another component.
Their respective value $f(c)$ and $f(c')$, with $f(c) \lt f(c')$, are respectively called its \emph{birth} and its \emph{death}.
The 2D point $\big(f(c), f(c')\big)$ summarizes this topological feature and is called a minimum-saddle persistence pair.
Its \emph{persistence} is given by $f(c') - f(c)$ and is often seen as a measure of importance of the given feature.
Note that the first connected component that appeared, at the global minimum, never dies since it never merges with another one, in practice its death value is cropped to the global maximum.
This process can be \revisionTVCGminor{similarly defined} for saddle-maximum pairs using the super-level set $\superlevelset{{f}}(\isovalue)=\{p \in \revisionTVCGminor{\domain} \mid f(p) \geq \isovalue\}$.

\revisionTVCGminor{By considering} $\complex$ \revision{as} a simplicial complex (\revision{for example}, the \revisionTVCG{Freudenthal} triangulation of a regular grid \cite{freudenthal42, kuhn60}) and $f$ to be injective on its vertices (easily achieved \revisionTVCGminor{in practice} with symbolic perturbations \cite{edelsbrunner90}), this sub-level set process is done through the so-called \emph{lower star filtration} \cite{edelsbrunner09, guillou_tvcg23} which intuitively processes each vertex one at a time in function order (i.e. in the increasing order of its function value).


A persistence pair is often visualized with a segment connecting the values of its two critical points in the $y$-axis, whose length 
corresponds to its persistence (\autoref{fig_tas} top insets).
\revisionTVCGminor{The set of persistence pairs of a PL scalar field $f$ is called the persistence diagram $\diagram(f)$. It}
provides a visual summary of \revisionTVCGminor{the} features of interest \revisionTVCGminor{of $f$}, where salient features stand out from the diagonal (\autoref{fig_tas}b top inset, colored pairs) while pairs likely corresponding to noise are 
near the diagonal (\autoref{fig_tas}b top inset, white pairs).
The diagonal in this 2D birth/death space consists of features with zero persistence, i.e. non-existent structures of the data. 
\revisionTVCGminor{Formally, $\diagram(f)$ also consists of}
an infinite number of points \revisionTVCGminor{on} the diagonal with infinite multiplicity.

\subsection{Merge trees}
\label{sec_mt}

In addition to encoding the presence and range of the topological features of $f$, as persistence diagrams do, merge trees also capture their connectivity.
The minimum-saddle pairs of $f$ are captured by its \emph{join} tree while the saddle-maximum pairs by its \emph{split} tree (\autoref{fig_tas}, center insets).
Each of these two trees is called a \emph{merge tree}, noted $\mergetree(f)$.
The nodes of such a tree are critical points of $f$, the root of a join tree is the global maximum and its leaves are all the local minima, while the split tree has the global minimum as a root and all local maxima as leaves.
For a PL $d$-manifold $\revisionTVCGminor{\domain}$ with $d \gt 1$, the inner nodes of a merge tree are the saddle critical points of the persistence pairs it represents.
The edges of the join tree inform about which connected components of the sub-level set $\sublevelset{{f}}(\isovalue)$ merged with each other as $\isovalue$ increases (similarly for the split tree with the super-level set $\superlevelset{{f}}(w)$ as $\isovalue$ decreases).

The join tree $\jointree(f)$ is defined in a continuous manner as the quotient space $\jointree\revisionTVCG{(f)} = \revisionTVCGminor{\domain} / \revisionTVCG{\sim_f}$ by the equivalence relation $\revisionTVCG{\sim_f}$ stating that two points $p_1$ and $p_2$ in $\revisionTVCGminor{\domain}$ are equivalent if $f(p_1) = f(p_2)$ and if $p_1$ and $p_2$ belong to the same connected component of the sub-level set $\sublevelset{{f}}\big(f(p_1)\big)$.
The split tree $\splittree(f)$ is defined in a similar manner using the super-level set $\superlevelset{{f}}$.
Their discretization corresponds to the objects that were described in the previous paragraph.

Merge trees are often visualized using a persistence-driven \emph{branch decomposition} \cite{pascucci_mr04} \revision{representing} the persistence pairs captured by these trees.
A \emph{persistent branch} is a monotone path on the tree connecting the two nodes of a same persistence pair. 
Then, each persistent branch is represented as a vertical segment and their ancestor relations with a horizontal one (\autoref{fig_tas}, center insets).
The persistence-driven \emph{branch decomposition tree} (BDT), noted $\branchtree$, is a dual representation of the merge tree where each persistent branch is collapsed to a node and its edges encode their hierarchy (\autoref{fig_tas}, bottom insets).
Specifically, if one of the two nodes of a persistent branch $b_i$ is itself on the path of another persistent branch $b_j$ (the other node of $b_i$ being a leaf), then $b_i$ will be \revision{a child} of $b_j$ in $\branchtree$.
Each node of a BDT corresponds therefore to a persistence pair and is associated with its birth and death values.

To tackle instabilities called \emph{saddle swaps}, merge trees can be pre-processed \cite{SridharamurthyM20} to merge adjacent interior nodes (being saddle points if $d \gt 1$) if the difference in their function value is below a threshold determined by the input parameter \mbox{$\epsilon_1 \in [0, 1]$}.
\revisionTVCG{Specifically, in practice, two interior nodes $v_i$ and its parent $v_j$ in a merge tree are \emph{merged} if $|f(v_i) - f(v_j)| \leq \epsilon_1 \Delta_f$ with $\Delta_f = \max f - \min f$, i.e. the node $v_i$ is deleted and its children become children of $v_j$.}
This has no effect when $\epsilon_1 = 0$, and as $\epsilon_1$ increases, more nodes are merged.
In the BDT, this parameter has the effect to bring up pairs in the tree hierarchy.
\revisionTVCG{Moreover, in practice, a node in a BDT corresponding to a persistent branch $b_i$ is still attached with its original birth and death values even if one of its two nodes is merged due to $\epsilon_1$.}

\subsection{Wasserstein metric space}
\label{sec_wassersteinMetric}

We 
\revision{now define} the Wasserstein distance between persistence diagrams \cite{edelsbrunner09} 
\revision{and} its generalization to merge trees \cite{pont_vis21}.

The Wasserstein distance between two persistence diagrams seeks an assignment between their points.
A point of a diagram can either be assigned to \revision{one} point\revision{, and exactly one}, in the other diagram, representing two features that are matched together, or to its projection on the diagonal of the other diagram, intuitively representing the deletion or insertion of a feature.
Two points 
$p_i = (x_i, y_i) \in \diagram(f_i)$ and $p_j = (x_j, y_j) \in \diagram(f_j)$, 
can be assigned to each other with a cost corresponding to their distance in the birth/death space, called the \emph{ground metric}:

\begin{equation}
\label{eq_groundMetric}
\pointMetric_{q}(p_i, p_j) = \big(|x_i - x_j|^q + |y_i - y_j|^q\big)^{1/q} = \| p_i - p_j \|_q,
\end{equation}

\noindent
with \revision{$q \geq 1$}.
In the 
case where both $p_i$ and $p_j$ are 
located on the diagonal (i.e. $x_i = y_i$ and $x_j = y_j$), $\pointMetric_q(p_i, p_j)$ is set to zero (such that these dummy features do not contribute in the distance).

\revisionTVCG{Let \revisionTVCGminor{$\phi$ be} a partial mapping between a subset of the off-diagonal points in $\diagram(f_i)$ and 
one in $\diagram(f_j)$, let $P_i$ be the 
points in $\diagram(f_i)$ mapped by $\phi$ to 
points in $\diagram(f_j)$, let $\overline{P_i}$ 
and $\overline{P_j}$ \revisionTVCGminor{be} the off-diagonal points in respectively $\diagram(f_i)$ and $\diagram(f_j)$ left unmapped by $\phi$, 
considered to be mapped to the diagonal.
}

\revisionTVCG{Following the definition in \cite{pont_vis21}, simplifying the transition to the case of merge trees but also to our method (\autoref{sec_regionWassersteinDistance}), the} $L^q$-Wasserstein distance $\wasserstein{q}$ 
\revisionTVCG{can be}
defined as:

\revisionTVCG{
\begin{eqnarray}
\label{eq_wasserstein}
\wasserstein{q}\big(\diagram(f_i), \diagram(f_j)\big)
= 
& \hspace{-0.3cm} \underset{\phi \in \Phi}\min \hspace{-0.2cm} &
\Big(
\sum_{p_i \in P_i} \pointMetric_q\big(p_i, \phi(p_i)\big)^q \\
\label{eq_wasserstein_destruction}
& + & \sum_{p_i \in \overline{P_i}} \pointMetric_q\big(p_i, \projection(p_i)\big)^q \\
\label{eq_wasserstein_creation}
& + & \sum_{p_j \in \overline{P_j}} \pointMetric_q\big(\projection(p_j), p_j\big)^q
\Big)^{1/q}
\end{eqnarray}
}

\noindent
where $\Phi$ is the set of all possible assignments $\phi$ mapping 
\revisionTVCG{a} 
point 
$p_i \in \diagram(f_i)$ 
to \revision{exactly one} point 
$p_j \in \diagram(f_j)$.
\revisionTVCG{Lines \ref{eq_wasserstein_destruction} and \ref{eq_wasserstein_creation}, \revisionTVCGminor{in which} $\projection(p_i) = (\frac{x_i+y_i}{2},\frac{x_i+y_i}{2})$ is the diagonal projection of an off-diagonal point $p_i = (x_i, y_i)$\revisionTVCGminor{,} indicate the deletion or the appearance of a feature, respectively.}
\revisionTVCG{$\wasserstein{q}$} 
can be 
\revisionTVCG{computed}
using traditional assignment problem solvers \cite{Bertsekas81, Kerber2016, Munkres1957}.
\revisionTVCG{The Bottleneck distance \cite{edelsbrunner02} is recovered when $q \rightarrow \infty$, in which case only the maximum matching cost is retained.}

\begin{figure}
    \centering
    \includegraphics[width=\linewidth]{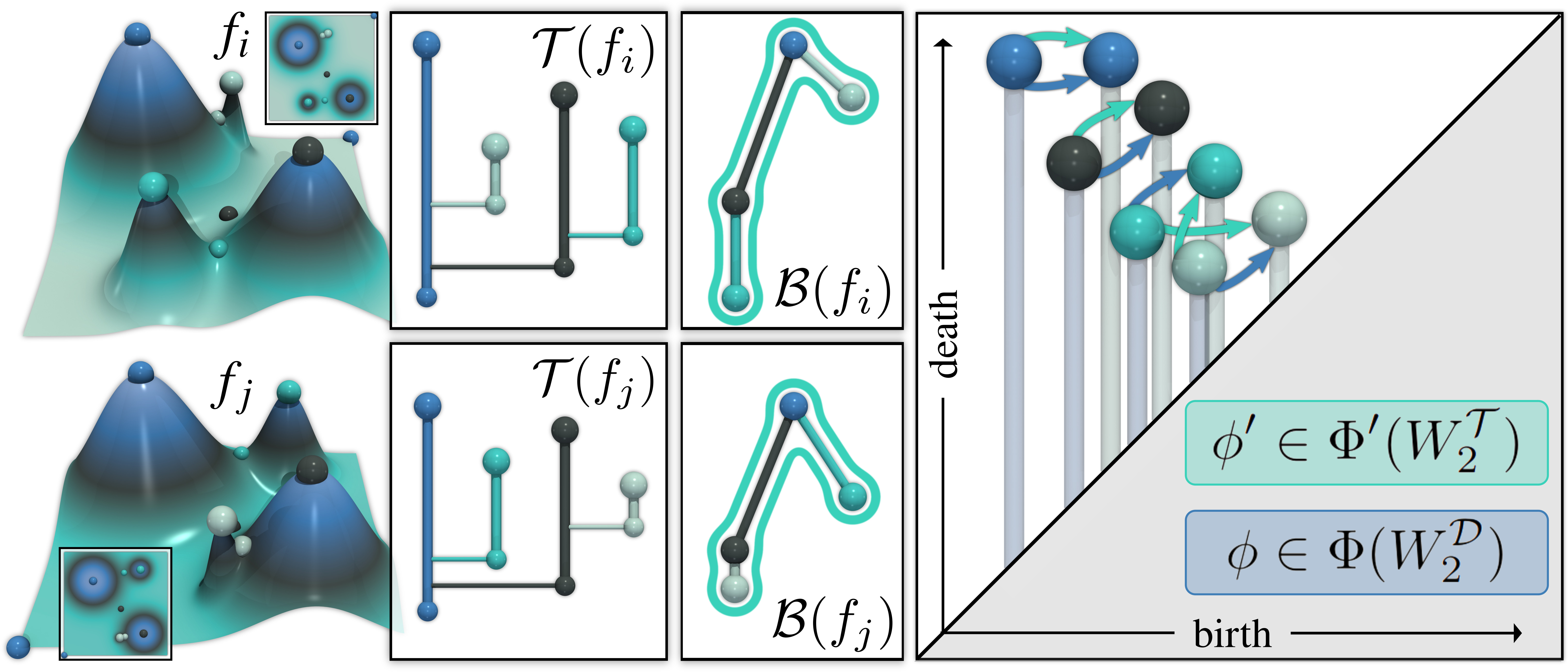}
    \caption{
The Wasserstein distance $\wassersteinTree$ between two BDTs $\branchtree(f_i)$ and $\branchtree(f_j)$ optimizes an assignment between their pairs in the birth/death space (green arrows, right) with a search space constrained by the structure of the trees.
It amounts to search for a common contiguous structure between the BDTS (cyan halo).
When $\epsilon_1 = 1$, the tree structures do not constrain the search space anymore and $\wassersteinTree$ get back to the Wasserstein distance between persistence diagrams $\wasserstein{2}$ (black arrows, right).
}
    \label{fig_wassersteinMetric}
\end{figure}

\revisionTVCG{For merge trees, a classical alternative is the interleaving distance \cite{morozov14}, \revisionTVCGminor{providing} a stable notion of comparison based on structure-preserving maps between trees, being however challenging to compute in practice. To tackle this,} The Wasserstein distance between persistence diagrams was later generalized to BDTs \cite{pont_vis21}.
The expression of this distance, noted $\wassersteinTree\big(\branchtree(f_i), \branchtree(f_j)\big)$,  \revisionTVCG{\revisionTVCGminor{corresponds} to} \revisionTVCGminor{Eqs.~\ref{eq_wasserstein}--\ref{eq_wasserstein_creation}} with $q = 2$ \revisionTVCG{and $P_i$, $\overline{P_i}$ and $\overline{P_j}$ being sets of BDT nodes instead of persistence diagram points}.
The important difference relies on its possibly smaller search space of assignments, noted $\Phi' \subseteq \Phi$, that can 
\revisionTVCG{constrain the matchings using}
the structure of the trees.

The search space $\Phi'$ between $\branchtree(f_i)$ and $\branchtree(f_j)$ \revisionTVCG{is} the set of \revisionTVCG{their} rooted partial isomorphisms \cite{pont_vis21}.
\revisionTVCG{Formally, let $V_i$, $E_i$ and $r_i$ respectively be the nodes, edges and root of $\branchtree(f_i)$. 
Let $U_i$ be a subset of $V_i$ such that $r_i \in U_i$ and for every $u \in U_i$ all its ancestors also belong to $U_i$.
The rooted induced subtree $\branchtree(f_i)[U_i]$ is a connected rooted subtree of $\branchtree(f_i)$ with $U_i$ as nodes and the edges of $E_i$ connecting them, noted 
\mbox{$E_i[U_i] = E_i \cap (U_i \times U_i)$}, 
similarly for $\branchtree(f_j)$.
A mapping $\phi'$ belongs to $\Phi'$ if \emph{1)} $\phi' : U_i \rightarrow U_j$,
\emph{2)} $\phi'$ is a bijection, \emph{3)}~$\phi'(r_i) = r_j$ and \emph{4)} $\phi'$ is an isomorphism between $\branchtree(f_i)[U_i]$ and $\branchtree(f_j)[U_j]$, i.e $(u, v) \in E_i[U_i] \Leftrightarrow (\phi'(u), \phi'(v)) \in E_j[U_j]$.}
\revisionTVCGminor{In the expression of the distance (Eqs.~\ref{eq_wasserstein}--\ref{eq_wasserstein_creation}), $\Phi$ is replaced by $\Phi'$, $P_i$ by $U_i$, $\overline{P_i}$ by $V_i \setminus U_i$ and $\overline{P_j}$ by $V_j \setminus U_j$.}

\revisionTVCG{Intuitively, the matching is restricted to a} 
common contiguous subtree of \revisionTVCG{$\branchtree(f_i)$ and $\branchtree(f_j)$} including their root.
The 
assignment $\phi'$ 
\revisionTVCG{minimizing}
the distance (cyan halo on the BDTs of \autoref{fig_wassersteinMetric}) \revisionTVCG{is computed}
by recursively solving local
assignment problems in the tree hierarchies \revisionTVCG{with dynamic programming}.

\revisionTVCG{While being defined as an edit distance between BDTs}, it \revisionTVCG{generalizes} the Wasserstein distance between persistence diagrams thanks to the parameter $\epsilon_1$ (end of \autoref{sec_mt})\revisionTVCG{, used in a pre-processing step, controlling} how much of the BDT structures 
\revisionTVCG{will}
be taken into account in the distance. 
When $\epsilon_1 = 1$, all saddles in a merge tree are collapsed \revisionTVCG{resulting} in all the nodes of its BDT to be moved up in the hierarchy and \revisionTVCG{becoming} children of the root.
In that case, the various local and small assignment problems become one large and we have $\wassersteinTree\big(\branchtree(f_i), \branchtree(f_j)\big) = \wasserstein{2}\big(\diagram(f_i), \diagram(f_j)\big)$. 
This parameter allows to choose a trade-off between the stability of persistence diagrams and the discriminative power of merge trees.
 
\section{\revision{Region-aware Wasserstein distances}}

\begin{figure}[b]
    \centering
    \includegraphics[width=\linewidth]{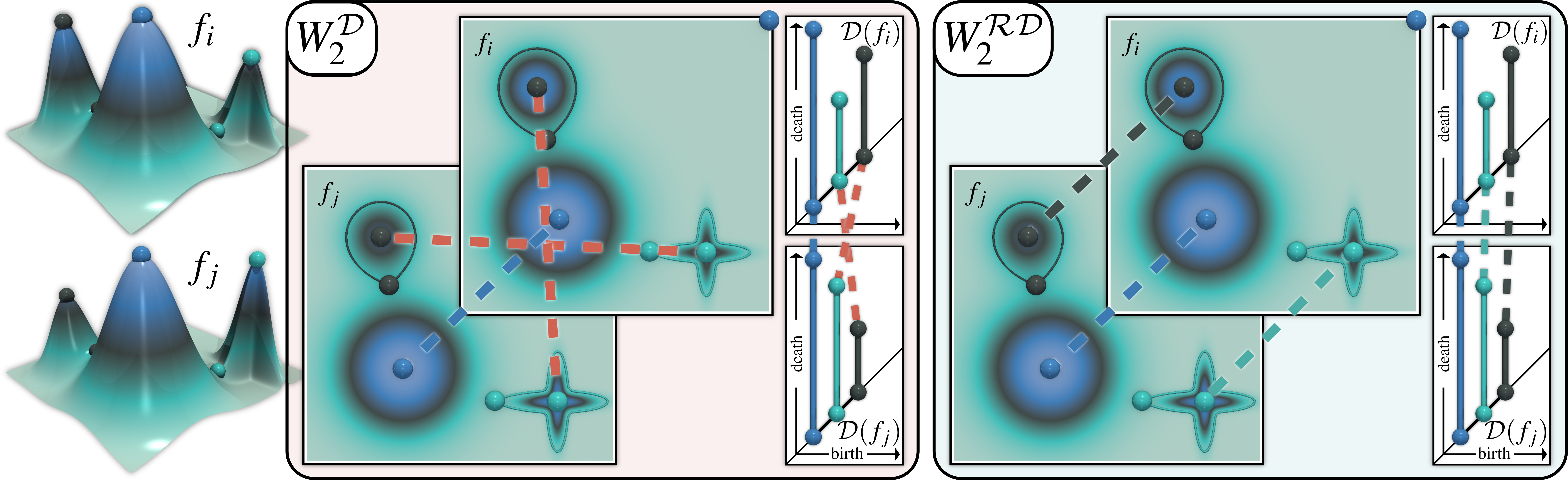}
    \caption{\revision{The Wasserstein distance $\wasserstein{2}$ matches topological features having overall close birth and death values.
In this example, the geometry of the features with black and green spheres are significantly different (circle shape and star shape).
For these two datasets, the death \revisionTVCGminor{values} of these features have been swapped resulting in their incorrect matching according \revisionTVCGminor{to} $\wasserstein{2}$ (red dashes).
In contrast, our metric $\regionWasserstein{2}$ takes into account their geometry through their regions in the input domain and \revisionTVCGminor{provides} a correct matching.
}}
    \label{fig_matching}
\end{figure}

This section introduces our \revision{generalization of the Wasserstein distance for} extremum \revisionTVCG{(0th)} persistence diagrams \revisionTVCG{(i.e. focusing on saddle-extrema persistence pairs)} and merge trees \cite{edelsbrunner09, pont_vis21}\revision{. It extends them by describing features not} only \revision{with} their minimum and maximum (i.e. their birth and their death) \revision{but with a vertex-level characterization of their regions in the input domain.}
An input parameter \revision{controls the impact}
of the regions' \revision{characterization in the distance,} 
\revision{making possible to not use}
them at all, getting back to the original Wasserstein distances 
\revisionTVCG{making them}
special cases of our method.

In the following, since persistence diagrams and merge trees can be represented as BDTs we will use this term to represent both objects.
We first formulate how persistence pairs can be augmented with their respective regions to define region-aware BDTs.
Then, we \revision{introduce} a new ground metric between region-aware persistence pairs, being at the core of our new distance \revision{for which we show theoretical properties (metric, stable and discriminative)}.
In \autoref{sec_algo} we present our \revision{algorithm for its computation} as well as two strategies to control the computation time and the memory footprint of the method.

\revision{\subsection{Motivation}

As shown in \autoref{fig_geom}, persistence diagrams and merge trees can lack in specificity, where two significantly different datasets, from a geometric point of view, have the same topological representation. 
In that case, their Wasserstein distance, being $0$, prevents to discriminate them.

Another failure case of current methods is illustrated in \autoref{fig_matching} showing an incorrect matching of the Wasserstein distance.
Two features being significantly different from a geometric point of view (the black and green features have respectively a circle and a star shape, \autoref{fig_matching}) will be incorrectly matched together by the Wasserstein distance (red dashes, \autoref{fig_matching}) because they have similar birth and death values.

Overall, this motivates the use of additional information on top of existing topological representations and the development of a metric taking that information into account to allow for the discrimination of geometrically different datasets.
For instance, in \autoref{fig_matching}, our novel distance metric uses a vertex-level characterization of the geometry of the features, that goes beyond the simple comparison of birth and death values, and therefore provides the correct matching.
}

\begin{figure}[t]
    \centering
    \includegraphics[width=\linewidth]{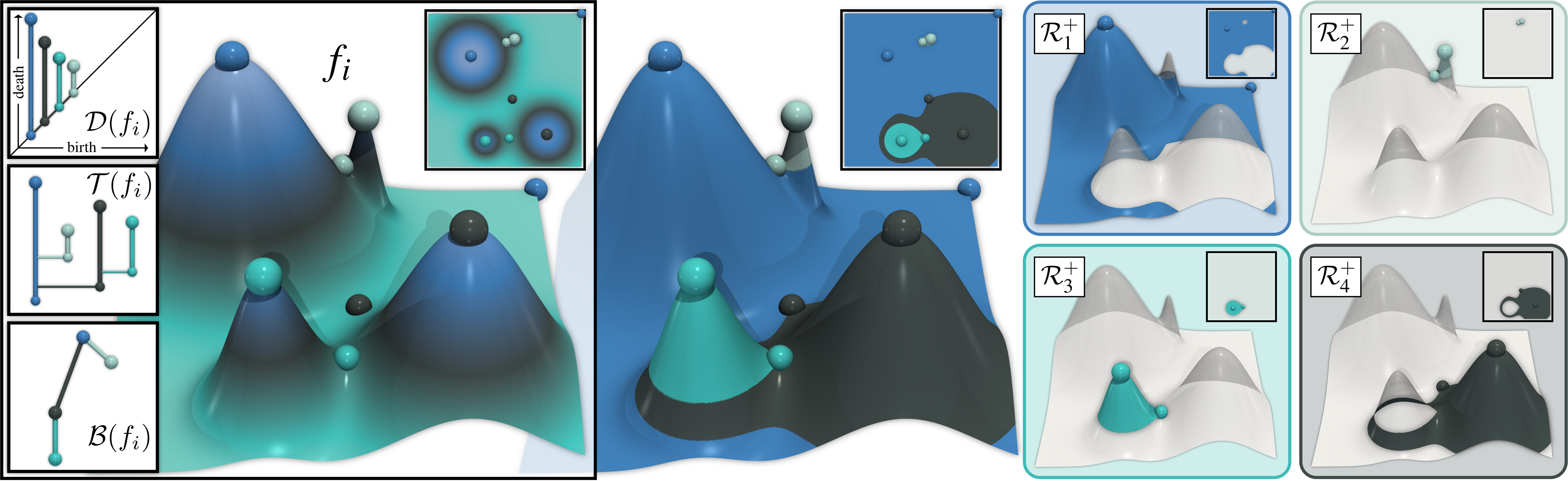}
    \caption{
\revisionTVCG{
The computation of the persistence diagram and the merge tree uniquely segments the domain of the function $f_i$ into regions, each 
\revisionTVCGminor{being}
a topological feature.
Each maximum region $\region_i^+$ with $i \in \{1, \dots, 4\}$ is a subset of the domain and contains a unique maximum and values \revisionTVCGminor{monotonically} decreasing outward from it.
The color code of the regions is the same \revisionTVCGminor{as} for the saddle-maximum persistence pairs in the topological representations.
}
}
    \label{fig_region}
\end{figure}

\subsection{Region-aware branch decomposition trees}
\label{sec_regionAwareBDTs}

We first formalize how 
persistence diagrams and merge trees allow to segment the domain into regions, each associated to a persistence pair.
Since we work on extremum persistence diagrams and merge trees, we will focus the explanation on minimum-saddle and saddle-maximum persistence pairs.
Then, we \revision{define the \emph{region-aware} persistence pairs, an extension of persistence pairs with a vertex-level characterization of their region, and use them to define} region-aware BDTs.

\revision{
Given
$f : \revisionTVCGminor{\domain} \rightarrow \range$, \revisionTVCG{let $\pi_f : \revisionTVCGminor{\domain} \rightarrow \mergetree(f) = \revisionTVCGminor{\domain} / \sim_f$ be the quotient map associated with the equivalence relation used to define the merge tree (\autoref{sec_mt}).}
Each
\revisionTVCG{node} $b_i$ \revisionTVCG{of the BDT} $\branchtree(f)$ \revisionTVCG{corresponds to a unique persistent branch $\gamma_i \subseteq \mergetree(f)$ (and vice versa) \revisionTVCGminor{and} these persistent branches form a partition of $\mergetree(f)$ (by taking them half-open at saddles).}
\revisionTVCGminor{The continuous}
region $\revisionTVCGminor{\continuousRegion}_i$ \revisionTVCG{associated with $b_i$ \revisionTVCGminor{is} the preimage of $\gamma_i$ under $\pi_f$: 
} 

\begin{equation}
  \nonumber
  \revisionTVCGminor{\continuousRegion}_i = \pi^{-1}_f(\revisionTVCG{\gamma_i}) = \{p \in \revisionTVCGminor{\domain} \mid \pi_f(p) \revisionTVCG{\in \gamma_i}\}.
\end{equation}

\revisionTVCG{Since} 
$\pi_f$ is \revisionTVCG{defined on all of \revisionTVCGminor{$\domain$} and the persistent branches $\{\gamma_i\}_i$ form a partition of $\mergetree(f)$,}
\revisionTVCG{then}
$\{\revisionTVCGminor{\continuousRegion}_i\}_i$ form a partition of the domain: $\bigcup_i \revisionTVCGminor{\continuousRegion}_i = \revisionTVCGminor{\domain}$ and $\revisionTVCGminor{\continuousRegion}_{i_1} \cap \revisionTVCGminor{\continuousRegion}_{i_2} = \emptyset$ for ${i_1} \neq {i_2}$.
\revisionTVCGminor{We define the vertex-level representation of such a continuous region as $\region_i = V(\complex) \cap \continuousRegion_i$ where $V(\complex)$ is the vertex set of $\complex$.}
Each 
$\region_i$ contains precisely one extremum \revisionTVCGminor{vertex}.
\revisionTVCG{It} is a \emph{maximum region}
$\region_i^+$
if $b_i$ is a saddle-maximum persistence pair (\autoref{fig_region}) or a \emph{minimum region}
$\region_i^-$ if $b_i$ is 
a minimum-saddle 
pair.
}

Given a persistence pair $b_i$ of the BDT $\branchtree(f)$, associated with a pair of values $(x_i, y_i)$ (its birth and its death), let $s_i$ the value of its saddle critical point (i.e. either its death $y_i$ if it is a minimum-saddle pair or its birth $x_i$ if it is a saddle-maximum one), and $e_i$ its \revision{value at the extremum} (the other value than $s_i$), we define its region-aware \revision{generalization} \revisionTVCG{as}:

\begin{equation}
  \nonumber
  b^\region_i = (\revisionTVCG{\region_i^f}, s_i),
\end{equation}

\noindent
where 
\revisionTVCG{$\region_i^f$} corresponds to the \revisionTVCG{region $\region_i$ \revisionTVCGminor{along with}} the function $f$. 
Note that by definition, the \revision{value at the extremum} $e_i$ of $b_i$ \revision{lies in the image of} $f_{|\region_i}$ \revisionTVCG{(i.e. the function $f$ restricted to the domain of $\region_i$)}, since its corresponding vertex belongs to $\region_i$, but not its saddle value $s_i$.
\revision{This generalization captures the feature's spatial support $\region_i$ with its values 
\revisionTVCG{as well as}
the saddle value $s_i$, providing a strictly richer characterization than the classical birth and death pair.}
Then, the region-aware BDT $\branchtree^\region(f)$ is defined as the BDT $\branchtree(f)$ where each of its pair $b_i$ is replaced by its region-aware \revision{generalization} $b^\region_i$.

\subsection{\revision{Definition}}
\label{sec_regionWassersteinDistance}

We
\revisionTVCG{now}
aim to define a distance that leverages 
\revisionTVCG{the vertex-level characterization of the regions associated with BDT features.}
\revision{For this, we generalize the original Wasserstein distances \cite{edelsbrunner09, pont_vis21} (\autoref{sec_wassersteinMetric}) to make them able to use such geometric properties 
\revisionTVCG{and}
overcome 
\revisionTVCG{the}
limitations regarding their lack of geometric considerations (see 
\autoref{fig_geom} and \autoref{fig_matching})}.

We start by defining a novel ground metric allowing to compare region-aware persistence pairs.
For this, we \revision{generalize} the \revision{classical} ground metric used for persistence diagrams and merge trees (\autoref{eq_groundMetric}) to use up to all the region values of such persistence pairs 
rather than just their minimum and maximum.
\revision{We then define the region-aware Wasserstein distance for which we show theoretical properties (metric, stable and discriminative).}
We \revision{additionally propose a parameter adjusting} the influence of the region properties on the distance\revision{, allowing} to ignore them entirely, reverting to the \revision{classical case (\autoref{sec_generalization})}. 
One benefit of this approach is its adaptability to different applications, depending on the relevance of region information.
\revision{Moreover, as} we will see in \autoref{sec_subsampling}, it also allows for controlling the computation time of the method.

\revisionTVCG{
\subsubsection*{\textbf{Continuous intuition}}
Given a smooth Morse function $f : \revisionTVCGminor{\domain} \rightarrow \range$, each extremum-saddle persistence pair induces a region $\revisionTVCGminor{\continuousRegion}_i \revisionTVCGminor{\subseteq} \revisionTVCGminor{\domain}$. 
A region-aware persistence pair is enriched with the restricted scalar field on its region in addition to the associated saddle value (\autoref{sec_regionAwareBDTs}).
Given two such features, we align their regions at their extrema and compare their scalar values over the aligned domains as well as their saddle values (\autoref{fig_region_distance}).
In the smooth setting, this amounts to an $L_q$-type discrepancy obtained by integrating the difference of the aligned scalar fields over their overlap, together with a penalty on the non-overlapping parts.
In the PL setting considered in this work, this continuous regional discrepancy is discretized as a sum over the sampled vertices of the aligned regions.

For $q = 1$, this corresponds to the accumulated absolute difference between the aligned scalar fields over their common support, 
while larger $q$ increasingly emphasize larger local 
mismatches.
In this sense, the geometry is captured implicitly through the spatial organization of scalar values over the region, rather than through explicit shape attributes.


}



\subsubsection{\revision{\textbf{Region-aware ground metric}}}
\label{sec_regionAwareGroundMetric}

\revision{
Let two BDTs $\branchtree(f_1)$ and $\branchtree(f_2)$ and two persistence pairs
$b_i = (e_i, s_i)$ \revisionTVCG{from} $\branchtree(f_1)$ and $b_j = (e_j, s_j)$ \revisionTVCG{from} $\branchtree(f_2)$.
The classical ground metric (\autoref{eq_groundMetric}) 
\revisionTVCG{is}
a two-term $L_q$ norm, one comparing the saddle values ($s_i$ and $s_j$) and the other the values of the extrema ($e_i$ and $e_j$).
The region-aware generalizations of $b_i$ and $b_j$, i.e. 
$b^\region_{i} = (\revisionTVCG{\region_{i}^{f_1}}, s_{i})$ and $b^\region_{j} = (\revisionTVCG{\region_{j}^{f_2}}, s_{j})$,
replace the values at their respective extrema ($e_i$ and $e_j$) by a vertex-level characterization of their regions 
(\revisionTVCG{$\region_{i}^{f_1}$} and \revisionTVCG{$\region_{j}^{f_2}$}).
Recall that no information is lost since $e_i \in \operatorname{Im}(f_{1|\region_{i}})$ and $e_j \in \operatorname{Im}(f_{2|\region_{j}})$ by definition.
Accordingly, we generalize the classical ground metric to region-aware persistence pairs by replacing the term comparing the values at their extremum by one taking into account the characterization of the regions.
We call this new term the \emph{regional discrepancy}, noted $\regionalDiscrepancy_q^\region$, and define the region-aware ground metric as:
}

\begin{figure}[t]
    \centering
    \includegraphics[width=\linewidth]{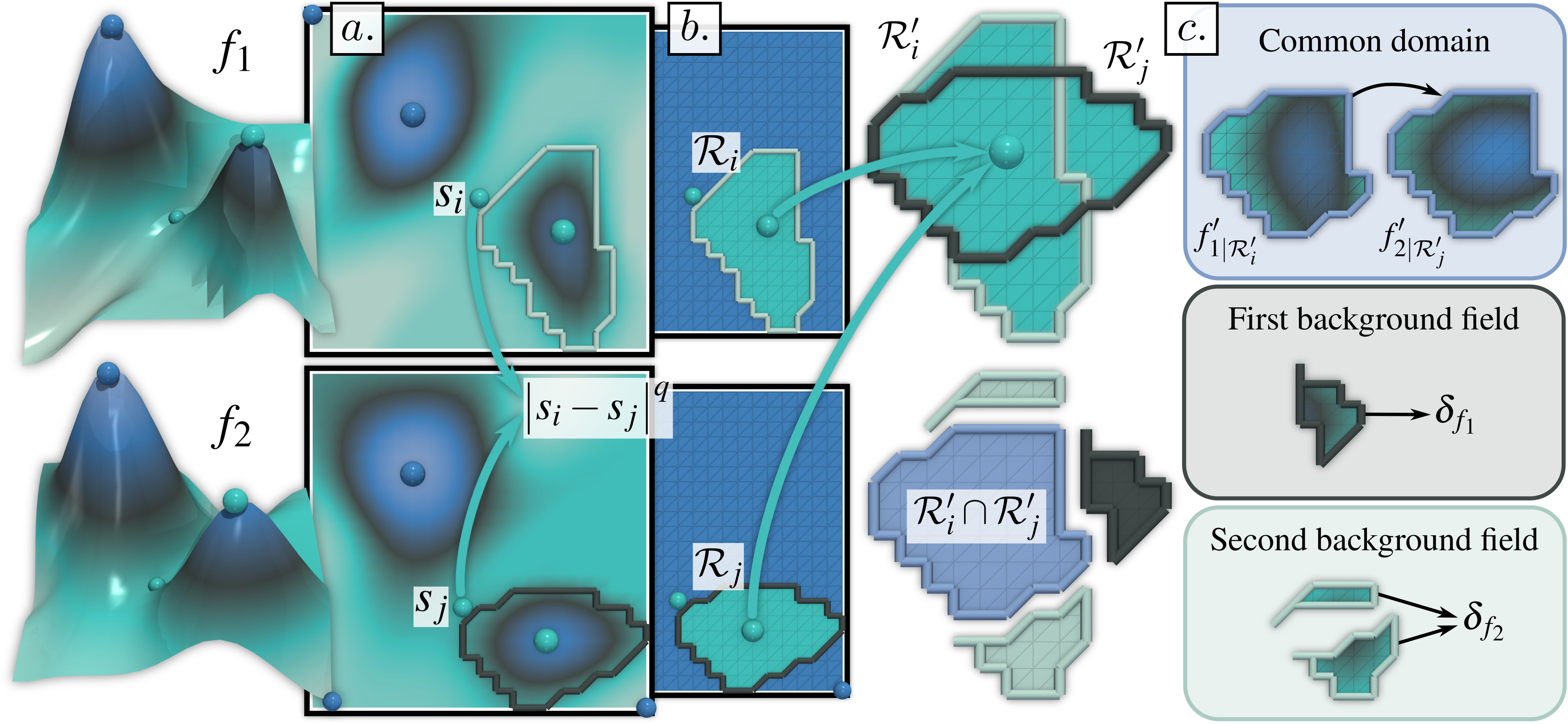}
    \caption{\revision{
\revisionTVCG{
The computation of the region-aware ground metric $d_q^\region$ consists in the comparison of the saddle values of two persistence pairs $(a)$, then, their regions $\region_i$ and $\region_j$ are aligned at their extrema $(b)$ and a distance is computed on their \revision{common} domain $\region'_i \cap \region'_j$ $(c)$. 
An \revision{additional} cost is used for the points \revision{outside} the \revision{common} domain (i.e. in $\region'_i \setminus \region'_j$ and $\region'_j \setminus \region'_i$) \revision{using the background fields $\projectionRegion_{f_1}$ and $\projectionRegion_{f_2}$ to make these comparisons well-defined}.
}}}
    \label{fig_region_distance}
\end{figure}

\begin{equation}
\label{eq_our_groundMetric}
  d_q^\region(b^\region_i, b^\region_j) = \big(|s_i - s_j|^q + \revision{\regionalDiscrepancy_q^\region(b^\region_i, b^\region_j)}\big)^{1/q}.
\end{equation}

\revision{If the regional discrepancy $\regionalDiscrepancy_q^\region$ is defined such that only the values of the extremum of $b^\region_i$ and $b^\region_j$ are compared (i.e. $\regionalDiscrepancy_q^\region(b^\region_i, b^\region_j) = |e_i - e_j|^q$) then $d_q^\region$ reduces to the original ground metric.
We now define it in order to use up to all the regions' information provided by region-aware persistence pairs, keeping the comparison of only the values of their extrema as a special case thanks to an input parameter (\autoref{sec_generalization}).}


\revision{
To compare the regions while preserving the comparison of the values of their extrema, we first \emph{align} (i.e. center) the regions at their respective extrema so that they coincide (\autoref{fig_region_distance}b), then we compare the values of the centered regions (\autoref{fig_region_distance}c).
We note $\region'_i$ and $\region'_j$ the centered regions and 
\revisionTVCG{$f'_1$ and $f'_2$ the corresponding centered} functions.
The intersection of 
$\region'_i \cap \region'_j$ (\autoref{fig_region_distance}b, violet area) contains at least their extrema.
On this overlap, both functions are defined allowing a pointwise comparison.
For each point outside this overlap, i.e. on the symmetric difference $\region'_i \triangle \region'_j = (\region'_i \setminus \region'_j) \cup (\region'_j \setminus \region'_i)$ (\autoref{fig_region_distance}b, white and black areas), only one region (either one of them depending on the point) provides a value with no counterpart in the other.
To make the comparisons well-defined everywhere, we embed each function
\revisionTVCG{$f'_1$ and $f'_2$}
into the union domain $\region'_i \cup \region'_j$ by extending them with a \emph{background} field, respectively $\projectionRegion_{f_1}$ and $\projectionRegion_{f_2}$, outside their original support.
This allows for points outside the common domain to also contribute.
We therefore define the regional discrepancy as:}

\begin{eqnarray}
\label{eq_our_regionalDiscrepancy}
  \revision{\regionalDiscrepancy_q^\region(b^\region_i, b^\region_j) =}
  &\hspace{-.35cm}
    \label{eq_our_groundMetric_mapping}
    \displaystyle \sum_{v \in \region'_i \cap \region'_j} &\hspace{-.2cm} 
    \big|\revisionTVCG{f'_1}(v) - \revisionTVCG{f'_2}(v)\big|^q\\
    \label{eq_our_groundMetric_deletion}
    &\hspace{-.35cm} + \displaystyle \sum_{x \in \region'_i \setminus \region'_j} &\hspace{-.2cm} 
    \big|\revisionTVCG{f'_1}(x)
     - \revision{\projectionRegion_{f_2}}(x)\big|^q\\
    \label{eq_our_groundMetric_insertion}
    &\hspace{-.35cm} + \displaystyle \sum_{y \in \region'_j \setminus \region'_i} &\hspace{-.2cm} \big|\revision{\projectionRegion_{f_1}}(y) - 
    \revisionTVCG{f'_2}(y)\big|^q,
\end{eqnarray}

\revision{\noindent intuitively measuring the cost of transforming one aligned region to another.
Line \ref{eq_our_regionalDiscrepancy} compares their functions pointwise where both aligned regions coincide.
Line \ref{eq_our_groundMetric_deletion} penalizes the parts of the first region that must be deleted and line \ref{eq_our_groundMetric_insertion} the parts of the second region that must be inserted.
These two lines penalize unmatched support via the background fields $\projectionRegion_{f_1}$ and $\projectionRegion_{f_2}$.
A natural background is the null field, $\projectionRegion_{f}(x) = 0$ for all $x$, which encodes the notion that a feature contributes no signal outside its own region and yields the intended interpretation of lines \ref{eq_our_groundMetric_deletion} and \ref{eq_our_groundMetric_insertion} as \revisionTVCGminor{deletions} and insertions of nonshared points.
To show the stability of our new metric (\autoref{sec_stability}), the background \revisionTVCG{field necessitates to be set} as the data itself, $\projectionRegion_{f}(x) = f(x)$ for all $x$, intuitively comparing the points outside the region overlap to what the other function is at those locations.
When regions degenerate to their extrema, the symmetric difference terms vanish and the overlap term reduces to $|e_i - e_j|^q$, recovering the classical case \revisionTVCG{(\autoref{eq_groundMetric})}.
}


\subsubsection{\revision{\textbf{Region-aware projection}}}
\label{sec_regionAwareProjection}

\revision{Before defining \revisionTVCG{the region-aware} Wasserstein distance, we
\revisionTVCG{now} generalize \revisionTVCG{the diagonal projection, and its cost,} to region-aware persistence pairs 
(representing a deletion or \revisionTVCG{an} insertion).
In the classical case (\autoref{sec_wassersteinMetric}), a persistence pair $b_i = (e_i, s_i)$ can be matched to its diagonal projection $(m_i, m_i)$ with $m_i = \frac{e_i + s_i}{2}$.
For a region-aware persistence pair 
$b^\region_i = (\revisionTVCG{\region_i^f}, s_i)$  
we consider a constant function}
\revisionTVCG{$f_\projection$} 
\revision{that} maps all points of $\region_i$ to the diagonal projection value $\revision{m_i}$ \revision{and define the region-aware projection:}
\begin{equation}
\nonumber
 \projection^\region(b^\region_i) = (\revisionTVCG{\region_i^{f_\projection}}, \revision{m_i}),
\end{equation}
\noindent
\revision{
The cost of matching $b^\region_i$ to the diagonal (i.e. deleting or inserting it) is then $d_q^{\projection}(b^\region_i) = d_q^\region\big(b^\region_i, \projection^\region(b^\region_i)\big)$.
When $\region_i$ collapses to its extremum the cost reduces to the classical point-to-diagonal distance, consistent with the classical metric.

\subsubsection{\textbf{Region-aware Wasserstein distance}}

Based on the definition of the Wasserstein distance between BDTs \cite{pont_vis21} we now define the region-aware Wasserstein distance between two region-aware BDTs $\branchtree^\region(f_1)$ and $\branchtree^\region(f_2)$.
It consists in the search \revisionTVCG{for} a mapping $\phi'$ from the nodes of $\branchtree^\region(f_1)$ to the nodes of $\branchtree^\region(f_2)$ having the lowest total cost measured by the region-aware ground metric $d_q^\region$ (\autoref{eq_our_groundMetric}).
Let $B_1$ \revisionTVCGminor{be} the nodes in $\branchtree^\region(f_1)$ being mapped by $\phi'$, let $\overline{B_1}$ (the complement of $B_1$) and $\overline{B_2}$ \revisionTVCGminor{be} the nodes 
not mapped in respectively $\branchtree^\region(f_1)$ and $\branchtree^\region(f_2)$.
The region-aware Wasserstein distance is defined as:

\begin{eqnarray}
\label{eq_our_distance}
 \hspace{-.3cm} \regionWassersteinTree[q]\big(\branchtree^\region(f_1), \branchtree^\region(f_2)\big) = 
  & \hspace{-.3cm} 
  \underset{\phi' \in \Phi'}\min  
    \Big(
    \label{eq_our_distance_mapping}
    \hspace{-0.25cm} & \sum_{b^\region_i \in B_1} d_q^\region\big(b^\region_i, \phi'(b^\region_i)\big)^q\\
    \label{eq_our_distance_destroying}
    & + & \sum_{b^\region_i \in \overline{B_1}} d_q^{\projection}\big(b^\region_i\big)^q \\
    \label{eq_our_distance_creating}
    & + & \sum_{b^\region_j \in \overline{B_2}} d_q^{\projection}\big(b^\region_j\big)^q
  \Big)^{1/q},
  \label{eq_editDistance}
\end{eqnarray}

}

\revision{\noindent where $\Phi'$ is the search space associated with the 
Wasserstein distance between BDTs (\autoref{sec_wassersteinMetric}).
}
\revisionTVCG{As the latter, $\regionWassersteinTree[q]$ is formulated as an edit distance between BDTs whose summed edit costs 
of lines \ref{eq_our_distance_mapping}, \ref{eq_our_distance_destroying} and \ref{eq_our_distance_creating} account for respectively the matchings, deletions and insertions of nodes in the BDTs.
The search for the optimal $\phi'$ therefore amounts to finding the minimum-cost edit operations under the combinatorial constraints induced by the BDT hierarchies.
}
When $\epsilon_1 = 1$, getting back to the persistence diagrams case, we note it $\regionWasserstein{q}$.

\subsection{\revision{Properties}}

\revisionTVCG{
\begin{theorem}[Metric property]
\label{sec_metric}
\textcolor{black}{$d_q^\region$ is a metric \revisionTVCG{on} the space of region-aware persistence pairs (after centering their regions and extending their centered functions with their background fields). Moreover, \revisionTVCG{$\regionWassersteinTree[q]$} (and therefore \revisionTVCG{$\regionWasserstein{q}$}) is a metric \revisionTVCG{on the space of} region-aware branch decomposition trees.
}
\end{theorem}

\begin{proof}
See \appendixMetricProof{} and \appendixMetricProofAny{}.
\end{proof}
}

We summarize here after the main elements of the proof in an intuitive manner and invite the reader to read \appendixMetricProof{} for more details.
Non-negativity and symmetry are satisfied by construction because, respectively, $d_q^\region$ is the $q$\textsuperscript{th} root of the sum of absolute values raised to a power of $q$ with $q \geq 1$, and deletion and insertion costs (lines \ref{eq_our_groundMetric_deletion} and \ref{eq_our_groundMetric_insertion}) are defined in a symmetric manner.
Identity of indiscernibles follows naturally after centering regions at their extrema and extending their functions with their background fields, pairs are considered identical if and only if their saddle values are \revisionTVCGminor{equal} as well as the extension of their functions on the union of their domain.
For the triangle inequality, we consider an $L_q$ norm on the common domain of three region-aware persistence pairs with their centered functions being extended by their background fields, allowing to use the Minkowski's inequality.

We now study the stability of the region-aware Wasserstein distance for persistence diagrams $\regionWasserstein{q}$ \revisionTVCG{and} show that
it can be bounded \revisionTVCGminor{from} above by the $L_q$ norm of the difference of the input scalar fields up to a constant independent of them.

\revisionTVCG{
\begin{theorem}[Stability property]
\label{sec_stability}
\textcolor{black}{When excluding extremum-swap instabilities (when the extrema of the input functions don't change order)
\revisionTVCG{and using} data background \revisionTVCG{(but not with the null background)}:
$\regionWasserstein{q}\big(\diagram^\region(f), \diagram^\region(g)\big) \leq 2^{1/q}\|f - g\|_q$.}
\end{theorem}
\begin{proof}
See \appendixStability{}.
\end{proof}
}

For this, we take inspiration of the proof by Skraba and Turner \cite{skrabaStability} for the 
Wasserstein distance 
for 
persistence diagrams that we extend to take into account the region properties.

\revisionTVCG{When multi-saddles are considered (not occurring in our experiments), 
using the Freudenthal triangulation of regular grids considered in our work,
the constant $2^{1/q}$ becomes 
$3^{1/q}$ in 2D and $6^{1/q}$ in 3D (see \appendixStabilitySameOrder{} for the general case).}


\revision{We now introduce}
an operator controlling the impact of the \revisionTVCGminor{regions'} properties of the features on the distance.
Given a \revision{region-aware} persistence pair 
$b^\region_{i} = (\revisionTVCG{\region_{i}^f}, s_{i})$, 
we define the operator $\subsampleRegion(b_i^\region, \subsampleParameter)$ with $\subsampleParameter \in [0, 1]$ returning a modified version of $b_i^\region$ where the size of its region decreases as long as $\subsampleParameter$ increases.
Specifically, when $\subsampleParameter = 0$ the region of $b_i^\region$ is not modified.
Then, \revision{the} more $\subsampleParameter$ increases\revision{, the} more points of its region will be removed.
Finally, when $\subsampleParameter = 1$, only the extremum of the region is kept.
More formally, let $\region^\subsampleParameter_i$ \revisionTVCGminor{be} the modified version of $\region_i$ by $\subsampleRegion(b_i^\region, \subsampleParameter)$, then the following properties must be satisfied: \revision{$\region^{\subsampleParameter_1}_i \supseteq \region^{\subsampleParameter_2}_i$} for \revision{$\subsampleParameter_1 \lt \subsampleParameter_2$} and $\region^\subsampleParameter_i = \{\revision{e}\}$ when $\subsampleParameter = 1$ where \revision{$e$} is the extremum $b^\region_{i}$.
We provide an example of such operator in \autoref{sec_subsampling} inspired by subsampling,
computed in a pre-processing step.

\revisionTVCG{
\begin{proposition}[Generalization property]
\label{sec_generalization}
When $\subsampleParameter = 1$, it follows that $\regionWassersteinTree[q]\big(\branchtree^\region(f_1), \branchtree^\region(f_2)\big) = \wassersteinTree[q]\big(\branchtree(f_1), \branchtree(f_2)\big)$ and similarly for $\regionWasserstein{q}$ and $\wasserstein{q}$.
\end{proposition}

\begin{proof}
\textcolor{black}{By definition, when \revision{$\subsampleParameter = 1$} only the extremum remains in a region, therefore $d_q^\region(\subsampleRegion(b^\region_i, \subsampleParameter), \subsampleRegion(b^\region_j, \subsampleParameter))$ will only compare the extrema and saddle values as in the simple case, getting back to the original ground metric of \autoref{eq_groundMetric}.
In that case, we have $\revisionTVCG{\regionWassersteinTree[q]}\big(\branchtree^\region(f_1), \branchtree^\region(f_2)\big) = \revisionTVCG{\wassersteinTree[q]}\big(\branchtree(f_1), \branchtree(f_2)\big)$.}
\revisionTVCG{Taking $\epsilon_1 = 1$ translates this result to $\regionWasserstein{q}$ and $\wasserstein{q}$.}
\end{proof}
}


Using the operator $\subsampleRegion(b_i^\region, \subsampleParameter)$ we now show that the region-aware Wasserstein distance $\revisionTVCG{\regionWassersteinTree[q]}$ is more discriminative than the 
Wasserstein distance $\revisionTVCG{\wassersteinTree[q]}$ \revisionTVCG{and similarly for $\regionWasserstein{q}$ and $\wasserstein{q}$.}

\revisionTVCG{
\begin{proposition}[Discriminative property]
When $\subsampleParameter \lt 1$, it follows that $\regionWassersteinTree[q]\big(\branchtree^\region(f_1), \branchtree^\region(f_2)\big) \geq \wassersteinTree[q]\big(\branchtree(f_1), \branchtree(f_2)\big)$ and similarly for $\regionWasserstein{q}$ and $\wasserstein{q}$.
\end{proposition}
\begin{proof}
{\color{black}\revision{Let $b_{i,\subsampleParameter}^\region = \subsampleRegion(b_i^\region, \subsampleParameter)$ and $\region^\subsampleParameter_i$ its region.
Recall that the more $\subsampleParameter$ decreases the more $\region^\subsampleParameter_i$ contains points, i.e. \mbox{$\region^{\subsampleParameter_1}_i \supseteq \region^{\subsampleParameter_2}_i$} for $\subsampleParameter_1 \lt \subsampleParameter_2$.
When comparing two pairs with the region-aware ground metric (\autoref{eq_our_groundMetric}), it implies that every sum in the regional discrepancy $\regionalDiscrepancy_q^\region$ for $\subsampleParameter_2$ is taken over a subset of the indices used for $\subsampleParameter_1$, since all terms are added and non-negative it implies that $\regionalDiscrepancy_q^\region(b_{i,\subsampleParameter_1}^\region, b_{j,\subsampleParameter_1}^\region) \geq \regionalDiscrepancy_q^\region(b_{i,\subsampleParameter_2}^\region, b_{j,\subsampleParameter_2}^\region)$.
The saddle values of the two pairs being unchanged we therefore have $d_q^\region(b^\region_{i,\subsampleParameter_1}, b^\region_{j,\subsampleParameter_1}) \geq d_q^\region(b^\region_{i,\subsampleParameter_2}, b^\region_{j,\subsampleParameter_2})$.
}
\revision{Let $\branchtree^\region_\subsampleParameter(f)$ \revisionTVCGminor{be} the region-aware BDT $\branchtree^\region(f)$ on which the operator $\subsampleRegion(\cdot, \subsampleParameter)$ has been applied to all its pairs.
For any matching $\phi' \in \Phi'$, let $\mathcal{C}_\subsampleParameter(\phi')$ \revisionTVCGminor{be} its associated cost between $\branchtree^\region_\subsampleParameter(f_1)$ and $\branchtree^\region_\subsampleParameter(f_2)$. 
We have $\mathcal{C}_{\subsampleParameter_1}(\phi') \geq \mathcal{C}_{\subsampleParameter_2}(\phi')$, taking the minima over $\phi' \in \Phi'$ yields:

\begin{equation}
\nonumber
 \revisionTVCG{\regionWassersteinTree[q]}\big(\branchtree^\region_{\subsampleParameter_1}(f_1), \branchtree^\region_{\subsampleParameter_1}(f_2)\big) \geq \revisionTVCG{\regionWassersteinTree[q]}\big(\branchtree^\region_{\subsampleParameter_2}(f_1), \branchtree^\region_{\subsampleParameter_2}(f_2)\big).
\end{equation}

Taking $\subsampleParameter_2 = 1$ (getting back to the original Wasserstein distance, \autoref{sec_generalization}) and $\subsampleParameter_1 = 0$ (being the region-aware Wasserstein distance without preprocessing by $\subsampleRegion$) we get:} $\revisionTVCG{\regionWassersteinTree[q]}\big(\branchtree^\region(f_1), \branchtree^\region(f_2)\big) \geq \revisionTVCG{\wassersteinTree[q]}\big(\branchtree(f_1), \branchtree(f_2)\big)$. 
\revisionTVCG{Similarly, taking $\epsilon_1 = 1$ translates this result to $\regionWasserstein{q}$ and $\wasserstein{q}$.}
}
\end{proof}
}

\section{Algorithm}
\label{sec_algo}

\revision{
We compute $\regionWassersteinTree[q]$ by extending the dynamic-programming scheme of Pont et al. \cite{pont_vis21}, which efficiently searches over rooted partial isomorphisms of two input BDTs.
In our setting, the input trees are the region-aware BDTs (\autoref{sec_regionAwareBDTs}), possibly preprocessed with $\subsampleRegion(\cdot, \subsampleParameter)$ (\autoref{sec_generalization}), every comparison of two pairs is done with the region-aware ground metric $d_q^\region$ (\autoref{sec_regionAwareGroundMetric}) and deletions/insertions are handled via the region-aware projection $\projection^\region(\cdot)$ (\autoref{sec_regionAwareProjection}).
Since $d_q^\region$ is a metric (\autoref{sec_metric}), the optimality of the algorithm of Pont et al. \cite{pont_vis21} holds and will indeed find the assignments minimizing the region-aware Wasserstein distance (\autoref{eq_our_distance}).
}

The algorithm 
takes $\mathcal{O}(|\branchtree|^2)$ steps in practice\revision{\mbox{\cite{pont_vis21}}}, with $|\branchtree|$ the number of nodes in the input BDTs.
The computation of $d_q^\region$ necessitates $\mathcal{O}(n + m)$ steps in the worst case, where $n$ and $m$ are respectively the number of vertices in the two regions to be compared.
\revision{The overall complexity of $\regionWassersteinTree[q]$ is therefore $\mathcal{O}\big(|\branchtree|^2(N + M)\big)$ with $N$ and $M$ the total number of vertices in the input scalar fields.}
For a region-aware BDT, the values of all its regions are stored without duplication in a single structure corresponding to the original data.
In this work we focus on regular grids, each region-aware persistence pair then stores the maximal bounds of its region within the original data, it allows to easily convert a local index of the region to a global index in the original data.
It additionally stores a boolean mask indicating whether or not a specific index in its maximal bounds indeed belongs to the region.

We then present two strategies to control both computation time and memory footprint of the method.
First, in \autoref{sec_subsampling}, we provide an example of the operator $\subsampleRegion$ (\autoref{sec_regionWassersteinDistance}) allowing to choose how much of the \revisionTVCGminor{regions'} properties should be taken into account,
\revision{having a direct impact on execution time}.
Second, in \autoref{sec_compression}, we present compression strategies in order to obtain low-memory representations of the regions.

\subsection{Subsampled regions comparison}
\label{sec_subsampling}

We now \revision{define an} operator $\subsampleRegion(b_i^\region, \subsampleParameter)$ (\autoref{sec_regionWassersteinDistance}) based on subsampling, \revision{allowing} to use only a subset of \revision{a region} $\region_i$ of a region-aware persistence pair 
$b_i^\region$
in the computation.
When the input domain is a regular grid, the subsampling-based operator $\subsampleRegion(b_i^\region, \subsampleParameter)$ selects a subset of points with a stride of $n$ in each dimension (one point over $n$) starting from the extremum of $\region_i$ ensuring its inclusion.
We use $n = \subsampleParameter m + 1$, rounded to the nearest integer where $m$ is the maximum number of points in the dimension containing the most.
When $\subsampleParameter = 0$ the stride of 1 ensures that indeed all the points in the region are kept (as defined in \autoref{sec_regionWassersteinDistance} for this operator).
When $\subsampleParameter = 1$ the stride of $m + 1$ ensures that no points except the extremum of the region is kept.
All the ensembles considered in this work are regular grids and the members within a same ensemble have the same number of points in each dimension.
This ensures that the points of two subsampled regions being aligned at their respective extremum will also be aligned.

\subsection{Regions compression}
\label{sec_compression}

The region-aware Wasserstein distance \revisionTVCGminor{$\regionWassersteinTree[q]$} requires access to the feature regions in the original data. 
However, a key advantage of 
\revisionTVCGminor{topological representations}
is their significantly lower memory footprint compared to the data themselves, enabling 
\revisionTVCGminor{analyses}
without storing the full dataset. 
To compute 
\revisionTVCGminor{$\regionWassersteinTree[q]$}
without retaining the original data on disk, compression methods can be considered.
Specifically, we propose to compress the original data \revision{(to avoid to fully store it on disk)} given an input parameter $\compressionParameter \in [0, 1]$ such that the size of the compressed data is a ratio $\compressionParameter$ of the original data size.

When $\compressionParameter = 1$ the original data is kept. 
\revision{The} more $\compressionParameter$ decreases, \revision{the} more the size of the compressed data decreases as well.
When $\compressionParameter = 0$, no parameters are allowed for the compressed data and therefore no information about the regions are considered.
\revision{We additionally store the region membership of each original vertex, such that the decompressed data has the same \revisionTVCGminor{membership as} the original one.}
The 
parameter $\compressionParameter$ allows to choose the memory overhead of our method compared to \revisionTVCGminor{that} of the original persistence diagrams and merge trees.
By doing so, the user can choose how much memory storage to use depending on its capacity and its use-cases.

We therefore focus on compression methods that allow to easily control the size of the compressed representation.
Most compression methods focus however on an important but different use-case, where the error is controlled, and not necessarily the size.
Still, some methods can be noted \revision{including} ZFP \cite{zfp}, neural fields \cite{neural_fields}, and B-splines \cite{b_splines}.
We show in \appendixCompression{} how given $\compressionParameter$, the parameters of these models need to be defined in order to be as close as possible to the target size.
In \autoref{sec_frameworkQuality} we show how this compression affects our distance and propose guidelines to choose between the different models given user requirements such as compression size and execution time.

\section{Results}
\label{sec_results}

This section presents experimental results obtained on a computer with two AMD EPYC 7453 CPUs (2.75 GHz, 28 cores each) and 512 GB of RAM.
The input persistence diagrams and merge trees were computed with FTM \cite{gueunet_tpds19} and pre-processed to remove low persistent features (with a 
threshold of 0.5 \% of the data range).
We have implemented our method in C++ as an extension of the Wasserstein distance module of persistence diagrams and merge trees in TTK \cite{ttk17} \revisionTVCG{and plan to integrate it in the official TTK distribution}.

Our experiments were performed on a variety of simulated and acquired 2D and 3D ensembles inspired by \cite{pont_vis21} \revisionTVCGminor{that} used ensembles extracted from various SciVis contests \cite{scivis2018, scivis2017, scivis2016, scivis2015, scivis2014, scivis2008, scivis2006, scivis2004} or from previous \revisionTVCGminor{work} \cite{favelier2018}.
We have extended the SciVis contests ensembles used in \cite{pont_vis21} in order to create more challenging ones (described in \autoref{tab_timings}).
For most of them, it consists of all the available data (for a same set of simulation parameters when applicable) instead of only a small subset of it.
\revisionTVCGminor{The preprocessing scripts are available at: \url{https://github.com/MatPont/regionWassersteinData}}.

\begin{figure*}
    \centering
    \includegraphics[width=\linewidth]{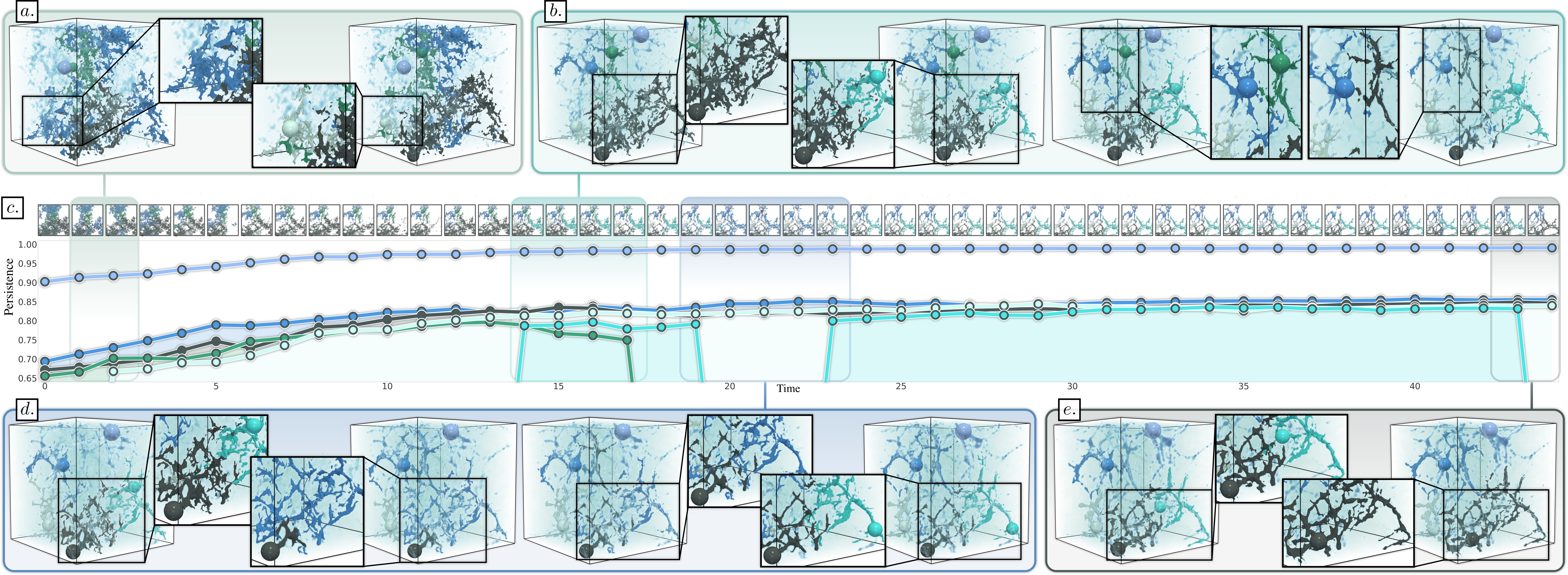}
    \caption{
\revisionTVCG{
Visual analysis of the \emph{Dark Sky} ensemble (dark matter density) with the region-aware Wasserstein distance of merge trees.
Members at different points in time are shown in $(a)$, $(b)$, $(d)$ and $(e)$.
We track the evolution of the topological features using the assignments of our method for consecutive time steps (matched features having the same color).
The temporal persistence curves in $(c)$ allow to visually detect important changes in the ensemble by representing the evolution of the persistence of the matched features over time (only curves for the highest persistent features are shown with \revisionTVCGminor{the} same coloring \revisionTVCGminor{as} in the input domain).
In this example it allows to track the appearance, the split and the merge of the filament structures of the cosmic web.
For instance, in the first time window $(a)$, we see the white feature emerging from the blue one.
In $(d)$, both the cyan and blue features \revisionTVCGminor{merge} (first zoom), and split later (second zoom), the same color is used after split.
Finally, in $(e)$ it merges with the black one.
}}
    \label{fig_curve2}
\end{figure*}

\subsubsection{\textbf{\revision{Baseline methods and parameter selection}}}
We compare our method to the classical Wasserstein distances $\wasserstein{2}$ and $\wassersteinTree$ \cite{edelsbrunner09, pont_vis21} and to two methods that took a first step in the direction of using geometrical information by focusing on \revision{coarse} properties.
First, the so-called \emph{geometrical lifting}\revision{\mbox{\cite{soler_ldav18, lin_geometryAware23, mingzheTracking}}}, i.e. the use of the coordinates in the domain of the critical points in the distance computation, that we note $\liftingWasserstein{2}$ and $\liftingWassersteinTree$ for respectively persistence diagrams and merge trees.
Second, the use of the volume of the regions \cite{SaikiaSW14_branch_decomposition_comparison}, i.e. the number of vertices, that we respectively note $\volumeWasserstein{2}$ and $\volumeWassersteinTree$.
We formally define these methods in \appendixGeometric{} \revision{and use the implementations available in TTK \cite{ttk17}}.
The last two methods are subject to parameters controlling \revision{the impact of} the geometrical properties (\appendixGeometric), the weights $w_\lifting$ for the coordinates distance and $w_\volume$ for the volume difference.
Our method is subject to the subsampling parameter $\subsampleParameter$ (\autoref{sec_subsampling}).

We use the values $w_\lifting = 0.5$, $w_\volume = 0.2$ and $\subsampleParameter = 0.1$ since they gave the best results for our applications \revision{averaged} over all the \revisionTVCG{considered} ensembles (see \autoref{sec_dimRed}) \revisionTVCG{and recommend using them by default.
For larger ensembles or resource-constrained settings, $\subsampleParameter$ can be increased to further reduce the cost in computation time of region comparisons, at the expense of accuracy, while stronger compression can be used when memory is the main bottleneck (see \autoref{sec_frameworkQuality} for the recommended models).
Unless stated otherwise, the null background field is used 
as it helps interpretability by naturally considering absent the values outside a feature region and make them contributing as deletion or insertion costs, \revisionTVCGminor{although stability is shown for data background (\autoref{sec_stability})}.}
When processing merge trees, we use the \revisionTVCGminor{recommended} value of $\epsilon_1 = 0.05$ given by Pont et al. \cite{pont_vis21}.
\revision{Finally, for the ground metric (\autoref{eq_groundMetric} and \autoref{eq_our_groundMetric}), we use $q = 2$, the standard Euclidean metric, consistent with prior works for visualization applications \cite{vidal_vis19, pont_vis21, pont_tvcg23, pont_tvcg24, sisouk_techrep23}.}

\subsection{Time performance}

\begin{table}[b]
\caption{Comparison of the running times (in seconds), between our method and the baselines, of the distance matrix computation for each ensemble using persistence diagrams and merge trees.
}
\label{tab_timings}
\centering
\scalebox{0.6}{
  \begin{tabular}{|l|r|r||r|r|r|r||r|r|r|r|}
    \hline
    \rule{0pt}{2.25ex}  \textbf{Dataset} & $\ensembleSize$ & $|\branchtree|$ & \multicolumn{4}{c||}{Persistence Diagrams} & \multicolumn{4}{c|}{Merge Trees} \\
       & & & $\wasserstein{2}$ & $\liftingWasserstein{2}$ & $\volumeWasserstein{2}$ & $\regionWasserstein{2}$ & $\wassersteinTree$ & $\liftingWassersteinTree$ & $\volumeWassersteinTree$ & $\regionWassersteinTree$ \\
        &  &  &  &  &  & \textbf{(ours)} &  &  & & \textbf{(ours)} \\
    \hline
      Asteroid Impact (3D)    &  68 &  854 &  33.3 &  31.7 &    34.2 &    35.9 &  20.4 &  19.8 &  20.5 &    23.2 \\
      Cloud Processes (2D)    &  91 & 2880 & 419.5 & 387.9 &   435.2 &   503.4 & 350.7 & 312.3 & 359.8 &   416.7 \\
      Viscous Fingering (3D)  & 120 &   48 &  13.2 &   9.9 &    10.2 &     5.8 &   7.2 &   8.6 &   6.1 &     7.3 \\
      Dark Matter (3D)        &  99 & 4646 & 973.9 & 984.4 & 1,187.9 & 1,492.1 & 759.9 & 755.2 & 789.7 & 1,138.8 \\
      Volcanic Eruptions (2D) & 200 &  828 & 102.2 &  99.2 &   109.9 &   143.1 &  45.2 &  49.2 &  54.1 &    66.3 \\
      Ionization Front (2D)   & 200 &   68 &  20.8 &  23.8 &    20.4 &    17.4 &  15.4 &  19.8 &  18.1 &    18.1 \\
      Ionization Front (3D)   & 200 & 1228 & 218.2 & 236.5 &   235.8 &   354.8 & 236.9 & 265.2 & 262.4 &   298.5 \\
      Earthquake (3D)         & 212 &  614 &  56.4 &  61.8 &    61.4 &    95.6 &  36.8 &  44.6 &  40.1 &    64.3 \\
      Isabel (3D)             &  48 & 1832 &  66.6 &  58.9 &    67.2 &    68.4 &  54.6 &  52.9 &  56.6 &    62.3 \\
      Starting Vortex (2D)    &  12 &   39 &   0.1 &   0.1 &     0.1 &     0.1 &   0.1 &   0.1 &   0.1 &     0.1 \\
      Sea Surface Height (2D) &  48 & 1453 &  40.9 &  37.9 &    41.6 &    45.3 &  30.6 &  28.4 &  29.7 &    32.2 \\
      Vortex Street (2D)      &  45 &    9 &   0.2 &   0.2 &     0.3 &     0.2 &   0.4 &   0.4 &   0.5 &     0.4 \\
      Heated Cylinder (2D)    &  60 &   12 &   0.9 &   0.5 &     0.6 &     0.6 &   0.5 &   0.9 &   0.6 &     0.7 \\
    \hline
  \end{tabular}
}
\end{table}

\begin{figure*}[hp!]
    \centering
    \includegraphics[width=0.95\linewidth]{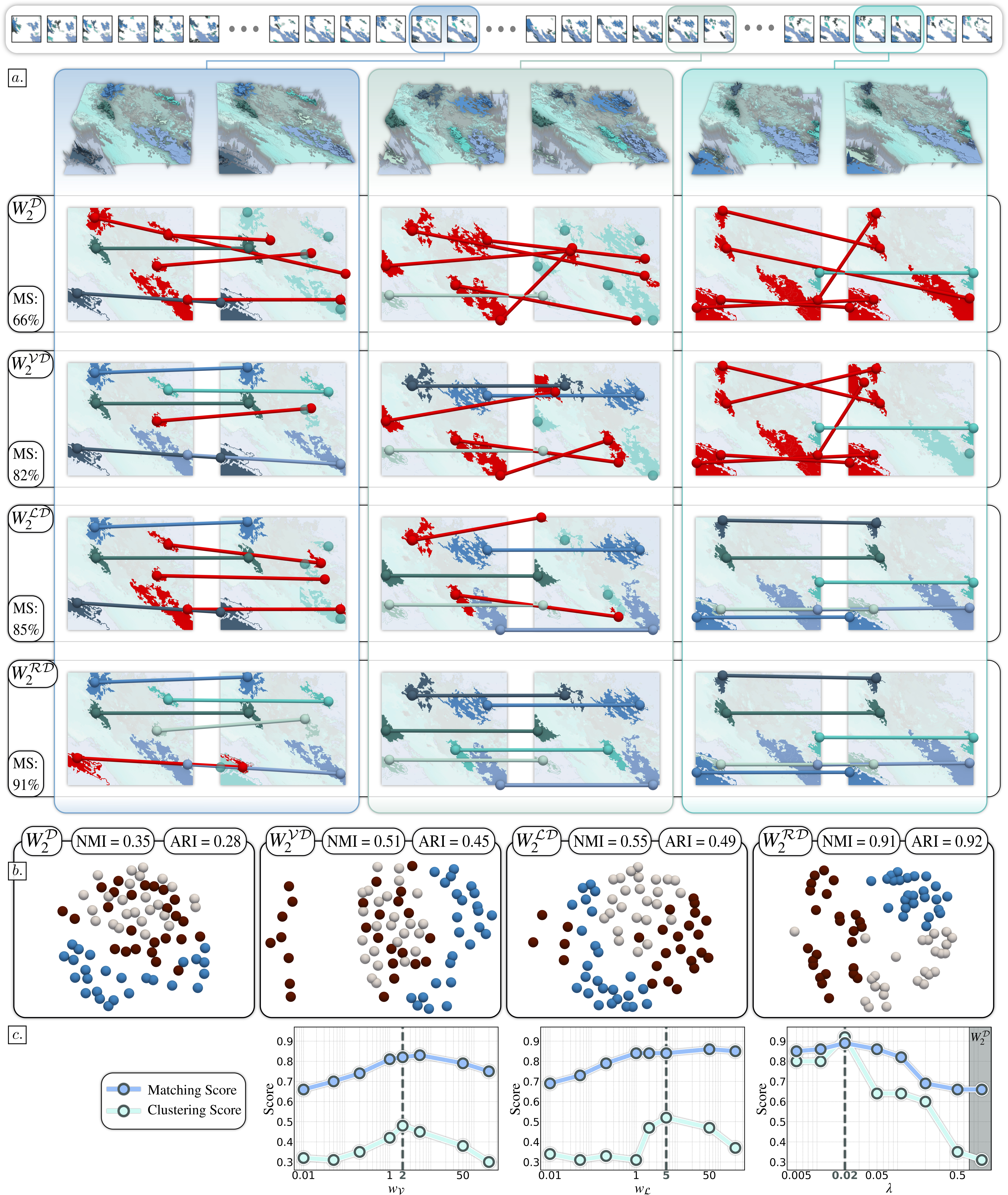}
    \caption{\revision{
Pairwise feature tracking and dimensionality reduction of the Cloud ensemble (cloud pressure in Pa), using the baseline methods ($\wasserstein{2}$, $\volumeWasserstein{2}$ and $\liftingWasserstein{2}$) and ours ($\regionWasserstein{2}$).
In \emph{(a)}, three matchings are selected to highlight where $\volumeWasserstein{2}$, $\regionWasserstein{2}$ and $\liftingWasserstein{2}$ (respectively for each column) perform the best comparing to the other methods.
No matchings of $\wasserstein{2}$ manage to outperform the other methods, hence no column specifically for this method.
We see that $\liftingWasserstein{2}$ can fail when the extrema of a same region at different time steps change position (first column in \emph{(a)}).
We also evaluate the accuracy of the matchings, \emph{matching score} (MS), given a ground-truth of the features that should be matched together, our method provides a 91\% accuracy, outperforming the baseline methods.
In \emph{(b)}, we show the MDS planar embeddings of the four methods, our method outperforms greatly the others with a NMI and ARI of 0.91 and 0.92 respectively and have clusters that are visually more separated.
In \emph{(c)}, we show the matching score and the clustering score (average of NMI and ARI) given the parameter that needs to be tuned for each method (none for $\wasserstein{2}$).
For each method, the parameter was selected to maximize both matching and clustering scores.
Note that the chosen $w_\lifting$ value for $\liftingWasserstein{2}$ can be not high enough to match some features together (second column in \emph{(a)}) while it still provides the best results overall.
For $\regionWasserstein{2}$ and $\subsampleParameter = 1$ we get back to $\wasserstein{2}$ (gray area in the chart), for any other values of $\subsampleParameter$ we see that $\regionWasserstein{2}$ outperforms $\wasserstein{2}$.
}}
    \label{fig_cloud}
\end{figure*}

\autoref{tab_timings} evaluates the time performance for both persistence diagrams and merge trees of our method compared to \revision{the} three baselines.
We observe that the execution time of each method is indeed a function of the number of ensembles members ($\ensembleSize$) and the average number of persistence pairs ($|\branchtree|$).
$\regionWasserstein{2}$ and $\regionWassersteinTree$ are $1.3\revisionTVCGminor{\times}$ slower \revisionTVCGminor{on} average than $\wasserstein{2}$ and $\wassersteinTree$, it is an expected result since they use the regions' properties in the computation and compare more values than only the birth and the death.
Overall, the lifting distances $\liftingWasserstein{2}$ and $\liftingWassersteinTree$ and the volume distances $\volumeWasserstein{2}$ and $\volumeWassersteinTree$ have execution times between those of the classical and region-aware Wasserstein distances.
Averaged over all ensembles, it takes approximately 3 minutes to compute the distance matrix of an ensemble using $\regionWasserstein{2}$ and $\regionWassersteinTree$.
\revision{Note that the same topological representations are used as input for all methods. $\liftingWasserstein{2}$ and $\liftingWassersteinTree$ additionally use the coordinates of the extrema, $\volumeWasserstein{2}$ and $\volumeWassersteinTree$ the volume of the features and $\regionWasserstein{2}$ and $\regionWassersteinTree$ the regions' properties.
The subsampling parameter $\subsampleParameter$ (\autoref{sec_subsampling}) and the compression parameter $\compressionParameter$ (\autoref{sec_compression}) only \revisionTVCGminor{impact} the additional information used by our method and not the input topological representations.}

\revisionTVCGminor{In \appendixMemory{}, we compare the running time and peak memory of the baseline methods and our method on a more usual workstation-class configuration, whose specifications are detailed therein.
We observe a mean per-ensemble slowdown of $5.33\times$ compared to the reference configuration used for \autoref{tab_timings} and a mean per-ensemble peak memory overhead of $3.49\times$ when comparing $\regionWasserstein{2}$ to $\wasserstein{2}$ due to the need to load and access the input data.
Across both configurations, 
all methods typically require only a few seconds and a few GB of peak memory for the smallest ensembles. 
For the largest ones, the computations can reach tens of minutes to a few hours depending on the 
configuration, while peak memory remains moderate, on the order of a few tens of GB.
Overall, in the largest reported settings, 
these 
methods are mainly suitable for batch analysis rather than interactive use, and should be computed as a preprocessing step before interactive analysis.
}



\subsection{Topological matching for feature tracking}
\label{sec_featureTracking}

\begin{figure*}[t]
    \centering
    \includegraphics[width=\linewidth]{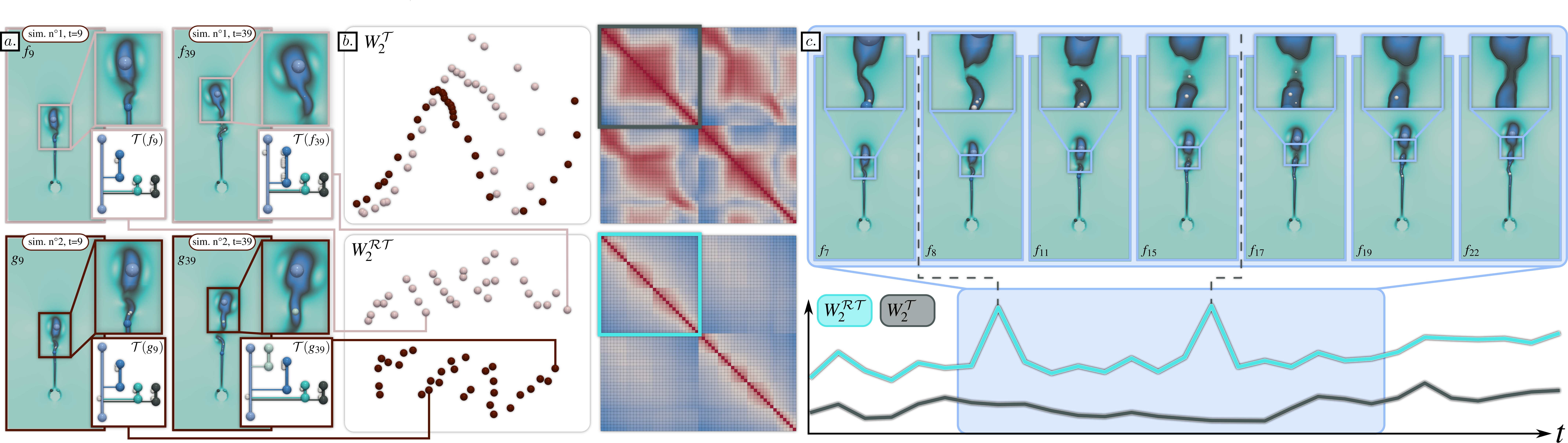}
    \caption{
\revisionTVCG{
Dimensionality reduction \revision{and phases detection} of the heated cylinder ensemble.
\revision{It consists} of two time-varying simulations that differ regarding the geometry of the \emph{front} of their heated fluid, \revision{one being tilted to the right, the other to the left} (examples of each are shown in $(a)$).
Since merge trees encode each feature independently of its geometry, the Wasserstein distance $\wassersteinTree$ \revisionTVCGminor{cannot} differentiate between the two simulations.
This can be seen in the 2D MDS embedding of the distance matrix of $\wassersteinTree$ \revision{($(b)$, top)} where the points of the two simulations overlap (the color code indicating the simulation membership).
On the contrary, the region-aware Wasserstein distance $\regionWassersteinTree$ can identify this specificity and correctly differentiate \revision{and separate the points of} the two simulations \revision{($(b)$, bottom)}.
In $(c)$ the \revision{distances} between consecutive time steps of the first simulation are shown for both $\wassersteinTree$ (black) and $\regionWassersteinTree$ (green).
The latter allows to detect important changes in the simulation (peaks in the green distance curve in $(c)$) that $\wassersteinTree$ \revisionTVCGminor{cannot}.
These changes correspond to the moments when the tail of the heated fluid separates from the front (first peak) and when it stops moving away to starts merging again with it (second peak).
}}
    \label{fig_heatedCylinder}
\end{figure*}

\revision{The computation of} our distance consists in the optimization of assignments between the persistence pairs of input BDTs.
Since each persistence pair corresponds to a feature in the input data, these assignments allow to establish relationships between them.
When a time-varying ensemble is considered, a distance can be computed for each consecutive time steps and the global evolution of each feature can be tracked over the whole ensemble.
Overall, the matchings provide visual hints to the users to help them relate features with each other.
\revisionTVCG{We focus here on evaluating topological methods as they provide feature-level abstractions of scalar fields together with interpretable \revisionTVCGminor{correspondences} between their structures.}

\revision{An example of such matchings is \revisionTVCG{illustrated} in \autoref{fig_cloud} where we compare the matchings provided by the three baseline methods and ours on the cloud ensemble.
While each column in \autoref{fig_cloud}a corresponds to the best result for respectively each method (except for $\wasserstein{2}$ that does not outperform any other method), our distance still provides visually meaningful matchings for each of them.
\revisionTVCGminor{Moreover}, we quantitatively evaluate such matchings according \revisionTVCGminor{to} a ground-truth of the features that should be matched together, \revisionTVCGminor{on} average our method increases by 17\% the matchings accuracy on this example compared to the other methods considered.
}

We introduce the \emph{temporal persistence curves}, a plot of the evolution of the persistence of the features being matched together given time.
It consists \revision{of} various curves, one for each feature, visually representing how its persistence evolves over time, permitting to detect when it appears, grows or disappears.
Such curves act as a visual summary of an ensemble, enabling the visual identification of important changes within it in a single static view.
It can help the user to select pertinent time steps to be visualized and therefore facilitate data exploration.
\revision{A related but different object, the \emph{vineyards}, has been introduced by Cohen-Steiner et al. \cite{vinesVineyards}.
They represent a series of persistence diagrams as 3D curves, each representing the birth and the death of features being matched together over time. Temporal persistence curves instead are 2D curves encoding the evolution of persistence, facilitating their visualization on a plane, providing a high-level and concise visual summary of a time-series for change detection and time-step selection.}

\revision{The use of temporal persistence curves} is illustrated in \autoref{fig_teaser} with the \emph{Isabel} ensemble (wind velocity of a hurricane), and in \autoref{fig_curve2} with the \emph{Dark Sky} ensemble (dark matter density in a cosmology simulation), with our distances $\regionWasserstein{2}$ and $\regionWassersteinTree$.
Since persistence \revisionTVCGminor{diagrams} and merge trees encode features regardless of their geometry, the Wasserstein distances for these objects, $\wasserstein{2}$ and $\wassersteinTree$, can produce mismatches between the features (red arrows in the red insets of \autoref{fig_teaser}).
It prevents their correct tracking over time and impacts the subsequent \revisionTVCGminor{interpretation} of the curves.
In contrast, our method correctly tracks the features over the whole ensemble \revision{and the resulting curves allow to automatically detect key changes within it}.



\subsection{Distance matrices for dimensionality reduction}
\label{sec_dimRed}

Given an ensemble, our method \revision{can be used} to compute a distance matrix for all its members.
Such distance matrices \revision{are} then processed by typical dimensionality reduction algorithms such as MDS \cite{kruskal78}.
This method produces an embedding in $d$ dimensions such that the distance matrix between the embedded points is as close as possible \revisionTVCGminor{to} the given distance matrix.
In order to visually represent the members of an ensemble in a plane, we have chosen $d = 2$.
Overall, such embeddings provide visual hints to the users to help to understand the relation between members of a same ensemble.

\begin{figure}[t]
    \centering
    \includegraphics[width=\linewidth]{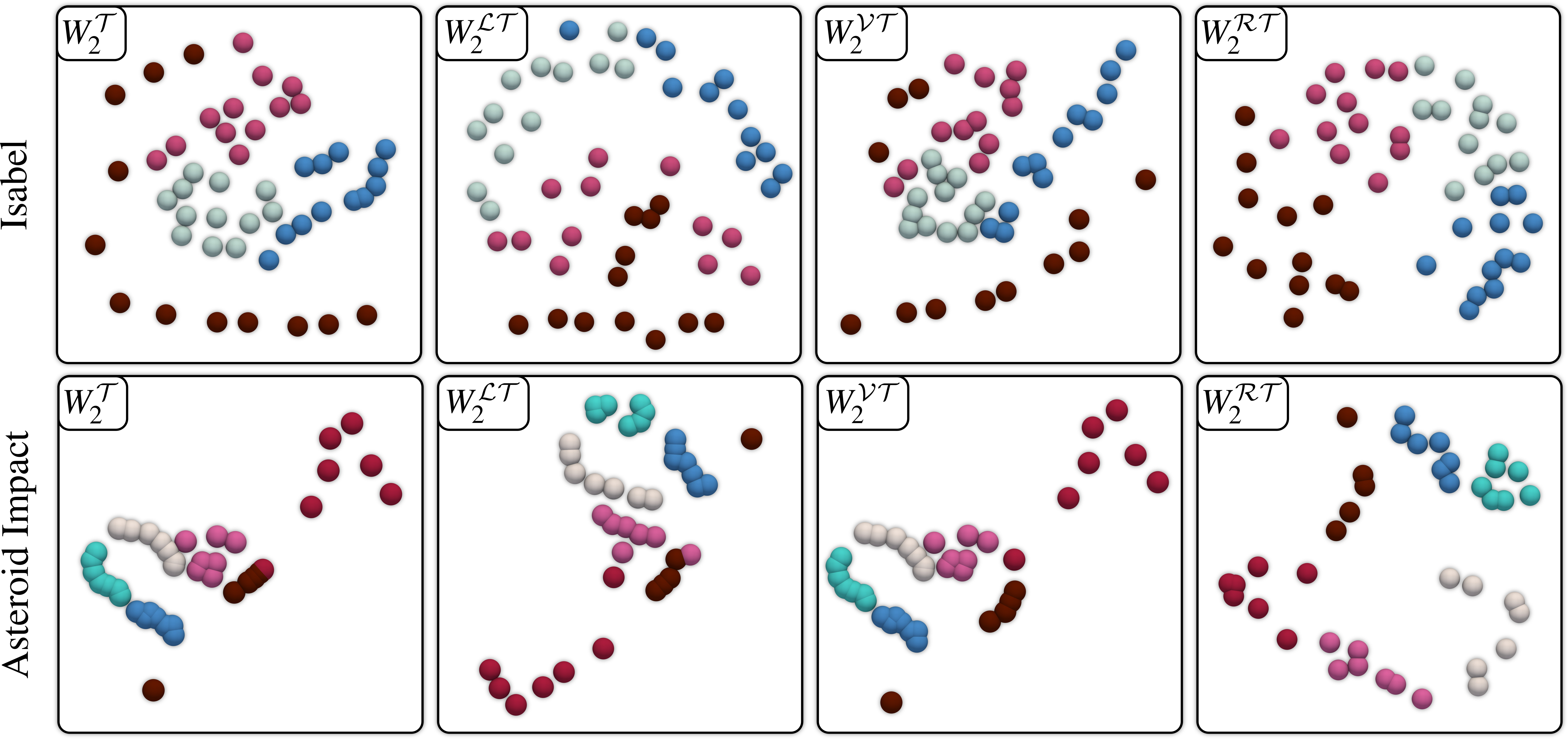}
    \caption{
Comparison of the MDS planar embeddings of all methods for two input ensembles.
The color encodes the classification ground-truth.
\revision{On these examples, only our method ($\regionWassersteinTree$, right column) provides embeddings where the classes are correctly separated ($NMI = ARI = 1$).}
}
    \label{fig_distMat2}
\end{figure}

We show some examples of such embeddings\revision{, first\revisionTVCGminor{,}} in \autoref{fig_heatedCylinder}b \revision{where the classical Wasserstein distance $\wassersteinTree$ fails to separate the two simulations (embeddings overlap), whereas our method $\regionWassersteinTree$ cleanly splits them, demonstrating sensitivity to geometric differences that topology alone cannot capture.}
\revision{Second, in} \autoref{fig_distMat2} \revision{where, 
\revisionTVCGminor{for these} examples,
our method is the only one among all the methods considered that provides embeddings where the classes are correctly separated (given a ground-truth classification).}
\revisionTVCG{We extend this evaluation in \appendixDimRedExperiments{} where embeddings generated using the classical $L_2$ distance between the input scalar fields are compared against those using the considered topological methods.
This shows their benefits when the features of interest of two fields are not in the same position in the domain, preventing the $L_2$ distance to compare them directly and instead compare them to unrelated locations in other fields, resulting in high distances between some members and to embeddings where clusters are not clearly separated.
Note that the $L_2$ distance can still be competitive on smoothly evolving time series or when topological instabilities occur \cite{wetzelsStability}.
}
We additionally show that the distance matrices generated by our \revision{method} permit to detect key changes in a time-varying ensemble that the classical Wasserstein distance \revisionTVCGminor{cannot} (\autoref{fig_heatedCylinder}c). 

\begin{table}[t]
  \caption{
  Comparison of \revision{MDS planar embeddings} quality scores averaged over \emph{all} ensembles for persistence diagrams and merge trees, bold: best values.
  }
  \centering
  \scalebox{0.8}{
    \begin{tabular}{|l||r|r|r|r||r|r|r|r|}
      \hline
      \rule{0pt}{2.25ex}  \textbf{Indicator} & \multicolumn{4}{c||}{PDs Ensembles Average} & \multicolumn{4}{c|}{MTs Ensembles Average} \\
       & $\wasserstein{2}$ & $\liftingWasserstein{2}$ & $\volumeWasserstein{2}$ & $\regionWasserstein{2}$ & $\wassersteinTree$ & $\liftingWassersteinTree$ & $\volumeWassersteinTree$ & $\regionWassersteinTree$ \\
       &  &  &  & \textbf{(ours)} &  &  & & \textbf{(ours)} \\
      \hline
      NMI & 0.66 & 0.69 & 0.69 & \textbf{0.78} & 0.66 & 0.69 & 0.69 & \textbf{0.79} \\
      ARI & 0.59 & 0.63 & 0.62 & \textbf{0.72} & 0.60 & 0.66 & 0.64 & \textbf{0.73} \\
      \hline
    \end{tabular}
  }
  \label{tab_qualityIndicators}
\end{table}

\autoref{tab_qualityIndicators} compares 
\revisionTVCGminor{quantitatively}
the MDS embeddings 
\revisionTVCGminor{averaged over}
all 
ensembles
using our method and the baselines.

As indicators, we consider the normalized mutual information (\emph{NMI})\revision{\mbox{\cite{strehl2002cluster,vinh2010information}}} and the adjusted rand index (\emph{ARI})\revision{\mbox{\cite{hubert1985comparing}}} that evaluate how much the clusters are well preserved in the 2D embedding compared to a ground truth classification associated with each ensemble (such as the distinct phases of a time-varying phenomenon, the different simulation parameters etc.).
\revision{We define these indicators in \appendixDimRedIndicators{}}.
\autoref{tab_qualityIndicators} averages these scores over all persistence diagrams and merge trees ensembles.
We have optimized the parameters of the various methods in order to get the best scores for each of them.
Specifically, we have optimized the weight $w_\lifting$ for the geometrical lifting\revision{\mbox{\cite{soler_ldav18, lin_geometryAware23, mingzheTracking}}}, the weight $w_\volume$ for the volume difference \cite{SaikiaSW14_branch_decomposition_comparison} (\appendixGeometric) and the subsampling parameter $\subsampleParameter$ of our method (\autoref{sec_subsampling}).
In practice, we use $w_\lifting = 0.5$, $w_\volume = 0.2$ and $\subsampleParameter = 0.1$.
The use of the coordinates distance ($\liftingWasserstein{2}$ and $\liftingWassersteinTree$)\revision{\mbox{\cite{soler_ldav18, lin_geometryAware23, mingzheTracking}}} and the volume difference ($\volumeWasserstein{2}$ and $\volumeWassersteinTree$) \cite{SaikiaSW14_branch_decomposition_comparison} slightly improve the scores compared to the classical Wasserstein distances.
In contrast, our method provides a way better improvement of these scores with an approximately \revision{21\%} increase compared to the classical Wasserstein distances \revision{and an approximately 16\% increase compared to all the}
\revisionTVCG{baseline methods considered}.

\revision{Another example is shown in \autoref{fig_cloud} where we compare the embeddings provided by the three baseline methods and ours on the cloud ensemble.
We quantitatively evaluate such embeddings according \revisionTVCGminor{to} a ground-truth classification, \revisionTVCGminor{on} average our method increases by 109\% the scores on this example  compared to the other methods considered.}

\subsection{Framework quality}
\label{sec_frameworkQuality}


\begin{figure}[b]
    \centering
    \includegraphics[width=\linewidth]{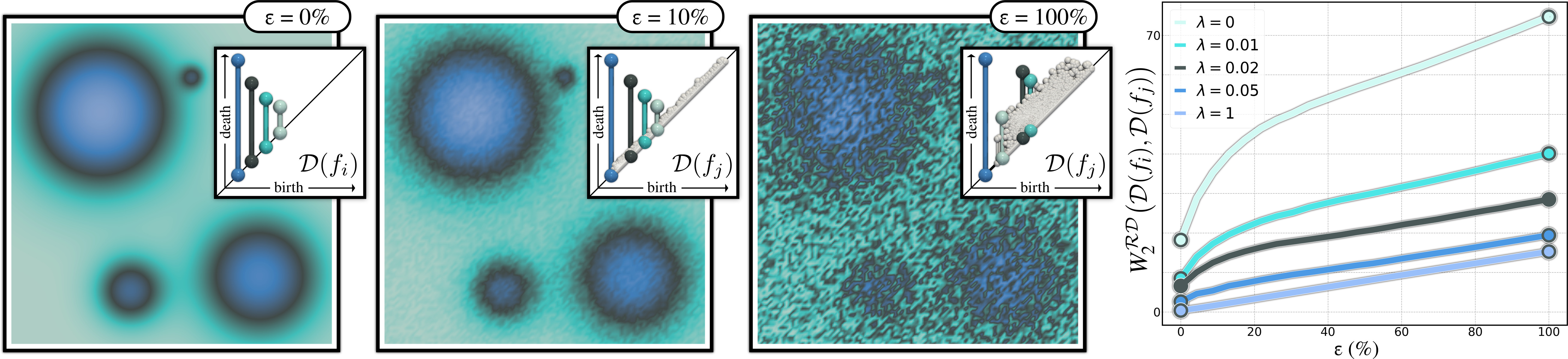}
    \caption{
Empirical stability evaluation of $\regionWasserstein{2}$. 
A noisy version $f_j$ of an input scalar field $f_i$ is created by adding a noise of increasing amplitude $\epsilon$.
The evolution of $\regionWasserstein{2}\big(\diagram(f_i), \diagram(f_j)\big)$ given $\epsilon$ and for different values of the subsampling parameter $\subsampleParameter$ shows no sudden spikes or drops that would be related to high instabilities, but rather a smooth evolution.
}
    \label{fig_stability}
\end{figure}

\subsubsection{\revision{\textbf{Empirical stability}}}  We empirically evaluate the stability of the region-aware Wasserstein distance \revision{$\regionWasserstein{q}$} in \autoref{fig_stability} \revision{using null background (\autoref{sec_regionWassersteinDistance}).
Recall that in \appendixStability{} we have proven the stability of $\regionWasserstein{q}$ using data background (\autoref{sec_regionWassersteinDistance}) when excluding extremum-swap instabilities}.
Given a scalar field $f_i$, we create noisy variants $f_j$ of it with a noise of increasing amplitude $\epsilon$, i.e. $\| f_i - f_j \|_{\infty} \leq \epsilon$ \revisionTVCG{(shown in \autoref{fig_stability} as a percentage of the scalar range of $f_i$)}.
Then, we compute the distance $\regionWasserstein{2}\big(\diagram^\region(f_i), \diagram^\region(f_j)\big)$ and observe its evolution given $\epsilon$ for different values of the subsampling parameter $\subsampleParameter$.
When $\subsampleParameter = 1$ we have \mbox{$\regionWasserstein{2}\big(\diagram^\region(f_i), \diagram^\region(f_j)\big) = \wasserstein{2}\big(\diagram(f_i), \diagram(f_j)\big)$} which is proven to be stable \cite{CohenSteinerEH05} and we indeed observe its evolution to be linear (blue line at the bottom in \autoref{fig_stability}).
\revision{The} more $\subsampleParameter$ decreases, \revision{the} more we see $\regionWasserstein{2}$ increasing as more values of the regions are considered in the distance.
In that case, we can observe that $\regionWasserstein{2}$ increases rapidly for low noise amplitude until arriving to a similar linear slope, this shift happens for around 20\% of noise amplitude.
Overall, for the various values of $\subsampleParameter$ we observe a smooth evolution of the distance and no sudden spikes or drops that would be related to high instabilities.
\revisionTVCG{We extend this evaluation in \appendixStabilityEvaluation{}, where we first evaluate the stability as in \autoref{fig_stability} but on a real-life data instance, showing the same overall message but making it valid beyond synthetic datasets for this example. 
Then, we present another evaluation of the stability, this time not only using a single data instance but on a whole real-life ensemble.
It overall shows that for the recommended value $\subsampleParameter = 0.1$, $\regionWasserstein{2}$ shares similar robustness behavior \revisionTVCGminor{as} $\wasserstein{2}$ for this example.
Moreover, $\regionWasserstein{2}$ remains relatively stable for high noise amplitudes and even when $\subsampleParameter$ decreases (getting closer to the pure region-aware Wasserstein distance).
}


\begin{figure}[t]
    \centering
    \includegraphics[width=\linewidth]{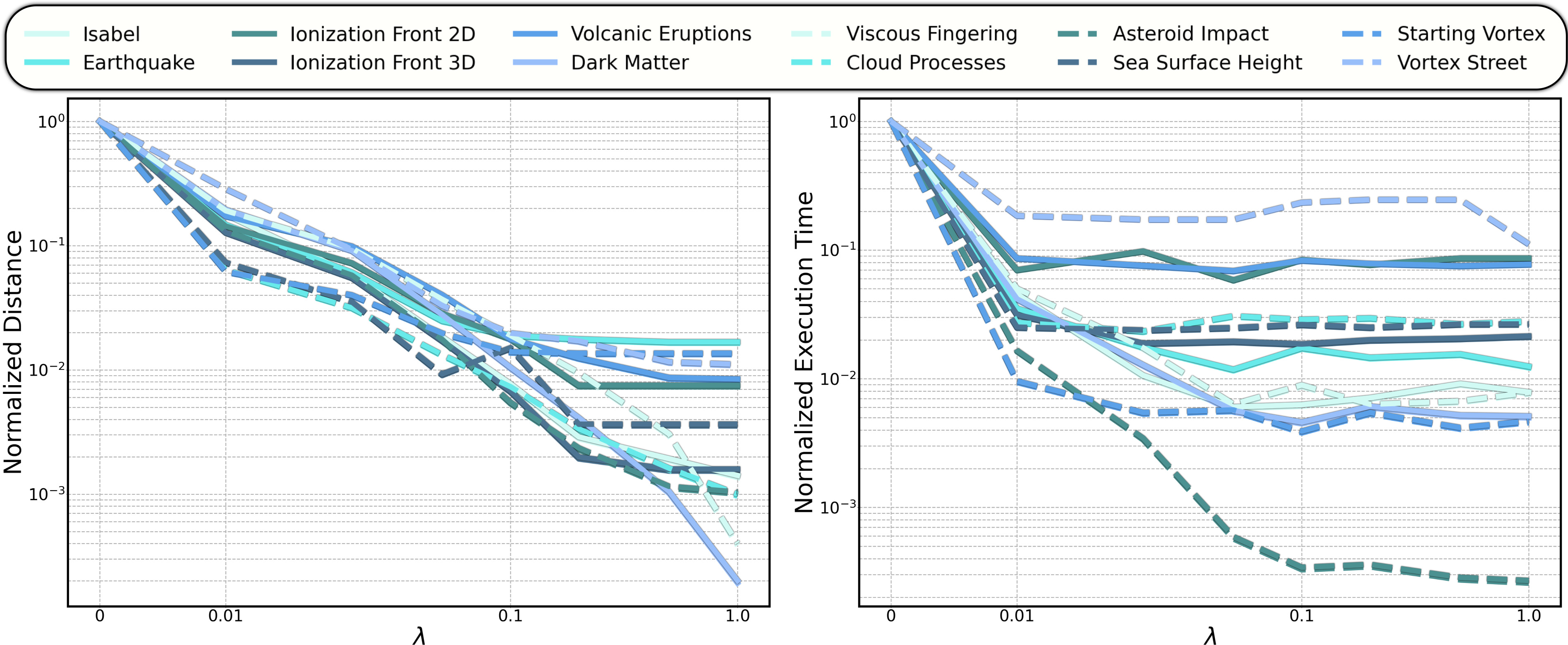}
    \caption{
Evolution of the normalized distance and execution time given the subsampling parameter $\subsampleParameter$ for all persistence diagram ensembles.}
    \label{fig_subsampling_exp}
\end{figure}

\subsubsection{\revision{\textbf{Impact of subsampling}}} We then evaluate in \autoref{fig_subsampling_exp} how the subsampling parameter $\subsampleParameter$ (\autoref{sec_subsampling}) impacts the region-aware Wasserstein distance and its execution time.
Interestingly, all normalized distance curves (one for each ensemble) \revisionTVCGminor{follow} the same kind of trend, they decrease approximately linearly as $\subsampleParameter$ increases in the log-log plot, implying an approximately power-law relationship between the normalized distance and $\subsampleParameter$.
Overall, this shows that there is a smooth transition between the classical Wasserstein distance at $\subsampleParameter = 1$ and the region-aware variant at $\subsampleParameter = 0$, following a common pattern.
As expected, the execution time decreases as $\subsampleParameter$ increases.
In that case, less values of the regions are used in the computation, naturally decreasing the execution time.


\begin{figure}[b]
    \centering
    \includegraphics[width=\linewidth]{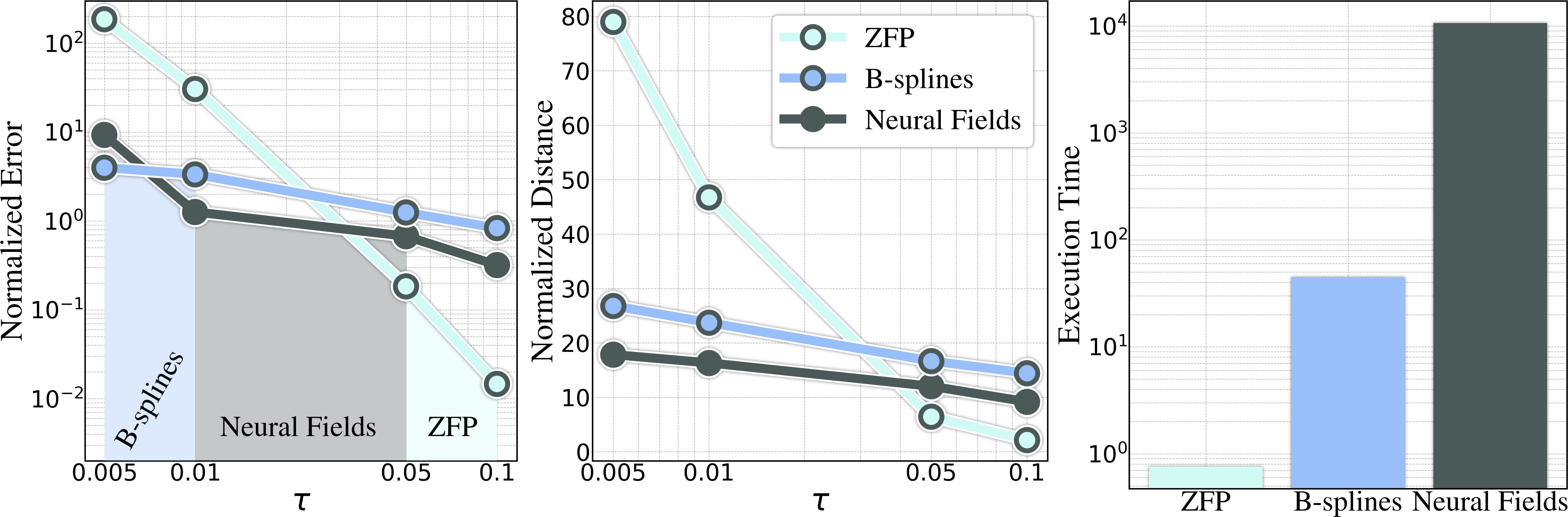}
    \caption{
Evolution of the normalized compression error (left) and distance (middle) given the compression parameter $\compressionParameter$ averaged over \emph{all} ensembles, and average execution time (right) for each compression method.
}
    \label{fig_compression}
\end{figure}

\subsubsection{\revision{\textbf{Impact of compression}}} In \autoref{fig_compression} we evaluate the impact of the compression of the regions on our method.
\revision{We aim} to better understand the benefits and drawbacks of each compression method considered \revision{in our context}.
We evaluate each of them for different values $\compressionParameter \in \{0.005, 0.01, 0.05, 0.1\}$ (\autoref{sec_compression}), ranging compression ratios from 200 : 1 to \mbox{10 : 1}.
We explain in \appendixCompression{} how the parameters of each method are set to give the best results.
We first evaluate the compression error of each method averaged over all ensembles for each $\compressionParameter$ value (\autoref{fig_compression}, left).
We observe that B-splines gave the best results for high compression ratios ($\compressionParameter = 0.005$), close to neural fields that are more effective for intermediary compression ratios ($\compressionParameter = 0.01$).
Above these values, ZFP seems to outperform the other two methods.
We then evaluate the impact on our distance $\regionWasserstein{2}$ (\autoref{fig_compression}, center) by computing a distance between an original region-aware persistence diagram $\diagram^\region(f_i)$ and the same diagram with its regions compressed.
We see that the distance follows \revisionTVCGminor{the} same kind of trend \revisionTVCGminor{as} the compression error.
Finally, the execution time of each method averaged over all ensembles and all $\compressionParameter$ values (\autoref{fig_compression}, right) indicates very fast compressions for ZFP, with less than 1 second \revisionTVCGminor{on} average per member, to \revisionTVCGminor{a} few minutes for B-splines, and is relatively slow for neural fields (few hours).

\subsection{Limitations}

Our method necessitates the points of extrema-aligned regions to coincide with each other for their fast comparison\revisionTVCG{, which is naturally satisfied by the regular grids considered in our work}.
However, our method \revisionTVCGminor{cannot} be directly applied to unstructured \revisionTVCGminor{grids} as it is\revisionTVCG{, in that case, a standard interpolation or resampling step onto a common regular grid would be needed before comparison. Although such preprocessing is common in practice, we did not perform in this work a quantitative analysis of the approximation error or runtime overhead it introduces and leave that for future work.}

Persistence diagrams and merge trees can be visualized in the plane to effectively summarize the topological features of a scalar field.
It is harder to do such visualizations of the regions of these features, especially for 3D ones.
In our work, we have decided to tackle this drawback by directly visualizing the features in the original scalar field (\autoref{fig_teaser} and \autoref{fig_curve2}).
We leave for future work an investigation of concise and expressive visualizations representing the \revisionTVCGminor{features'} geometry.

\revisionTVCG{We also recall that our method can only focus on 0th persistence diagrams (and merge trees) but not on higher dimensional topological features such as cycles and voids. Moreover, when using merge trees, the application of our method focused on persistence-based BDTs but more work would be needed to explore its use for other kinds of BDTs.}


\revisionTVCG{Because our method aligns regions only through their extrema and then compares them pointwise, it is inherently sensitive to geometric deformations such as rotations, stretching, or shear. This constitutes a limitation of the current formulation, although our empirical results suggest that the method remains practically robust on several real-world datasets. 
\revisionTVCGminor{However, this sensitivity can prevent a feature from being matched across time or ensemble members when it undergoes substantial spatial transformations, despite semantically representing the same structure.}
A transformation-aware metric could improve such cases, at the price of additional computational cost.}

%
\revision{Finally, the proven stability of $\regionWasserstein{q}$ (\autoref{sec_stability})
needs to exclude extremum-swap instabilities due to the alignment of regions to their extrema when comparing them.
\revisionTVCGminor{To the best of our knowledge, there is currently no practical method to guarantee this condition in general, and quantifying how frequently it occurs in practice is itself non-trivial. Developing a parameter similar to $\epsilon_1$ for such instabilities remains an open but interesting problem. Another promising direction}
to tackle this is the development of a new metric that aligns regions not simply on their extrema but to minimize a mismatch term.
}

\section{Conclusion}

In this paper, we presented \revision{a generalization of the Wasserstein distances} for persistence diagrams and merge trees\revision{\mbox{\cite{edelsbrunner09, pont_vis21}} that uses a richer characterization of the topological features and redefines their} comparison as a distance between their extrema-aligned regions.
\revision{A parameter is defined to control the amount of information about the regions to use, making the original Wasserstein distances} special cases of our method.
\revision{We have shown some theoretical properties: it is a metric, it is stable under certain conditions and it is more discriminative than the original distances it generalizes.}
We have presented two strategies to control both computation time and memory footprint of the method by respectively using only subsets of the regions in the computation and to compress the regions' properties to obtain low-memory representations.

A natural direction for future work is the definition of more advanced analysis methods using this novel distance metric such as barycenters\revision{\mbox{\cite{pont_vis21}}} (to compute a region-aware BDT representative of a set) or variance analysis methods \revision{like} principal geodesic analysis\revision{\mbox{\cite{pont_tvcg23}}}.
An orthogonal direction \revisionTVCGminor{to} the latter consists in the improvement of the region-aware ground metric.
Instead of simply aligning the regions at their extrema, 
\revision{they}
could 
be 
deformed to 
minimize a fitting criterion.
\revisionTVCGminor{Extending the method to cycles and voids is promising, but since they do not naturally define canonical regions as extremum-saddle pairs do, it remains to explore stable representatives \cite{representativeCycles, representativeVolumes} and how to align and compare them.}
\revisionTVCGminor{The extension to unstructured meshes would require 
a systematic evaluation of interpolation or resampling strategies and their impact on region definitions, distance values, runtime and memory.}
\revisionTVCG{From a performance point of view, the method’s reliance on vertex-wise comparisons makes it particularly well suited to GPU acceleration and could be explored.}
Overall, this work is a first step towards the extension and generalization of the Wasserstein distances for persistence and merge trees by incorporating region-aware comparisons.

\bibliographystyle{abbrv-doi}
\bibliography{template}


\section{Biography Section}

\vspace{-33pt}
\begin{IEEEbiography}[{\includegraphics[width=1in,height=1.25in,clip,keepaspectratio]{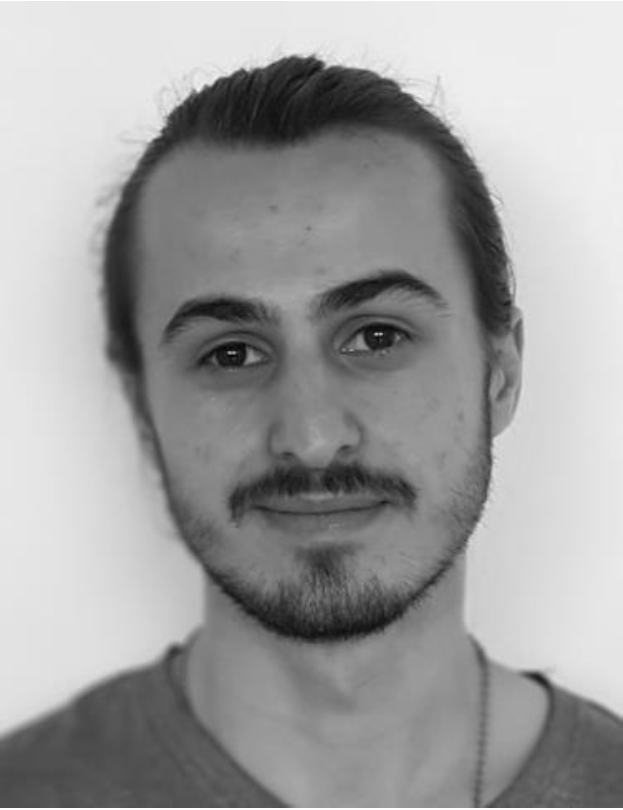}}]{Mathieu Pont}
received the Ph.D. degree in Computer Science from Sorbonne Université in 2023.
He has now a post-doctoral position at the RPTU Kaiserslautern-Landau.
He is an active contributor to the Topology ToolKit (TTK), an open source library for topological data analysis. 
His notable contributions to TTK include distances, geodesics and barycenters of merge trees, for feature tracking and ensemble clustering.
\end{IEEEbiography}

\vspace{-33pt}
\begin{IEEEbiography}[{\includegraphics[width=1in,height=1.25in,clip,keepaspectratio]{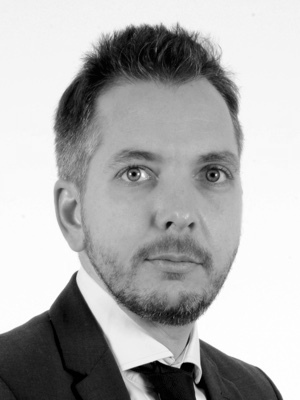}}]{Christoph Garth}
received the PhD degree in computer science from Technische Universit\"at (TU) Kaiserslautern in 2007. 
After four years as a postdoctoral researcher with the University of California, Davis, he rejoined TU Kaiserslautern where he is currently a full professor of computer science. 
His research interests include large scale data analysis and visualization, in situ visualization, topology-based methods, and interdisciplinary applications of visualization.
\end{IEEEbiography}

\clearpage
\appendices
\section*{Appendix}
\begin{center}
  \includegraphics[width=0.97\linewidth]{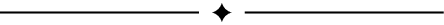}
\end{center}
\section{Metric proof \revision{using common background field}}
\label{metric_proof}

We show in this section that the region-aware ground distance $d_q^\region$ proposed in this work is indeed a metric and that therefore the region-aware Wasserstein distance \revision{$\regionWassersteinTree[q]$} (and \revision{$\regionWasserstein{q}$}) also is.

We first recall the expression of $d_q^\region$ (with $q \geq 1$):

\begin{eqnarray}
\label{eq_our_groundMetric_appendix}
  d_q^\region(b^\region_i, b^\region_j) = \big(|s_i - s_j|^q + \revision{\regionalDiscrepancy_q^\region(b^\region_i, b^\region_j)}\big)^{1/q},
\end{eqnarray}

\revision{\noindent with $\regionalDiscrepancy_q^\region$, the regional discrepancy, defined as:}

\begin{eqnarray}
\label{eq_our_regionalDiscrepancy_appendix}
  \revision{\regionalDiscrepancy_q^\region(b^\region_i, b^\region_j) =}
  &\hspace{-.35cm}
    \label{eq_our_groundMetric_mapping}
    \displaystyle \sum_{v \in \region'_i \cap \region'_j} &\hspace{-.2cm} 
    \big|\revisionTVCG{f'_1}(v) - \revisionTVCG{f'_2}(v)\big|^q\\
    \label{eq_our_groundMetric_deletion}
    &\hspace{-.35cm} + \displaystyle \sum_{x \in \region'_i \setminus \region'_j} &\hspace{-.2cm} 
    \big|\revisionTVCG{f'_1}(x)
     - \revision{\projectionRegion_{f_2}}(x)\big|^q\\
    \label{eq_our_groundMetric_insertion}
    &\hspace{-.35cm} + \displaystyle \sum_{y \in \region'_j \setminus \region'_i} &\hspace{-.2cm} \big|\revision{\projectionRegion_{f_1}}(y) - 
    \revisionTVCG{f'_2}(y)\big|^q,
\end{eqnarray}

\noindent
where 
$b^\region_{i} = (\revisionTVCG{\region_{i}^{f_1}}, s_{i})$ \revisionTVCG{from} $\branchtree^\region(f_1)$ and \mbox{$b^\region_{j} = (\revisionTVCG{\region_{j}^{f_2}}, s_{j})$} \revisionTVCG{from} $\branchtree^\region(f_2)$ 
are two \revision{region-aware} persistence pairs in two BDTs.
$\region'_i$ and $\region'_j$ are the centered regions at the respective extremum of $b^\region_{i}$ and $b^\region_{j}$, and 
\revisionTVCG{$f'_1$} and \revisionTVCG{$f'_2$ the corresponding centered} functions.
\revision{$\projectionRegion_{f_1}$ and $\projectionRegion_{f_2}$ are the background fields against which the values outside the intersection domain are compared.}
\revision{We will restrict the explanation to common background \revisionTVCGminor{fields}, i.e. $\projectionRegion_{f_1} = \projectionRegion_{f_2}$, such as the null background (\secDistanceDefinitionGroundMetric{}) and show the general case in Appendix \ref{metric_proof_general_case}.}

We now want to show that it is a metric.
For this, we need to show four properties: \emph{(i)} non-negativity, \emph{(ii)} identity of indiscernibles, \emph{(iii)} symmetry and \emph{(iv)} triangle inequality.

\revision{Given a region-aware persistence pair 
$b^\region_i = (\revisionTVCG{\region_i^f}, s_i)$,
let $\region'_i$ \revisionTVCGminor{be} the centered region at the extremum of $b^\region_{i}$ and 
\revisionTVCG{$f'$ its corresponding centered function.}
We define $\tilde{f}_i$ as the function 
\revisionTVCG{$f'$}
extended with the background field $\projectionRegion_f$ outside $\region'_i$.
We call $\tilde{f}_i$ the \emph{extended} function, defined as:
}

\begin{equation}
\label{eq_extendFunction}
  \tilde{f}_i\revision{(v)} =
    \begin{cases}
      \revisionTVCG{f'}(v)
      & \text{if $v \in \region'_i$}\\
      \revision{\projectionRegion_f(v)} & \text{if $v \notin \region'_i$}.
    \end{cases}
\end{equation}

\revision{
We adopt the convention that whenever two region-aware persistence pairs 
$b^\region_{i} = (\revisionTVCG{\region_{i}^{f_1}}, s_{i})$ \revisionTVCG{from} $\branchtree^\region(f_1)$ and \mbox{$b^\region_{j} = (\revisionTVCG{\region_{j}^{f_2}}, s_{j})$} \revisionTVCG{from} $\branchtree^\region(f_2)$ 
are compared, their  extended functions are understood to be defined on the common domain $\Omega = \region_i' \cup \region_j'$. 
All equalities $\tilde{f}_i = \tilde{f}_j$ are to be interpreted on $\Omega$.
In this section we show that $d_q^\region$ is a metric for the space of region-aware persistence pairs with the equivalence relation stating that $b^\region_{i} \sim b^\region_{j} \iff s_i = s_j \; \text{and} \; \tilde{f}_i = \tilde{f}_j \; \text{on} \; \Omega = \region_i' \cup \region_j'$.}

\textbf{\emph{(i)} Non-negativity:} 
each term of the definition of $d_q^\region$ is non-negative, each being an absolute value \revisionTVCGminor{raised} to a power of $q$. 
The $q$\textsuperscript{th} root (with $q \geq 1$) of the sum of non-negative numbers is non-negative therefore $d_q^\region(b^\region_i, b^\region_j) \geq 0$.

\textbf{\emph{(ii)} Identity of indiscernibles:}
\revision{as stated before, we identify region-aware persistence pairs modulo $\sim$, we write $b^\region_i = b^\region_j$ to mean $b^\region_i \sim b^\region_j$ and we keep the same symbol $d_q^\region$ for the induced metric on the quotient.}
We want to show that two points are indistinguishable in distance if and only if they are actually the same, i.e. $d_q^\region(b^\region_i, b^\region_j) = 0 \iff b^\region_i = b^\region_j$.

First, we show that if $b^\region_i = b^\region_j$ then $d_q^\region(b^\region_i, b^\region_j) = 0$.
If two pairs are identical then their saddle values \revisionTVCGminor{satisfy} $s_i = s_j$ therefore $|s_i - s_j|^q = 0$.
The functions \revision{of their aligned regions} 
\revisionTVCG{$f'_1$}
and
\revisionTVCG{$f'_2$}
coincide on the intersection domain making 
$\big|\revisionTVCG{f'_1}(v) - \revisionTVCG{f'_2}(v)\big|^q = 0$ 
for all \revisionTVCG{$v \in \region'_i \cap \region'_j$}.
\revisionTVCGminor{Moreover}, since the regions are identical, there are no points outside the intersection domain, i.e. $\region'_i \setminus \region'_j = \region'_j \setminus \region'_i = \emptyset$.

Second, we \revision{want to} show that if $d_q^\region(b^\region_i, b^\region_j) = 0$ then \mbox{$b^\region_i = b^\region_j$}.
The term $|s_i - s_j|^q = 0$ forces $s_i = s_j$.
The sum over the intersection domain $\region'_i \cap \region'_j$ being 0 implies that for every point $v \in \region'_i \cap \region'_j$ then 
$\revisionTVCG{f'_1}(v) = \revisionTVCG{f'_2}(v)$.
For any point $x \in \region'_i \setminus \region'_j$, the deletion cost 
$\big|\revisionTVCG{f'_1}(x) - \projectionRegion_{f_2}(x)\big|^q = 0$
forces 
$\revisionTVCG{f'_1}(x) = \projectionRegion_{f_2}(x)$.
Similarly, for \revision{any point} $y \in \region'_j \setminus \region'_i$, the \revisionTVCG{value} 
$\revisionTVCG{f'_2}(y)$
must be \revision{equal to $\projectionRegion_{f_1}(y)$}.
By considering the extended functions $\tilde{f}_i$ and $\tilde{f}_j$ (\autoref{eq_extendFunction}), we see that the two functions agree everywhere on the combined domain. 
Combined with the equality $s_i = s_j$ and the fact that the alignment procedure makes the comparison of regions well-defined, we conclude that the two region-aware \revision{persistence} pairs are identical.

Therefore, $d_q^\region(b^\region_i, b^\region_j) = 0 \iff b^\region_i = b^\region_j$.

\textbf{\emph{(iii)} Symmetry:}
The absolute difference $|s_i - s_j|$ is symmetric.
The intersection $\region'_i \cap \region'_j$ is symmetric with respect to the two regions.
The deletion cost for $b^\region_i$ relative to $b^\region_j$ and the insertion cost for $b^\region_j$ relative to $b^\region_i$ swap roles if we reverse the order. In other words, when we exchange $b^\region_i$ and $b^\region_j$, the sums in lines for deletion and insertion exchange but their contributions remain the same.
Thus, $d_q^\region(b^\region_i, b^\region_j) = d_q^\region(b^\region_j, b^\region_i)$.

\textbf{\emph{(iv)} Triangle inequality:}
Let $b^\region_i$, $b^\region_j$ and $b^\region_k$ be three region-aware persistence pairs.
We want to show that the direct path between two points is never longer than going through an intermediate point, i.e. \mbox{$d_q^\region(b^\region_i, b^\region_k) \leq d_q^\region(b^\region_i, b^\region_j) + d_q^\region(b^\region_j, b^\region_k)$}.
For this, we will build on the definition of extended functions (\autoref{eq_extendFunction}) but this time for three region-aware persistence pairs, i.e. we define the three extended functions $\tilde{f}_i$, $\tilde{f}_j$ and $\tilde{f}_k$ like in \autoref{eq_extendFunction} but this time on $\Omega = \region'_i \cup \region'_j \cup \region'_k$.
Next, we can use the $L_q(\Omega)$ norm to measure the difference between these extended functions:

\begin{equation}
  \|\tilde{f}_i - \tilde{f}_j\|^q_{L_q(\Omega)} = \sum_{v \in \Omega} \big|\tilde{f}_i(v) - \tilde{f}_j(v)\big|^q,
\end{equation}

\noindent
allowing to rewrite the region-aware ground distance as:

\begin{equation}
  \label{metric_proof_ground_metric_rewrite}
  d_q^\region(b^\region_i, b^\region_j)^q = |s_i - s_j|^q + \|\tilde{f}_i - \tilde{f}_j\|^q_{L_q(\Omega)}.
\end{equation}

The \revision{$L_q$} norm (for $q \geq 1$) satisfies the triangle inequality by Minkowski's inequality: \mbox{$\|\tilde{f}_i - \tilde{f}_k\|_{L_q(\Omega)} \leq \|\tilde{f}_i - \tilde{f}_j\|_{L_q(\Omega)} + \|\tilde{f}_j - \tilde{f}_k\|_{L_q(\Omega)}$}.
Similarly, the absolute difference on the real numbers satisfies the triangle inequality: $|s_i - s_k| \leq |s_i - s_j| + |s_j - s_k|$.
This separation of the contribution of both terms leads to the definition of a vector in $\mathbb{R}^2$: 

\begin{equation}
    \label{eq_metric_vector}
    v_{ij} = \Big(|s_i - s_j|, \; \|\tilde{f}_i - \tilde{f}_j\|_{L_q(\Omega)}\Big),
\end{equation}

\noindent
then, the overall distance is simply the \revision{$L_q$} norm of this vector:

\begin{equation}
\label{eq_groundMetricVectorNorm}
    d_q^\region(b^\region_i, b^\region_j) = \|v_{ij}\| = \Big(|s_i - s_j|^q + \|\tilde{f}_i - \tilde{f}_j\|^q_{L_q(\Omega)}\Big)^{1/q}.
\end{equation}

Using the Minkowski inequality for the \revision{$L_q$} norm in $\mathbb{R}^2$, we know that $\|v_{ik}\| \leq \|v_{ij}\| + \|v_{jk}\|$.
Translating back, this inequality becomes:

\begin{equation}
\begin{aligned}
\nonumber
    \hspace{-0.12cm}
    \Big(|s_i - s_k|^q + \|\tilde{f}_i - \tilde{f}_k\|^q_{L_q(\Omega)}\Big)^{1/q} \hspace{-0.1cm} \leq & \Big(|s_i - s_j|^q + \|\tilde{f}_i - \tilde{f}_j\|^q_{L_q(\Omega)}\Big)^{1/q} \\
     + & \Big(|s_j - s_k|^q + \|\tilde{f}_j - \tilde{f}_k\|^q_{L_q(\Omega)}\Big)^{1/q},
\end{aligned}
\end{equation}

\noindent
using the definition of $d_q^\region$ in \autoref{eq_groundMetricVectorNorm}, it directly implies the triangle inequality for $d_q^\region$, i.e. \mbox{$d_q^\region(b^\region_i, b^\region_k) \leq d_q^\region(b^\region_i, b^\region_j) + d_q^\region(b^\region_j, b^\region_k)$}.

Finally, we want to show that the region-aware Wasserstein distance \revision{$\regionWassersteinTree[q]$} (and its special case for persistence diagrams \revision{$\regionWasserstein{q}$}) is a metric.
It consists of the Wasserstein distance between merge trees $\wassersteinTree$ using the region-aware ground metric $d_q^\region(b^\region_i, b^\region_j)$.
In their metric proof, Pont et al. \cite{pont_vis21} \revisionTVCGminor{show} that their distance is indeed a metric if the ground distance is also a metric.
Specifically, the mapping edit cost $\gamma\big(b_i \rightarrow \phi'(b_i)\big)$ in their proof must be replaced by $d_q^\region(b^\region_i, b^\region_j)$, and the deletion and insertion costs, $\gamma(b_i \rightarrow \emptyset)$ and $\gamma(\emptyset \rightarrow b_j)$, by respectively $d_q^\region\big(b^\region_i, \projection^\region(b^\region_i)\big)$ and $d_q^\region\big(\projection^\region(b^\region_j), b^\region_j, \big)$.
Therefore \revision{$\regionWassersteinTree[q]$} (and the special case \revision{$\regionWasserstein{q}$}) is a metric.

\revision{
\section{Metric proof using any background field}
\label{metric_proof_general_case}

In this section, we extend the proof of Appendix \ref{metric_proof} for the use of non-common background fields, i.e. $\projectionRegion_{f_1} \neq \projectionRegion_{f_2}$ given two region-aware BDTs $\branchtree^\region(f_1)$ and $\branchtree^\region(f_2)$.
In that case the triangle inequality shown in Appendix \ref{metric_proof} can break.
Specifically, \autoref{metric_proof_ground_metric_rewrite} is only valid for common background.
Indeed, recall that in \autoref{metric_proof_ground_metric_rewrite} we have $\Omega = \region'_i \cup \region'_j \cup \region'_k$ and rewriting the ground distance (\autoref{eq_our_groundMetric_appendix}) as \autoref{metric_proof_ground_metric_rewrite} necessitates that for every $x \in \Omega' = \Omega \setminus (\region'_i \cup \region'_j)$ we have $\|\tilde{f}_i - \tilde{f}_j\|^q_{L_q(\Omega')} = 0$ which is not the case if $\projectionRegion_{f_1} \neq \projectionRegion_{f_2}$.

\textbf{\emph{(i)} Triangle inequality counter-example in the base case:} Considering the data background fields (i.e. $\projectionRegion_f(x) = f(x)$) and three scalar fields $f = [m, a, \varepsilon, b, 2\varepsilon, 0, d]$, \mbox{$g = [m, a, \varepsilon, b, 0, 2\varepsilon, d]$}, $h = [m, a, \varepsilon, b, -c, 2\varepsilon, d]$, with $m \lt -c \lt 0 \lt \varepsilon \lt b \lt d \lt a$ and $\varepsilon$ very small, near $0$.
We also consider $q = 1$ for simplicity but the example works for every valid \revisionTVCGminor{value} of $q$.
There are three regions with same saddle values and with region values 
$[b, 2\varepsilon]$ \revisionTVCG{for $\region_1^f$}, $[b]$ \revisionTVCG{for $\region_1^g$} and $[b]$ \revisionTVCG{for $\region_1^h$},  
let respectively $b_f^\region$, $b_g^\region$ and $b_h^\region$ \revisionTVCGminor{be} the region-aware persistence pairs associated with these region values.
When comparing $b_f^\region$ to $b_h^\region$, the cost is $|2\varepsilon + c|$ (the vertex having value $2\varepsilon$ in $f$ is part of the symmetric difference and is compared, due to the use of the data background fields, to $-c$ in $h$).
However, from $b_f^\region$ to $b_g^\region$ we have a cost of $2\varepsilon$ and for $b_g^\region$ to $b_h^\region$ a cost of $0$.
Since $|2\varepsilon + c|$ can be arbitrarily large compared to $2\varepsilon$ we have $d_1^\region(b^\region_f, b^\region_h) \gt d_1^\region(b^\region_f, b^\region_g) + d_1^\region(b^\region_g, b^\region_h)$.

\textbf{\emph{(ii)} Triangle inequality for any background field:} 
Since the region-aware ground distance uses a partner-dependent domain, it can therefore \revisionTVCGminor{break} the triangle inequality when the values outside this common domain do not coincide (i.e. for non-common background fields).
Specifically, let three region-aware persistence pairs $b_f^\region$, $b_g^\region$ and $b_h^\region$ and their respective regions $\region_1$, $\region_2$ and $\region_3$.
When comparing $b_f^\region$ to $b_h^\region$ the domain used is $\region_1 \cup \region_3$, while being $\region_1 \cup \region_2$ for $b_f^\region$ to $b_g^\region$ and $\region_2 \cup \region_3$ for $b_g^\region$ to $b_h^\region$.
For a vertex $v \in \region_1 \cup \region_3$, it is possible that either $x \notin \region_1 \cup \region_2$ or $x \notin \region_2 \cup \region_3$ (but not both), therefore, the Minkowski's inequality \revisionTVCGminor{cannot} be applied.

To solve this, we redefine the comparison of two regions to be done on a fixed-size domain $\Lambda$.
Let a region-aware persistence pair 
$b^\region_{i} = (\revisionTVCG{\region_{i}^f}, s_{i})$ 
\revisionTVCGminor{of}
$\branchtree^\region(f)$  
\revisionTVCGminor{with}
$f : \revisionTVCGminor{\domain} \rightarrow \range$ \revisionTVCGminor{defined on a PL $d$-manifold $\domain$ triangulated by a simplicial complex $\complex$. For simplicity, we identify $\domain$ with the underlying space $|\complex|$. $f$ is specified on the vertices of $\complex$ and extended to the simplices of higher dimension using barycentric interpolation}.
We rewrite its extended function (\autoref{eq_extendFunction}) as:

\begin{equation}
\label{eq_extendFunction2}
  \tilde{f}_i(v) =
    \begin{cases}
      \revisionTVCG{f'}(v)
      & \text{if $v \in \region'_i$}\\
      \projectionRegion_f(v) & \text{if $v \notin \region'_i$ and $v \in \revisionTVCGminor{V(\complex}')$}\\
      0 & \text{if $v \notin \revisionTVCGminor{V(\complex}')$},
    \end{cases}
\end{equation}

\noindent
with $\tilde{f}_i : \Lambda \rightarrow \range$\revisionTVCGminor{,} $\revisionTVCGminor{\complex}'$ the 
\revisionTVCGminor{simplicial complex whose vertex coordinates are centered} at the extremum of $b^\region_{i}$ \revisionTVCGminor{and $V(\complex')$ its vertex set}.
Note the need of defining a default value for vertices not in $\revisionTVCGminor{V(\complex')}$ (third line).
Indeed, some vertices of the common fixed-size domain $\Lambda$ around the extremum could possibly not belong to $\revisionTVCGminor{\complex}'$ and not be a preimage under $\projectionRegion_f$.

Additionally, since $\Lambda$ can be arbitrarily large (possibly being the full domain in practice), vertices being far from a region can be part of $\Lambda$ when centering the region to its extremum.
In order to restore a notion of locality about the regions, the values of the extended function can be weighted (such as a decaying weight given the distance to the extremum), giving the weighted extended function $\Psi_i^f(v) = w_i(v)\tilde{f}_i(v)$.
Moreover, to avoid the multiple contribution of a vertex (being part of many $\Lambda$ after aligning various regions at their extremum), the weights of a same vertex $v$ can be normalized to enforce a partition of unity such that $\sum_i w_i(v) = 1$ for all regions $\region_i$.

Let 
$b^\region_{i} = (\revisionTVCG{\region_{i}^{f_1}}, s_{i})$ \revisionTVCG{from} $\branchtree^\region(f_1)$ and $b^\region_{j} = (\revisionTVCG{\region_{j}^{f_2}}, s_{j})$ \revisionTVCG{from} $\branchtree^\region(f_2)$ 
\revisionTVCGminor{be} two region-aware persistence pairs
and $\Psi_i^{f_1}$ the weighted extended function for the pair $i$ given $f_1$ (same for $\Psi_j^{f_2}$).
We can rewrite \autoref{eq_our_regionalDiscrepancy_appendix} as $\regionalDiscrepancy_q^\region(b^\region_i, b^\region_j) = \|\Psi_i^{f_1} - \Psi_j^{f_2}\|^q_{L_q(\Lambda)}$ and the region-aware ground metric (\autoref{eq_our_groundMetric_appendix}) as:

\begin{equation}
  d_q^\region(b^\region_i, b^\region_j) = \big(|s_i - s_j|^q + \|\Psi_i^{f_1} - \Psi_j^{f_2}\|^q_{L_q(\Lambda)}\big)^{1/q}.
\end{equation}

From this, the triangle inequality can be proven in the same way \revisionTVCGminor{as} in Appendix \ref{metric_proof} starting from \autoref{eq_metric_vector}.
Moreover, non-negativity, identity of indiscernibles and symmetry can be proven in the same way \revisionTVCGminor{as} in Appendix \ref{metric_proof}.
}

\revision{
\section{Stability of $\regionWasserstein{q}$}

In this section we study the stability of the region-aware Wasserstein distance for persistence diagrams $\regionWasserstein{q}$.
We 
prove, by excluding extremum-swap instabilities 
and by using data background, that $\regionWasserstein{q}$ can be bounded \revisionTVCGminor{from} above by the $L_q$ norm of the difference of the input scalar fields multiplied by a constant independent \revisionTVCGminor{of} the input data, specifically $\regionWasserstein{q}\big(\diagram^\region(f), \diagram^\region(g)\big) \leq 2^{1/q}\|f - g\|_q$.

For this, we take inspiration of the proof by Skraba and Turner \cite{skrabaStability} for the stability of the original Wasserstein between persistence diagrams $\wasserstein{q}$, that we extend to take into account the region properties.
The main idea of their proof is to consider a straight line homotopy between the input scalar fields $f$ and $g$ and to partition it into finitely many sub-intervals over which the vertex ordering induced by the scalar function remains unchanged.
In this way, they establish a local result on each sub-interval that can be \emph{combined} together to arrive at the final inequality $\wasserstein{q}\big(\diagram(f), \diagram(g)\big) \leq \|f - g\|_q$.

\subsection{
Homotopy and partition
}
Let $f : \revisionTVCGminor{\domain} \rightarrow \range$ and $g : \revisionTVCGminor{\domain} \rightarrow \range$ \revisionTVCGminor{be} two scalar fields \revisionTVCGminor{defined on a PL $d$-manifold $\domain$ triangulated by} a simplicial complex $\complex$. \revisionTVCGminor{For simplicity, we identify $\domain$ with the underlying space $|\complex|$. $f$ and $g$ are specified on the vertices of $\complex$} and \revisionTVCGminor{extended to} the simplices of higher dimension \revisionTVCGminor{using barycentric interpolation}.
Without loss of generality, they are considered injective on the vertices using symbolic perturbations \cite{edelsbrunner90} that break ties.
Let 
$\diagram^\region(f)$ and $\diagram^\region(g)$ \revisionTVCGminor{be} their region-aware persistence diagrams computed using the lower-star filtration.
Let a straight line homotopy between
$f$ and $g$ be $h_t = (1 - t) f + t g$ with $t \in [0, 1]$, i.e. their linear interpolation as $t$ varies.
There are finitely many times \mbox{$0 \lt t_1 \lt t_2 \lt \dots \lt t_n \lt 1$} at which two vertex values become equal, i.e. the order changes by a \emph{vertex swap}.
Specifically, let $t_0 = 0$ and $t_{n+1} = 1$, for each $t_k$ with $k \in \{1, \dots,  n\}$ there exist two vertices $\simplex_i$ and $\simplex_j$ such that $h_t(\simplex_i) \lt h_t(\simplex_j)$ (respectively $h_t(\simplex_i) \gt h_t(\simplex_j)$) for $t \in (t_{k-1}, t_k)$, $h_t(\simplex_i) = h_t(\simplex_j)$ for $t = t_k$ and $h_t(\simplex_i) \gt h_t(\simplex_j)$ (respectively $h_t(\simplex_i) \lt h_t(\simplex_j)$) for $t \in (t_{k}, t_{k+1})$.
On each open interval $I_k = (t_k, t_{k+1})$ the vertex order is fixed for all $h_t$ with $t \in I_k$.

\subsection{
Stability of $\regionWasserstein{q}$ for functions with same vertex order
}
\label{sec_stability_same_order}

We first consider the simple case where functions with same vertex order are considered, such as functions within the same interval $I_k = (t_k, t_{k+1})$ of the straight line homotopy $h_t$.
Let $a \in I_k$ and $b \in I_k$ and $a \lt b$, given that $h_a$ and $h_b$ have the same vertex order, the critical vertices, the persistence pairing and the region partition $\{R_i\}_i$ are the same for $\diagram^\region(h_a)$ and $\diagram^\region(h_b)$, they both contain the same number of region-aware persistence pairs $b^\region_i$ for all $i \in \{1, \dots, m\}$ with $m = |\diagram^\region(h_a)| = |\diagram^\region(h_b)|$.
Consider the assignment $\phi'$
mapping each 
$b^\region_i = (\revisionTVCG{\region_i^{h_a}}, s_i) \in \diagram^\region(h_a)$  
to its equivalent pair 
$b^\region_j = (\revisionTVCG{\region_j^{h_b}}, s_j) \in \diagram^\region(h_b)$,  
i.e. $\phi'(b^\region_i) = b^\region_j$ and $b^\region_i$ and $b^\region_j$ have the same region partition ($\region_i = \region_j$) and critical vertices, only their values can differ (but not their order).
\revisionTVCG{We will note $h'_{a,i}$ the corresponding centered function $h_a$ of the centered region of $b^\region_i$, similarly for $h'_{b,j}$ with $b^\region_j$.}

The region-aware Wasserstein distance is the minimum over all possible assignments and therefore bounded \revisionTVCGminor{from} above by the associated cost of $\phi'$ (line \ref{eq_stability_order_1}).
Since the matched pairs given $\phi'$ have the same region, the regional \revisionTVCGminor{discrepancy} $\regionalDiscrepancy_q^\region$ (line \ref{eq_stability_order_2}) consists only in the $L_q$ norm on the common domain of the matched pairs (line \ref{eq_stability_order_3}), i.e. there are no vertices to compare in their symmetric difference.
To simplify notation, let $A = \diagram^\region(h_a)$ and $\mathcal{W}^q_{ab} = \regionWasserstein{q}\big(\diagram^\region(h_a), \diagram^\region(h_b)\big)^q$.

\begin{eqnarray}
    \hspace{-0.52cm} \mathcal{W}^q_{ab} \hspace{-0.2cm} & \leq \hspace{-0.2cm} & \displaystyle \sum_{b^\region_i \in A} d_q^\region\big(b^\region_i, \phi'(b^\region_i)\big)^q \label{eq_stability_order_1} \\
    \hspace{-0.2cm} & = \hspace{-0.2cm} & \displaystyle \sum_{b^\region_i \in A} \Big( \big|s_i - s_j\big|^q + \regionalDiscrepancy_q^\region\big(b^\region_i, \phi'(b^\region_i)\big)\Big) \label{eq_stability_order_2} \\
    \hspace{-0.2cm} & = \hspace{-0.2cm} & \displaystyle \sum_{b^\region_i \in A} \Big( \big|s_i - s_j\big|^q + \displaystyle \sum_{v \in \region'_i} 
    \big|\revisionTVCG{h'_{a,i}}(v) - \revisionTVCG{h'_{b,i}}(v)\big|^q\Big)  
    \label{eq_stability_order_3} \\
    \hspace{-0.2cm} & = \hspace{-0.2cm} & \displaystyle \sum_{b^\region_i \in A} \big|s_i - s_j\big|^q + \displaystyle \sum_{b^\region_i \in A} \sum_{v \in \region'_i} 
    \big|\revisionTVCG{h'_{a,i}}(v) - \revisionTVCG{h'_{b,j}}(v)\big|^q  
    \label{eq_stability_order_4}.
\end{eqnarray}

\revisionTVCGminor{Let $\complexVertexSet = V(\complex)$ be the vertex set of $\complex$. Considering each saddle being involved in only one pair (see \appendixStabilityMultiSaddle{}), that} every saddle is a vertex \revisionTVCGminor{of $\complex$} and that the matched saddles given $\phi'$ are the same vertices of $\revisionTVCGminor{\complex}$ then:

\begin{equation}
    \label{eq_stability_saddle_upper_bound}
    \sum_{b^\region_i \in A} |s_i - s_j|^q \leq \sum_{v \in \revisionTVCGminor{\complexVertexSet}} |h_a(v) - h_b(v)|^q.
\end{equation}

Then, since each vertex lies in exactly one region $\region_i$ and that the regions of all $b^\region_i$ form a partition of $\revisionTVCGminor{\complexVertexSet}$ we have:

\begin{equation}
\begin{aligned}
    \label{eq_stability_region_upper_bound}
    \displaystyle \sum_{b^\region_i \in A} \sum_{v \in \region'_i} 
    \big|\revisionTVCG{h'_{a,i}}(v) - \revisionTVCG{h'_{b,j}}(v)\big|^q 
    & = \displaystyle \sum_{b^\region_i \in A} \sum_{v \in \region_i} \big|h_a(v) - h_b(v)\big|^q \\
    & = \displaystyle \sum_{v \in \revisionTVCGminor{\complexVertexSet}} \big|h_a(v) - h_b(v)\big|^q.
\end{aligned}
\end{equation}

Note that the term is first rewritten by using the vertices of the region without alignment using the original scalar field.
Using \autoref{eq_stability_saddle_upper_bound} and \autoref{eq_stability_region_upper_bound} in \autoref{eq_stability_order_4} gives:

\begin{equation}
    \mathcal{W}^q_{ab} \leq 2 \sum_{v \in \revisionTVCGminor{\complexVertexSet}} |h_a(v) - h_b(v)|^q = 2 \|h_a - h_b\|^q_q.
\end{equation}

Hence:

\begin{equation}
    \label{eq_stability_same_order}
    \regionWasserstein{q}\big(\diagram^\region(h_a), \diagram^\region(h_b)\big) \leq 2^{1/q}\|h_a - h_b\|_q,
\end{equation}

\noindent
for any two functions $h_a$ and $h_b$ having the same vertex order. 

\revisionTVCG{
\textbf{\emph{(i)} Presence of multi-saddles:} 
We generalize the previous result to the presence of multi-saddles, i.e. when a saddle vertex is shared by more than one persistence pair.
While this case did not occur in our experiments, we state the more general result for completeness.
Specifically, \autoref{eq_stability_saddle_upper_bound} assumes that each saddle vertex appears in at most one persistence pair, i.e. the functions $h_a$ and $h_b$ only contain simple-saddles.
If a saddle vertex $v$ is shared by several pairs, then the quantity $|h_a(v) - h_b(v)|$ is counted once per pair having $v$ as a saddle.
Therefore, with multi-saddles, \autoref{eq_stability_saddle_upper_bound} becomes:

\begin{equation}
    \label{eq_stability_saddle_upper_bound_multi_saddle}
    \sum_{b^\region_i \in A} |s_i - s_j|^q \leq \mu \sum_{v \in \revisionTVCGminor{\complexVertexSet}} |h_a(v) - h_b(v)|^q,
\end{equation}

\noindent
with $\mu = \max_{v \in \revisionTVCGminor{\complexVertexSet}} m(v)$ and $m(v)$ denotes an upper bound, depending only on the \revisionTVCGminor{simplicial complex $\complex$}, on the number of persistence pairs that can have $v$ as a saddle.
In the 0D lower-star setting, 
it is related to 
the maximal number of connected components that a lower link of $v$ can contain: if 
it
can have at most $k(v)$ connected components, then the insertion of $v$ would merge 
them to create $k(v) - 1$ persistence pairs, the oldest one continuing by Elder rule, hence $m(v) = k(v) - 1$.
Looking only at $\revisionTVCGminor{\complex}$, the maximal number of connected components of a lower link of $v$ corresponds to the independence number (in the graph-theoretic sense) of its link graph $G_v$ (the 1-skeleton of its link), since the 1-skeleton of the lower link is an induced subgraph of $G_v$ and its number of connected components is maximized when its vertices form a maximum independent set of $G_v$.
For the Freudenthal triangulation \cite{freudenthal42, kuhn60} of regular grids considered in our work (each interior vertex having 6 neighbors in 2D and 14 in 3D), this can be computed exactly with exhaustive enumeration of all vertex subsets and is $3$ in 2D and $6$ in 3D, resulting in respectively $\mu = 2$ and $\mu = 5$.
Hence, when multi-saddles are considered, using \autoref{eq_stability_saddle_upper_bound_multi_saddle} and \autoref{eq_stability_region_upper_bound} in \autoref{eq_stability_order_4} the multiplicative constant of $2^{1/q}$ in \autoref{eq_stability_same_order} becomes in general $(\mu + 1)^{1/q}$, being $3^{1/q}$ for 2D Freudenthal triangulation of regular grids and $6^{1/q}$ in 3D.

}

\textbf{\emph{(\revisionTVCG{ii})} Fixed-size domain and vertex weights:} 
We now prove the stability of $\regionWasserstein{q}$ for functions with same vertex order when the region-aware ground metric uses a fixed-size domain and when vertex weights are considered (as described in Appendix \ref{metric_proof_general_case}).
We will consider weights that decay given the distance to the extremum of a region and are greater than or equal to 0.
Since $h_a$ and $h_b$ have the same vertex order they have the same extrema vertices, therefore, the weight for each vertex for each region are equal, i.e. $w^{h_a}_i(v) = w^{h_b}_i(v) = w_i(v)$. 

We consider the assignment $\phi'$ described in \autoref{sec_stability_same_order}. 
Let $A = \diagram^\region(h_a)$. 
To simplify notation, we write the sum of all regional discrepancy terms as $\mathcal{C}_{\phi'} = \sum_{b^\region_i \in A} \regionalDiscrepancy_q^\region\big(b^\region_i, \phi'(b^\region_i)\big)$ and $v'$ the vertex $v \in \revisionTVCGminor{\complexVertexSet}$ after aligning the domain to the extremum of the region $\region_i$.
Since the matched pairs given $\phi'$ shared the same extrema, the vertices of $\Lambda$ being compared using  their extended functions are therefore a subset of the vertices of $\revisionTVCGminor{\complex}$ allowing to bound line \ref{eq_stability_region_upper_bound_2_2} by line \ref{eq_stability_region_upper_bound_2_3}.
Then, since $\sum_{b^\region_i \in A} w_i(v) = 1$, that each $w_i(v) \geq 0$ and that $q \geq 1$ then $\sum_{b^\region_i \in A} w_i(v)^q \leq 1$ allowing to bound line \ref{eq_stability_region_upper_bound_2_4} by line \ref{eq_stability_region_upper_bound_2_5}. 

\begin{eqnarray}
    \mathcal{C}_{\phi'} \hspace{-0.2cm} & = \hspace{-0.2cm} & \displaystyle \sum_{b^\region_i \in A} \|\Psi_i^{h_a} - \Psi_j^{h_b}\|^q_{L_q(\Lambda)} \label{eq_stability_region_upper_bound_2_1} \\ 
    \hspace{-0.2cm} & = \hspace{-0.2cm} & \displaystyle \sum_{b^\region_i \in A} \sum_{v' \in \Lambda} \big|w_i(v) \tilde{h}_{a_i}(v') - w_i(v) \tilde{h}_{b_j}(v')\big|^q \label{eq_stability_region_upper_bound_2_2} \\
    \hspace{-0.2cm} & \leq \hspace{-0.2cm} & \displaystyle \sum_{b^\region_i \in A} \sum_{v \in \revisionTVCGminor{\complexVertexSet}} w_i(v)^q \big|h_a(v) - h_b(v)\big|^q \label{eq_stability_region_upper_bound_2_3} \\
    \hspace{-0.2cm} & = \hspace{-0.2cm} & \displaystyle \sum_{v \in \revisionTVCGminor{\complexVertexSet}} \sum_{b^\region_i \in A} w_i(v)^q \big|h_a(v) - h_b(v)\big|^q \label{eq_stability_region_upper_bound_2_4} \\
    \hspace{-0.2cm} & \leq \hspace{-0.2cm} & \displaystyle \sum_{v \in \revisionTVCGminor{\complexVertexSet}} \big|h_a(v) - h_b(v)\big|^q. \label{eq_stability_region_upper_bound_2_5}
\end{eqnarray}

Hence:

\begin{equation}
\label{eq_stability_region_upper_bound_2}
    \sum_{b^\region_i \in A} \regionalDiscrepancy_q^\region\big(b^\region_i, \phi'(b^\region_i)\big) \leq \sum_{v \in \revisionTVCGminor{\complexVertexSet}} |h_a(v) - h_b(v)|^q.
\end{equation}

Writing $\mathcal{W}^q_{ab} = \regionWasserstein{q}\big(\diagram^\region(h_a), \diagram^\region(h_b)\big)^q$, we have:

\begin{eqnarray}
    \hspace{-0.52cm} \mathcal{W}^q_{ab} \hspace{-0.2cm} & \leq \hspace{-0.2cm} & \displaystyle \sum_{b^\region_i \in A} d_q^\region\big(b^\region_i, \phi'(b^\region_i)\big)^q \label{eq_stability_order_2_1} \\
    \hspace{-0.2cm} & = \hspace{-0.2cm} & \displaystyle \sum_{b^\region_i \in A} \Big( \big|s_i - s_j\big|^q + \regionalDiscrepancy_q^\region\big(b^\region_i, \phi'(b^\region_i)\big)\Big) \label{eq_stability_order_2_2} \\
    \hspace{-0.2cm} & = \hspace{-0.2cm} & \displaystyle \sum_{b^\region_i \in A} \big|s_i - s_j\big|^q + \displaystyle \sum_{b^\region_i \in A} \regionalDiscrepancy_q^\region\big(b^\region_i, \phi'(b^\region_i)\big) \label{eq_stability_order_2_3}.
\end{eqnarray}

Using \autoref{eq_stability_saddle_upper_bound} and \autoref{eq_stability_region_upper_bound_2} in \autoref{eq_stability_order_2_3} gives:

\begin{equation}
    \mathcal{W}^q_{ab} \leq 2 \sum_{v \in \revisionTVCGminor{\complexVertexSet}} |h_a(v) - h_b(v)|^q = 2 \|h_a - h_b\|^q_q.
\end{equation}

Hence:

\begin{equation}
    \label{eq_stability_same_order_2}
    \regionWasserstein{q}\big(\diagram^\region(h_a), \diagram^\region(h_b)\big) \leq 2^{1/q}\|h_a - h_b\|_q,
\end{equation}

\noindent
for any two functions $h_a$ and $h_b$ having the same vertex order when using a fixed-size domain and vertex weights.
\revisionTVCG{When multi-saddles are considered, the constant $2^{1/q}$ becomes $(\mu + 1)^{1/q}$ in general (see \autoref{sec_stability_same_order}-(i)), being $3^{1/q}$ for 2D Freudenthal triangulation of regular grids and $6^{1/q}$ in 3D.}

\subsection{
Stability of $\regionWasserstein{q}$ for any functions
}

We have shown in the previous section a local stability result for scalar fields having the same vertex order, such as those within a same interval $I_k = (t_k, t_{k+1})$ of the partition of the homotopy $h_t$.
To prove the stability of $\regionWasserstein{q}$ for any functions (when no extremum-swap instabilities are considered) we have to \emph{combine} together all the local results over all intervals $I_k$ with $k \in \{0, \dots, n\}$ of the homotopy $h_t$.

Recall that all scalar fields $h_t$ with values of $t$ in the same interval $I_k = (t_k, t_{k+1})$ share a common vertex order.
Moreover, since $f$ and $g$ are considered to be injective on the vertices of $\revisionTVCGminor{\complex}$, the common vertex order within $I_k$ is a \emph{strict total} order (an order without equalities), because ties can only occur at the endpoints $t_k$ and $t_{k+1}$. 
At these endpoints, the value of two vertices become equal and the vertex orders of $h_{t_k}$ and $h_{t_{k+1}}$ become \emph{partial} (i.e. contain equalities).
A partial order can be extended to various strict total orders depending on how ties are broken.
Specifically, since $h_{t_k}$ and $h_{t_{k+1}}$ are the endpoints of $I_k$, 
the common strict total order within $I_k$ is one possible (\emph{linear}) extension of the partial orders of $h_{t_k}$ and $h_{t_{k+1}}$.
Let $\diagram^\region_k(h_{t_{k}})$ and $\diagram^\region_k(h_{t_{k+1}})$ \revisionTVCGminor{be} the diagram of respectively $h_{t_{k}}$ and $h_{t_{k+1}}$ computed using the common order within $I_k$.
Since they share the same order we can use the result of \autoref{sec_stability_same_order} (i.e. \autoref{eq_stability_same_order} or \autoref{eq_stability_same_order_2}) for these diagrams:

\begin{equation}
\begin{aligned}
    \label{eq_stability_interval_stability}
    \regionWasserstein{q}\big(\diagram^\region_k(h_{t_{k}}), \diagram^\region_k(h_{t_{k+1}})\big) & \leq 2^{1/q}\|h_{t_k} - h_{t_{k+1}}\|_q \\ 
    & = 2^{1/q} (t_{k+1} - t_k) \|f - g\|_q.
\end{aligned}
\end{equation}

The last line follows from rewriting the term $h_{t_k} - h_{t_{k+1}}$ as $(t_{k+1} - t_k) (f - g)$ and due to $\|c u\|_q = |c|\|u\|_q$ for any scalar $c$.

Now that we have a local stability result for the endpoints of an interval $I_k$, to prove the stability of $\regionWasserstein{q}$ we aim to combine such result over all intervals of the homotopy.
Specifically, recall that $h_{t_0} = f$ and $h_{t_{n+1}} = g$, we will use the triangle inequality to bound \revisionTVCGminor{from} above $\regionWasserstein{q}\big(\diagram^\region(f), \diagram^\region(g)\big)$ by the sum of the distances between the scalar fields of all consecutive $t_k$ with $k \in \{0, \dots, n+1\}$ that will be shown to be itself bounded \revisionTVCGminor{from} above by the $L_q$ norm of the difference between $f$ and $g$ (up to a factor independent of $f$ and $g$).

Given two intervals $I_k = (t_k, t_{k+1})$ and $I_{k+1} = (t_{k+1}, t_{k+2})$, 
the distance between $\diagram^\region_k(h_{t_k})$ and $\diagram^\region_{k+1}(h_{t_{k+2}})$, i.e. the diagram of the endpoints of these two intervals given the common order of respectively $I_k$ and $I_{k+1}$, can be bounded \revisionTVCGminor{from} above (using the triangle inequality) by:

\begin{align}
    \label{eq_stability_two_intervals}
    \hspace{-0.16cm} \regionWasserstein{q}\big(\diagram^\region_k(h_{t_k}), \diagram^\region_{k+1}(h_{t_{k+2}})\big) \leq & \regionWasserstein{q}\big(\diagram^\region_k(h_{t_k}), \diagram^\region_k(h_{t_{k+1}})\big) \\
    + & \regionWasserstein{q}\big(\diagram^\region_k(h_{t_{k+1}}), \diagram^\region_{k+1}(h_{t_{k+1}})\big) \label{eq_stability_two_intervals_2} \\
    + & \regionWasserstein{q}\big(\diagram^\region_{k+1}(h_{t_{k+1}}), \diagram^\region_{k+1}(h_{t_{k+2}})\big), \label{eq_stability_two_intervals_3}
\end{align}

\noindent
i.e. by the distance between the diagrams of the endpoints of $I_k$ given its common order (line \ref{eq_stability_two_intervals}), plus the distance between the diagrams of the endpoints of $I_{k+1}$ given its common order (line \ref{eq_stability_two_intervals_3}), plus a \emph{bridge} term corresponding to the distance between the diagram of $h_{t_{k+1}}$ given the common order within $I_k$ and the one given the common order within $I_{k+1}$ (line \ref{eq_stability_two_intervals_2}).

\textbf{Remark:} the choice of how ties are broken does not affect the resulting (classical) persistence diagram \cite{skrabaStability}, i.e. 
for any strict total order extending a given partial order, the birth–death pairs (intervals) are the same, even if the specific critical vertices can differ. 
Consequently, in the classical case, the bridge term at a swap time $t_k$ is $0$, and the proof closes using \autoref{eq_stability_proof} alone.
By contrast, for region-aware persistence diagrams, region memberships do depend on the tie-breaking (a vertex could change region depending on the order choice), so the diagrams at the swap time may differ. 
We therefore need to control this bridge term in what follows.

\textbf{\emph{(i)} No possible bound using null background:} 
We first show with a counter-example that $\regionWasserstein{q}$ with the null background \revisionTVCGminor{cannot} be bounded \revisionTVCGminor{from} above by the $L_q$ norm of the difference between the two input scalar fields (up to a factor).

Let $f$ and $g$ \revisionTVCGminor{be} two scalar fields described in \autoref{fig_stability_null_counter_example} having respectively the same two extrema $a$ and $b$.
We will use $q = 1$ for simplicity but the example works for every valid \revisionTVCGminor{value} of $q$.
The $L_1$ norm of the difference between $f$ and $g$ is \mbox{$\|f - g\|_1 = 4\varepsilon$} (with $\varepsilon$ very small).
They only differ by the swap of two vertices having very close values.
When comparing the regions of $b$ in respectively $f$ and $g$, the vertex of the region $b$ in $f$ with value $c + \varepsilon$ is part of the symmetric difference (it has no \revisionTVCGminor{corresponding} vertex in the region of $b$ in $g$), due to the null background, it is therefore compared to $0$.
The difference $|c+\varepsilon-0|$ can be arbitrarily large compared to $4\varepsilon$ hence there is no constant $C$ such that $\regionWasserstein{1}\big(\diagram^\region(f), \diagram^\region(g)\big) \leq C \|f - g\|_1$.

\revisionTVCGminor{Regarding} the stability \revisionTVCGminor{proof} of $\regionWasserstein{q}$, recall that at a swap time $t_{k+1}$ a vertex $x$ could change region, i.e. $x \in \region$ in $\diagram^\region_k(h_{t_{k+1}})$ but $x \notin \region$ in $\diagram^\region_{k+1}(h_{t_{k+1}})$ for a region $\region$ in both diagrams.
When comparing both $\region$ regions in \mbox{both diagrams}, the contribution of $x$, coming from the symmetric difference, is $|h_{t_{k+1}}(x) - 0|^q$\revisionTVCGminor{, being} a comparison to $0$ that can be arbitrarily large.
Hence, the bridge term (line \ref{eq_stability_two_intervals_2}) \revisionTVCGminor{does} not vanish and \revisionTVCGminor{cannot} be bounded by $C \|f - g\|_q$ for some constant $C$.
Consequently, summing the local interval bounds cannot yield a global stability bound for $\regionWasserstein{q}$ using the null background.

\begin{figure}
    \centering
    \begin{tikzpicture}[
      line cap=round, line join=round, thick,
      dot/.style={circle,fill,inner sep=1.6pt},
      smalldot/.style={circle,fill,inner sep=1.0pt},
      lab/.style={font=\footnotesize,inner sep=1pt}
    ]
    
    \definecolor{reg1}{rgb}{0.3,0.6,0.9}
    \definecolor{reg2}{rgb}{0.82,0.98,0.96}
    \definecolor{border}{rgb}{0.3,0.35,0.35}

    \path
      (0.0, 0.0) coordinate(L0)
      (0.7, 1.5) coordinate(L1)   
      (1.4, 0.5) coordinate(L3)   
      (2.1, 0.6) coordinate(L4)   
      (2.8, 1.2) coordinate(L5)   
      (3.5, 0.1) coordinate(L6);
    \draw (L0)--(L1)--(L3)--(L4)--(L5)--(L6);

    \node[dot,fill=reg1,draw=border] at (L1) {}; \node[lab,above] at ($(L1)+(0,+0.12)$) {$a$};
    \node[lab,above,font=\large] at ($(0.0,1.5)+(0,+0.12)$) {$f$};
    \node[dot,fill=reg2,draw=border] at (L5) {}; \node[lab,above] at ($(L5)+(0,+0.12)$) {$b$};
    \node[dot,fill=reg1,draw=border] at (L3) {}; \node[lab,below] at ($(L3)+(0,-0.12)$) {$c-\varepsilon$};
    \node[dot,fill=reg2,draw=border] at (L4) {}; \node[lab,below] at ($(L4)+(0,-0.12)$) {$c+\varepsilon$};
    \node[dot,fill=reg1,draw=border] at (L0) {}; \node[lab,left] at ($(L0)+(-0.12,0)$) {$0$};
    \node[dot,fill=reg1,draw=border] at (L6) {}; \node[lab,right] at ($(L6)+(+0.12,0)$) {$d$};

    \path
      (4.6, 0.0) coordinate(R0)
      (5.3, 1.5) coordinate(R1)  
      (6.0, 0.6) coordinate(R3)  
      (6.7, 0.5) coordinate(R4)  
      (7.4, 1.2) coordinate(R5)  
      (8.1, 0.1) coordinate(R6);
    \draw (R0)--(R1)--(R3)--(R4)--(R5)--(R6);

    \node[dot,fill=reg1,draw=border] at (R1) {}; \node[lab,above] at ($(R1)+(0,+0.12)$) {$a$};
    \node[lab,above,font=\large] at ($(4.6,1.5)+(0,+0.12)$) {$g$};
    \node[dot,fill=reg2,draw=border] at (R5) {}; \node[lab,above] at ($(R5)+(0,+0.12)$) {$b$};
    \node[dot,fill=reg1,draw=border] at (R3) {}; \node[lab,below] at ($(R3)+(0,-0.12)$) {$c+\varepsilon$};
    \node[dot,fill=reg1,draw=border] at (R4) {}; \node[lab,below] at ($(R4)+(0,-0.12)$) {$c-\varepsilon$};
    \node[dot,fill=reg1,draw=border] at (R0) {}; \node[lab,left] at ($(R0)+(-0.12,0)$) {$0$};
    \node[dot,fill=reg1,draw=border] at (R6) {}; \node[lab,right] at ($(R6)+(+0.12,0)$) {$d$};

    \end{tikzpicture}
    \caption{\revision{
The scalar fields considered for the counter-example that shows the impossibility for the null background to have a stability bound for $\regionWasserstein{q}$.
The scalar fields $f$ and $g$ are defined on the same one-dimensional domain.
The labels indicate the values of the corresponding vertices and their position in the y-axis indicate their value order (for instance $a \gt b$ and $\varepsilon$ is very small).
The nodes color represent the (maximum) region membership of the vertices to the two regions having respectively $a$ and $b$ as extremum.
}}
    \label{fig_stability_null_counter_example}
\end{figure}

\textbf{\emph{(ii)} Stability using data background:} 
Recall that no extremum-swap instabilities are considered, therefore, at a swap time $t_{k+1}$, the extrema vertices of $\diagram^\region_k(h_{t_{k+1}})$ and $\diagram^\region_{k+1}(h_{t_{k+1}})$ are the same (even if the region membership of various vertices can change due to the swap of the values of two saddle, and/or regular, vertices).
A notable exception is the possible appearance or disappearance (depending on the common order within $I_k$ and $I_{k+1}$) of a zero-persistence pair formed by the two vertices that tie at $t_{k+1}$ (having equal value).

Consider the assignment $\phi'$ that maps each pair 
\mbox{$b^\region_i = (\revisionTVCG{\region_i^{h_{t_{k+1}}}}, s_i) \in \diagram^\region_k(h_{t_{k+1}})$}  
to the pair having the same extremum vertex 
$b^\region_j = (\revisionTVCG{\region_j^{h_{t_{k+1}}}}, s_j) \in \diagram^\region_{k+1}(h_{t_{k+1}})$.
If applicable, $\phi'$ also maps to the diagonal the zero-persistence pair of the two vertices that tie with a cost of $0$.
Let $\Psi_i^{h_{t_{k+1}}}$ and $\Psi_j^{h_{t_{k+1}}}$ the weighted extended functions using the data background (Appendix \ref{metric_proof_general_case}) of respectively $b^\region_i$ and $b^\region_j$.
Since the extrema vertices are the same and the scalar values also are, then $\Psi_i^{h_{t_{k+1}}}$ and $\Psi_j^{h_{t_{k+1}}}$ coincide everywhere and the regional discrepancy is $\|\Psi_i^{h_{t_{k+1}}} - \Psi_j^{h_{t_{k+1}}}\|^q_{L_q(\Lambda)} = 0$.
Moreover, for each pair, either its saddle is unchanged for both orders or it is one of the two vertices that tie, in either case the saddle values $s_i$ and $s_j$ are also equal and $|s_i - s_j|^q = 0$.
We conclude that $d_q^\region(b^\region_i, b^\region_j) = 0$ for all pairs, hence $\regionWasserstein{q}\big(\diagram^\region_k(h_{t_{k+1}}), \diagram^\region_{k+1}(h_{t_{k+1}})\big) = 0$.

Coming back to the combination of two intervals (\autoref{eq_stability_two_intervals}), the bridge term (line \ref{eq_stability_two_intervals_2}) being $0$, we can rewrite \autoref{eq_stability_two_intervals} as:

\begin{equation}
\begin{aligned}
    \nonumber
    \hspace{-0.1cm} 
    \regionWasserstein{q}\big(\diagram^\region_k(h_{t_k}), \diagram^\region_{k+1}(h_{t_{k+2}})\big) \leq & \regionWasserstein{q}\big(\diagram^\region_k(h_{t_k}), \diagram^\region_k(h_{t_{k+1}})\big) \\
    + & \regionWasserstein{q}\big(\diagram^\region_{k+1}(h_{t_{k+1}}), \diagram^\region_{k+1}(h_{t_{k+2}})\big).
\end{aligned}
\end{equation}

We can now generalize this result to not only over 2 intervals $I_k$ and $I_{k+1}$ but over all intervals $I_k$ with $k \in \{0, \dots, n\}$:

\begin{equation}
    \nonumber
    \regionWasserstein{q}\big(\diagram^\region_0(h_{t_0}), \diagram^\region_{n}(h_{t_{n+1}})\big) \leq  \sum_{k=0}^n \regionWasserstein{q}\big(\diagram^\region_k(h_{t_k}), \diagram^\region_k(h_{t_{k+1}})\big).
\end{equation}

Note that here $\diagram^\region_0(h_{t_0})$ \revisionTVCGminor{corresponds} to the diagram of $h_{t_0}$ computed using the common order in $I_0$, not \revisionTVCGminor{to be confused} with the diagram of dimension $0$ (same remark for $\diagram^\region_{n}(h_{t_{n+1}})$).

Recall that $h_{t_0} = f$ and $h_{t_{n+1}} = g$, we can therefore write $\diagram^\region(f) = \diagram^\region_0(h_{t_0})$ and $\diagram^\region(g) = \diagram^\region_{n}(h_{t_{n+1}})$ (since vertex orders agree).
Moreover, to simplify notation, we write $\regionWasserstein{q}\big(\diagram^\region(h_{t_k}), \diagram^\region(h_{t_{k+1}})\big) = \regionWasserstein{q}\big(\diagram^\region_k(h_{t_k}), \diagram^\region_k(h_{t_{k+1}})\big)$.
Then, using \autoref{eq_stability_interval_stability}, we have:

\begin{equation}
\begin{aligned}
    \label{eq_stability_proof}
    \regionWasserstein{q}\big(\diagram^\region(f), \diagram^\region(g)\big) & \leq \sum_{k=0}^n \regionWasserstein{q}\big(\diagram^\region(h_{t_k}), \diagram^\region(h_{t_{k+1}})\big) \\ 
    & \leq \sum_{k=0}^n 2^{1/q} \|h_{t_k} - h_{t_{k+1}}\|_q \\
    & = \sum_{k=0}^n 2^{1/q} (t_{k+1} - t_k) \|f - g\|_q \\
    & = 2^{1/q} \|f - g\|_q,
\end{aligned}
\end{equation}

Hence:

\begin{equation}
    \regionWasserstein{q}\big(\diagram^\region(f), \diagram^\region(g)\big) \leq 2^{1/q}\|f - g\|_q,
\end{equation}

\noindent
for any two functions $f$ and $g$ 
\revisionTVCG{without}
extremum-swap instabilities.
\revisionTVCG{When multi-saddles are considered, the constant $2^{1/q}$ becomes $(\mu + 1)^{1/q}$ in general (\autoref{sec_stability_same_order}-(i)), being $3^{1/q}$ for 2D Freudenthal triangulation of regular grids and $6^{1/q}$ in 3D.}

}

\revisionTVCG{
\section{Empirical stability evaluation}

We now extend the empirical stability evaluation of our method (\secFrameworkQuality{} of the main manuscript) to real-life datasets. First, on a single scalar field, and then in an ensemble manner.

\subsection{Single field empirical stability}

\begin{figure}[t]
    \centering
    \includegraphics[width=\linewidth]{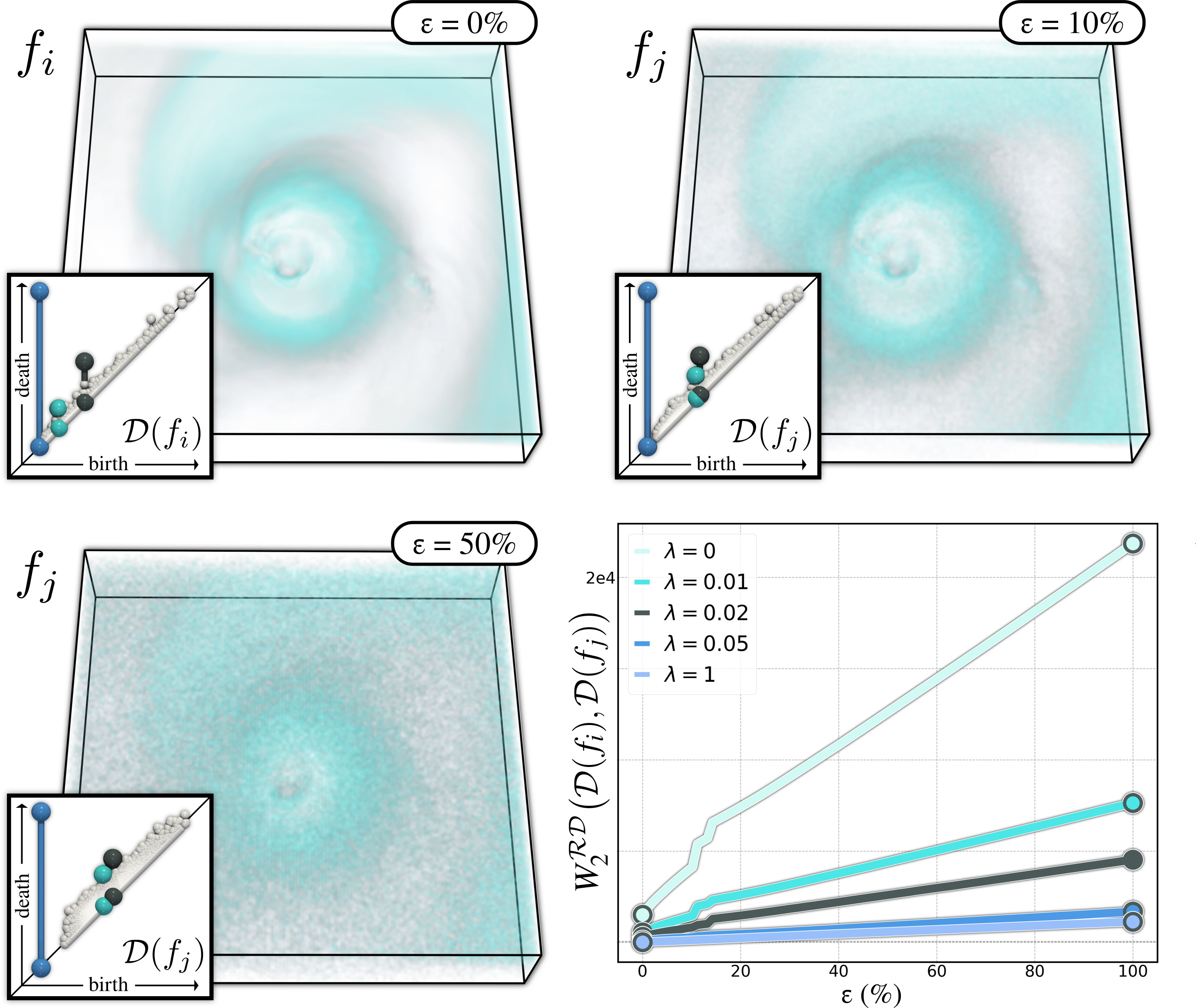}
    \caption{
\revisionTVCG{
Empirical stability evaluation of $\regionWasserstein{2}$. 
A noisy version $f_j$ of the 24th timestep of the isabel ensemble $f_i$ is created by adding a noise of increasing amplitude $\epsilon$.
The evolution of $\regionWasserstein{2}\big(\diagram(f_i), \diagram(f_j)\big)$ given $\epsilon$ and for different values of the subsampling parameter $\subsampleParameter$ shows no sudden spikes or drops that would be related to high instabilities, but rather a relatively stable evolution. 
Minor irregularities can be noticed between 10\% and 15\% of noise amplitude.
}
    }
    \label{fig_stability_isabel}
\end{figure}

\autoref{fig_stability_isabel} extends \figStability{} of the main manuscript by using a specific timestep of the isabel ensemble \cite{scivis2004} and the same evaluation protocol.
Noisy variants $f_j$ of the 24th timestep of the isabel ensemble $f_i$ (being in the middle of the time-series) are created with a noise of increasing amplitude $\epsilon$, i.e. $\| f_i - f_j \|_{\infty} \leq \epsilon$ (shown in the figure as a percentage of the scalar range of $f_i$).
Then, the distance $\regionWasserstein{2}\big(\diagram^\region(f_i), \diagram^\region(f_j)\big)$ is computed and its evolution displayed given $\epsilon$ for different values of the subsampling parameter $\subsampleParameter$.
Recall that when $\subsampleParameter = 1$ we have $\regionWasserstein{2}\big(\diagram^\region(f_i), \diagram^\region(f_j)\big) = \wasserstein{2}\big(\diagram(f_i), \diagram(f_j)\big)$ (\secDistancePropertiesGeneralization{} of the main manuscript), i.e. the region-aware Wasserstein distance get back to the original Wasserstein distance which is proven to be stable \cite{CohenSteinerEH05} and we indeed observe its evolution to be linear (blue line at the bottom in \autoref{fig_stability_isabel}).

For $\regionWasserstein{2}$, some minor irregularities can be noticed between approximately 10\% and 15\% of noise amplitude (e.g., light cyan curve for $\subsampleParameter = 0$ in \autoref{fig_stability_isabel}), likely reflecting the appearance of noise-induced features and modifications of existing ones, so that the assumptions underlying the stability result may no longer hold.
Importantly, these fluctuations remain limited and do not alter the overall smooth increasing trend. 
Before and after these noise amplitudes we observe the evolution of the distance to have a linear slope.
Overall, the message remains somewhat similar as for \figStability{}: for the various values of $\subsampleParameter$ we observe a stable evolution of the distance and no sudden spikes or drops that would be related to high instabilities.

\begin{figure*}
    \centering
    \includegraphics[width=\linewidth]{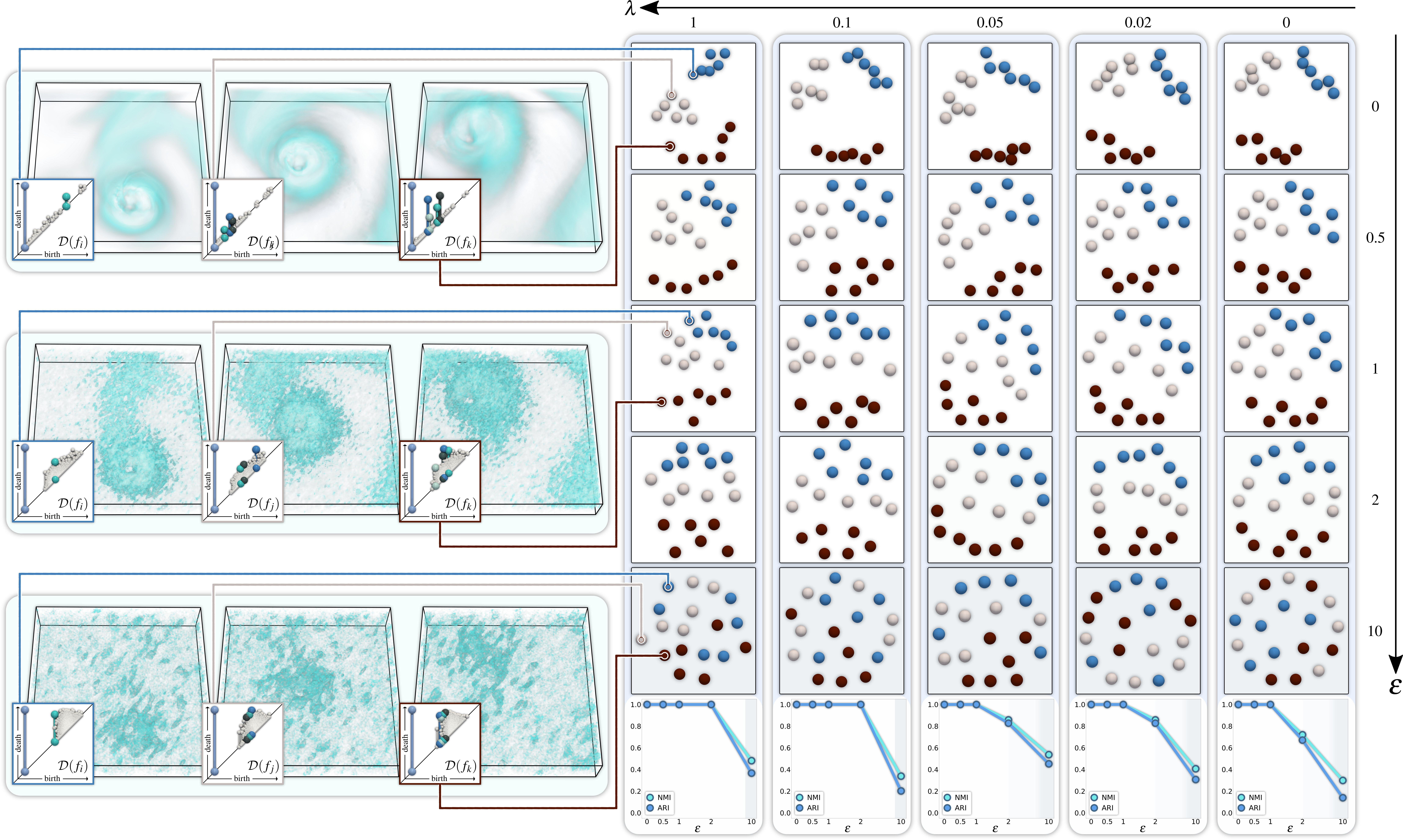}
    \caption{
\revisionTVCG{
Empirical stability evaluation of $\regionWasserstein{2}$ of the isabel ensemble for different values of the subsampling parameter $\subsampleParameter$ and different noise amplitudes $\epsilon$.
A subset of 18 members of the ensemble is selected to form three clusters, yielding a ground-truth classification (clusters in blue, white and dark red).
Then, various noisy versions of the ensemble are created by adding to each member a member-specific noise of increasing amplitude $\epsilon$, giving five versions of the ensemble (from $\epsilon = 0$ to $\epsilon = 10$, top to bottom).
For each ensemble, a 2D embedding is generated using MDS for various values of $\subsampleParameter$ (from the Wasserstein distance between persistence diagrams $\wasserstein{2}$ for $\subsampleParameter = 1$ to progressive blends towards the strict region-aware Wasserstein distance $\regionWasserstein{2}$ for $\subsampleParameter = 0$).
For each $\subsampleParameter$ value, the evolution of the quality scores NMI and ARI of the embedding given $\epsilon$ is shown (bottom curves).
In the 2D embedding and the quality scores curves, a grey background indicates clusters being totally mixed together.
While the quality scores slightly worsen overall when $\subsampleParameter$ decreases, we observe that the different methods share the same overall level of robustness, the clusters are progressively merged together but still visually distinguishable until $\epsilon = 2$. 
}
    }
    \label{fig_stability_ensemble}
\end{figure*}

\subsection{Ensemble empirical stability}

We now extend \figStability{} of the main manuscript and \autoref{fig_stability_isabel} to study the empirical stability of our method not only for a single field but for an ensemble of real-life data instances.

As for these two figures, the evaluation is done for various values of the subsampling parameter $\subsampleParameter$, specifically, \mbox{$\lambda \in \{1, 0.1, 0.05, 0.02, 0\}$}.
Recall that the proposed region-aware Wasserstein distance $\regionWasserstein{2}$ reduces to original Wasserstein distance $\wasserstein{2}$ when $\subsampleParameter = 1$ (\secDistancePropertiesGeneralization{} of the main manuscript).
A subset of 18 members of the isabel ensemble is made by selecting three distinct phases of the time-series.
It results in three clusters (6 members each) being relatively well separated in the noiseless MDS embeddings of all considered values of $\subsampleParameter$ (\autoref{fig_stability_ensemble}, top embeddings row) and providing a controlled ground-truth classification (clusters in blue, white and dark red in \autoref{fig_stability_ensemble}) for the stability evaluation.
From this ensemble, five different versions are created using a member-specific noise of increasing amplitude $\epsilon \in \{0, 0.5, 1, 2, 10\}$.
Specifically, for each member, a noise field is generated with an amplitude of $\epsilon$ times the maximum scalar range over the original noiseless ensemble.
A single noise realization is used for each ensemble member and scaled across amplitudes. 
This ensures that differences observed between noise levels are attributable to the perturbation magnitude itself, rather than to changes in the random noise pattern. 
At the same time, different noise realizations are used across members to avoid introducing artificial correlations within the ensemble.

For each combination of the parameters $\subsampleParameter$ and $\epsilon$, a distance matrix can be computed for the ensemble using $\regionWasserstein{2}$ and then be used as an input for MDS to create a 2D embedding, resulting in a total of 25 embeddings.
For each embedding we compute the NMI and ARI to evaluate how much the clusters are well separated and display their evolution given the noise amplitude $\epsilon$ for each $\subsampleParameter$ value (curves in \autoref{fig_stability_ensemble}, bottom).

As the noise amplitude $\epsilon$ increases, the three clusters of the MDS embeddings progressively merge together (\autoref{fig_stability_ensemble}, top to bottom, for each $\subsampleParameter$ value), with the members of the white cluster gradually moving between the other two until $\epsilon = 2$. 
After that, for $\epsilon = 10$, the three clusters strongly overlap and are no longer visually distinguishable from each other resulting in low NMI and ARI for all methods (\autoref{fig_stability_ensemble}, bottom curves).

\begin{figure*}
    \centering
    \includegraphics[width=0.96\linewidth]{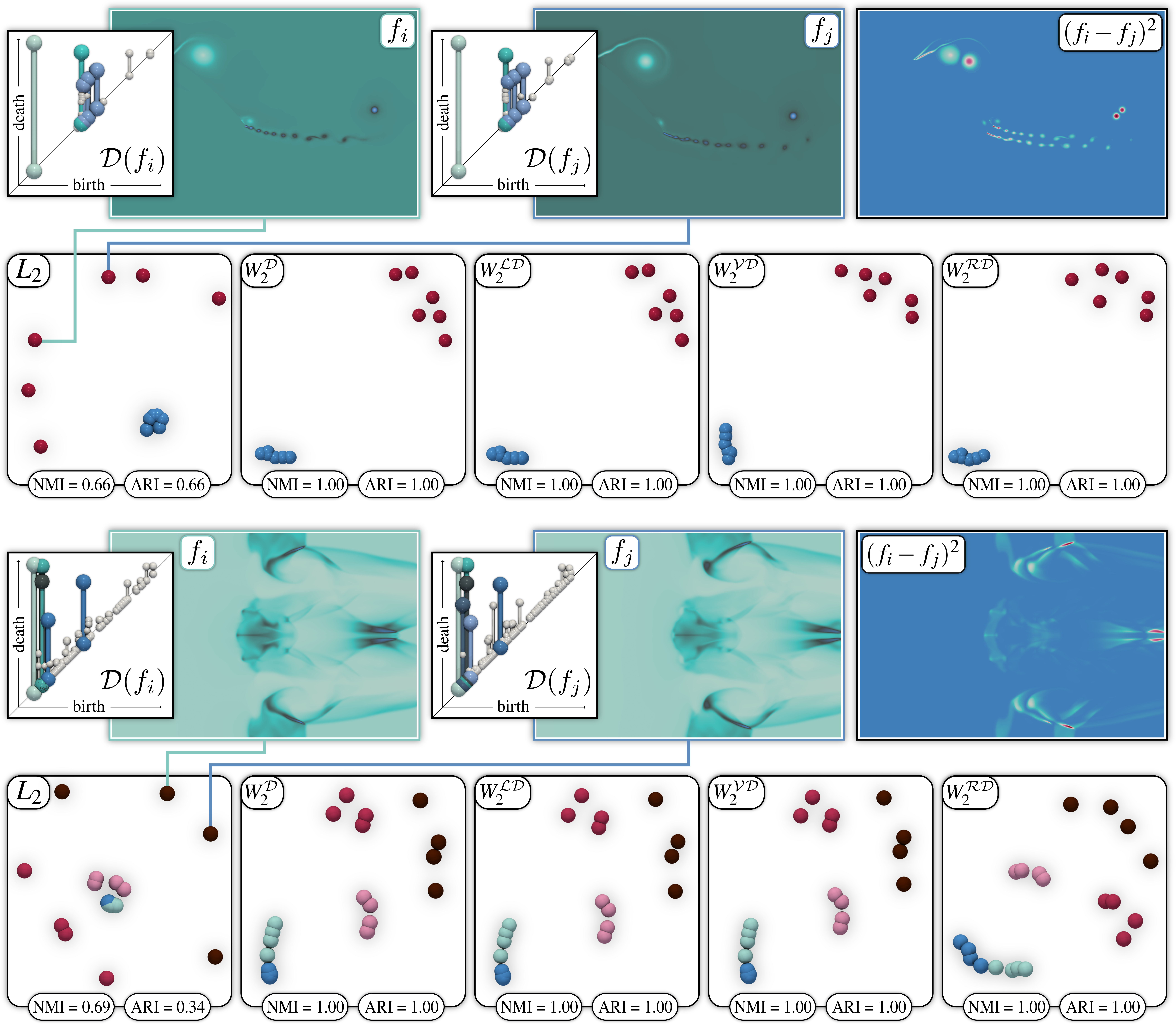}
    \caption{
\revisionTVCG{
Comparison of MDS planar embeddings of the methods considered in the main manuscript as well as the $L_2$ distance for two input ensembles: starting vortex (top) and ionization front 2D (bottom).
On these examples, only the topological methods ($\wasserstein{2}$, $\liftingWassersteinTree$, $\volumeWassersteinTree$ and $\regionWasserstein{2}$ in the four rightmost embeddings) provide embeddings where the classes are correctly separated ($NMI = ARI = 1$).
For each ensemble, two members $f_i$ and $f_j$ of a same cluster are visualized as well as their persistence diagrams $\diagram(f_i)$ and $\diagram(f_j)$.
The scalar field difference $(f_i - f_j)^2$ shows that the features of interest are not in the same position in the domain accross these members and therefore results in a high $L_2$ distance between them and to planar embeddings where the clusters are not well separated.
On the contrary, their persistence diagrams $\diagram(f_i)$ and $\diagram(f_j)$ encode the features no matter their position in the domain resulting in relatively similar representations for these members and a low topological distance, overall it makes the clusters of the topological methods well separated.
}
    }
    \label{fig_l2_comparison}
\end{figure*}

In this example, the various methods considered exhibit a relatively similar level of visual and qualitative robustness in their embeddings.
Quantitatively, however, the NMI and ARI indicators convey a slight degradation for high noise amplitudes $\epsilon \in \{2, 10\}$ as $\subsampleParameter$ decreases (i.e. when getting closer to the pure region-aware Wasserstein distance $\regionWasserstein{2}$).
Specifically, for $\epsilon = 2$, while these indicators are $1$ (being the best value) for $\subsampleParameter \in \{1, 0.1\}$ they are slightly lower for $\subsampleParameter \in \{0.05, 0.02, 0\}$ even if the visual disposition of the clusters in the MDS embeddings are relatively the same for all these values. 
This trend is consistent with the fact that decreasing $\subsampleParameter$ incorporates more region information into the distance and therefore possibly more perturbation-sensitive pointwise discrepancies, whereas $\subsampleParameter = 1$ corresponds to the original Wasserstein distance that is proven to be stable \cite{CohenSteinerEH05}.
At the same time, the absence of a sharp degradation before $\epsilon = 2$ suggests that the additional geometric information used by the region-aware variants still preserves useful discriminative cues under moderate noise.

In conclusion, we see in this experiment that using the recommended value of $\subsampleParameter = 0.1$ (see \secResults{} of the main manuscript), the robustness behavior of $\regionWasserstein{2}$ is relatively the same \revisionTVCGminor{as} for the original Wasserstein distance $\wasserstein{2}$ and that for all considered values of $\subsampleParameter$ the embeddings are relatively stable until even high noise amplitudes ($\epsilon = 2$).


\section{Dimensionality reduction experiments}

This section extends the dimensionality reduction experiments of \secDimRed{} of the main manuscript to also compare the considered methods to the $L_2$ distance.
Specifically, we show in \autoref{fig_l2_comparison} the MDS planar embeddings using the topological methods that \revisionTVCGminor{were} considered and using the $L_2$ distance.

In these examples, for the timesteps $f_i$ and $f_j$ being visualized, we can see in the pointwise squared difference $(f_i - f_j)^2$ that their features of interest are not in the same position in the domain.
The vertex-wise comparison of the $L_2$ distance \revisionTVCGminor{cannot} therefore compare them directly one to another, but instead compare them to unrelated locations in the other field, resulting in a high distance.
The persistence diagrams $\diagram(f_i)$ and $\diagram(f_j)$ look however very similar because they encode the features of interest whatever their position in the field.
This notably allows the comparison of these features even if they do not directly align in the original scalar fields, hence inducing in some way a position invariance of the comparison.
Overall, in these examples, this permits for the topological methods to provide better MDS planar embeddings than the $L_2$ distance.

We would still like to note that the $L_2$ distance is nevertheless a meaningful baseline and can be competitive on smoothly evolving time series for instance, in some cases it can provide better planar embeddings than topological methods that are not necessarily superior in all of them.
Topological methods have however the benefits, among others, of detecting features in the data and encoding them in concise, visual and informative representations and having a variety of analysis tools available.

}

\revisionTVCGminor{
\section{Workstation runtime and peak memory usage}

This section compares the running time and peak memory of the baseline methods and our method on the considered ensembles using:
\emph{1)} the machine used in the main manuscript and \emph{2)} a more usual workstation-class configuration.

Specifically, we compare \emph{1)} two AMD EPYC 7453 CPUs (2.75 GHz, 28 cores each, 112 threads in total) with 512 GB of RAM against \emph{2)} a workstation-class configuration with one Intel Core Ultra 7 165H (16 cores and 22 threads in total: 6 P-cores at 1.4 GHz with 2 threads each, 8 E-cores at 0.9 GHz and 2 LP E-cores at 0.7 GHz) with 64 GB of RAM.
We refer to the first configuration, used in the main manuscript, as the reference configuration, and to the second one, additionally used in this appendix, as the workstation configuration.

We observe in \autoref{tab_time} that for $\regionWasserstein{2}$ the mean per-ensemble slowdown of the workstation configuration compared to the reference one (defined as $\frac{1}{S} \sum_i \frac{w_i}{r_i}$, with $S$ the number of ensembles and $w_i$ and $r_i$ the execution times for the $i^{th}$ ensemble on the workstation and reference configurations, respectively) is $5.33\times$.
Moreover, averaged over all ensembles, it takes approximately 37 minutes to compute the distance matrix of an ensemble with $\regionWasserstein{2}$ using the workstation configuration compared to approximately 3 minutes for the reference one.
This makes the workstation $10.44\times$ slower in terms of the ratio of mean runtimes $\frac{1 / S \sum_i w_i}{1 / S \sum_i r_i}$.
The difference between these two factors comes from the fact that the ratio of mean runtimes mathematically gives more influence to ensembles that dominate the total runtime: $\frac{1 / S \sum_i w_i}{1 / S \sum_i r_i} = \sum_i \frac{r_i}{\sum_i r_i} \frac{w_i}{r_i}$, whereas the mean per-ensemble slowdown weights all ensembles equally and is therefore more representative of the slowdown observed for a typical ensemble.
Overall, in the largest reported settings, these methods (the baselines and ours) are primarily suitable for batch analysis rather than interactive use and should be computed as a preprocessing step before interactive analysis.

\begin{table}[b]
\caption{\revisionTVCGminor{
Running times (in seconds), using the two considered configurations, of our method and the baselines for the distance matrix computation of each ensemble using persistence diagrams.
}}
\label{tab_time}
\centering
\scalebox{0.56}{
  \begin{tabular}{|l|r|r||r|r|r|r||r|r|r|r|}
    \hline
    \rule{0pt}{2.25ex}  \textbf{Dataset} & $\ensembleSize$ & $|\branchtree|$ & \multicolumn{4}{c||}{2x AMD EPYC 7453} & \multicolumn{4}{c|}{Intel Core Ultra 7 165H} \\
     &  &  & \multicolumn{4}{c||}{512 GB of RAM} & \multicolumn{4}{c|}{64 GB of RAM} \\
       & & & $\wasserstein{2}$ & $\liftingWasserstein{2}$ & $\volumeWasserstein{2}$ & $\regionWasserstein{2}$ & $\wasserstein{2}$ & $\liftingWasserstein{2}$ & $\volumeWasserstein{2}$ & $\regionWasserstein{2}$ \\
        &  &  &  &  &  & \textbf{(ours)} &  &  & & \textbf{(ours)} \\
    \hline
      Asteroid Impact (3D)      &  68 &  854 & 33.3 & 31.7 & 34.2 & 35.9 & 204.3 & 84.4 & 101.4 & 142.5 \\
      Cloud Processes (2D)      &  91 & 2880 & 419.5 & 387.9 & 435.2 & 503.4 & 4,880.9 & 4,499.2 & 5,224.0 & 5,650.2 \\
      Viscous Fingering (3D)    & 120 &   48 & 13.2 & 9.9 & 10.2 & 5.8 & 0.5 & 0.9 & 1.0 & 2.3 \\
      Dark Matter (3D)          &  99 & 4646 & 973.9 & 984.4 & 1,187.9 & 1,492.1 & 10,868.3 & 9,984.3 & 11,316.8 & 15,782.7 \\
      Volcanic Eruptions (2D)   & 200 &  828 & 102.2 & 99.2 & 109.9 & 143.1 & 1,400.5 & 1,271.1 & 1,459.3 & 1,696.0 \\
      Ionization Front (2D)     & 200 &   68 & 20.8 & 23.8 & 20.4 & 17.4 & 2.5 & 3.0 & 2.9 & 4.0 \\
      Ionization Front (3D)     & 200 & 1228 & 218.2 & 236.5 & 235.8 & 354.8 & 2,575.6 & 2,719.3 & 2,624.5 & 3,976.1 \\
      Earthquake (3D)           & 212 &  614 & 56.4 & 61.8 & 61.4 & 95.6 & 692.9 & 627.9 & 713.9 & 1,089.9 \\
      Isabel (3D)               &  48 & 1832 & 66.6 & 58.9 & 67.2 & 68.4 & 497.1 & 176.4 & 255.4 & 290.3 \\
      Starting Vortex (2D)      &  12 &   39 & 0.1 & 0.1 & 0.1 & 0.1 & 8e-3 & 9e-3 & 7e-3 & 9e-3 \\
      Sea Surface Height (2D)   &  48 & 1453 & 40.9 & 37.9 & 41.6 & 45.3 & 293.0 & 131.0 & 151.4 & 177.9 \\
      Vortex Street (2D)        &  45 &    9 & 0.2 & 0.2 & 0.3 & 0.2 & 2e-2 & 2e-2 & 2e-2 & 1e-2 \\
      Heated Cylinder (2D)      &  60 &   12 & 0.9 & 0.5 & 0.6 & 0.6 & 3e-2 & 3e-2 & 4e-2 & 3e-2 \\
    \hline
  \end{tabular}
}
\end{table}

\begin{table}[t]
\caption{\revisionTVCGminor{
Peak memory (in GB), using the two considered configurations, of our method and the baselines for the distance matrix computation of each ensemble using persistence diagrams.
}}
\label{tab_memory}
\centering
\scalebox{0.592}{
  \begin{tabular}{|l|r|r||r|r|r|r||r|r|r|r|}
    \hline
    \rule{0pt}{2.25ex}  \textbf{Dataset} & $\ensembleSize$ & $|\branchtree|$ & \multicolumn{4}{c||}{2x AMD EPYC 7453} & \multicolumn{4}{c|}{Intel Core Ultra 7 165H} \\
     &  &  & \multicolumn{4}{c||}{512 GB of RAM} & \multicolumn{4}{c|}{64 GB of RAM} \\
       & & & $\wasserstein{2}$ & $\liftingWasserstein{2}$$^\dagger$ & $\volumeWasserstein{2}$$^\dagger$ & $\regionWasserstein{2}$ & $\wasserstein{2}$ & $\liftingWasserstein{2}$$^\dagger$ & $\volumeWasserstein{2}$$^\dagger$ & $\regionWasserstein{2}$ \\
        &  &  &  &  &  & \textbf{(ours)} &  &  & & \textbf{(ours)} \\
    \hline
      Asteroid Impact (3D)      &  68 &  854 & 15.07 & 37.18 & 37.49 & 37.47 & 4.71 & 27.26 & 27.19 & 27.04 \\
      Cloud Processes (2D)      &  91 & 2880 & 60.08 & 64.07 & 64.11 & 64.21 & 12.40 & 16.48 & 16.46 & 16.48 \\
      Viscous Fingering (3D)    & 120 &   48 & 0.64 & 15.37 & 15.37 & 15.38 & 0.57 & 12.68 & 12.68 & 12.68 \\
      Dark Matter (3D)          &  99 & 4646 & 174.07 & 177.74 & 177.91 & 176.05 & 38.76 & 47.59 & 47.30 & 46.95 \\
      Volcanic Eruptions (2D)   & 200 &  828 & 6.79 & 6.77 & 6.72 & 6.51 & 2.38 & 2.76 & 2.77 & 2.76 \\
      Ionization Front (2D)     & 200 &   68 & 0.66 & 1.02 & 1.02 & 1.02 & 0.58 & 0.93 & 0.93 & 0.93 \\
      Ionization Front (3D)     & 200 & 1228 & 29.69 & 41.67 & 41.56 & 41.22 & 7.79 & 19.73 & 19.65 & 19.62 \\
      Earthquake (3D)           & 212 &  614 & 8.71 & 17.98 & 17.92 & 18.00 & 4.25 & 13.66 & 13.68 & 13.72 \\
      Isabel (3D)               &  48 & 1832 & 30.87 & 31.50 & 32.27 & 31.27 & 7.53 & 9.50 & 9.59 & 9.53 \\
      Starting Vortex (2D)      &  12 &   39 & 0.50 & 0.79 & 0.78 & 0.79 & 0.48 & 0.78 & 0.78 & 0.78 \\
      Sea Surface Height (2D)   &  48 & 1453 & 15.23 & 15.67 & 15.27 & 15.46 & 3.94 & 4.61 & 4.61 & 4.59 \\
      Vortex Street (2D)        &  45 &    9 & 0.50 & 0.53 & 0.53 & 0.52 & 0.48 & 0.50 & 0.50 & 0.50 \\
      Heated Cylinder (2D)      &  60 &   12 & 0.48 & 0.48 & 0.48 & 0.48 & 0.52 & 0.55 & 0.55 & 0.55 \\
    \hline
\end{tabular}
}
\par\vspace{0.3em}
\noindent\begin{minipage}{\linewidth}
\revisionTVCGminor{\footnotesize $^\dagger$ The implementations of $\liftingWasserstein{2}$ and $\volumeWasserstein{2}$ load the input scalar fields to compute their required geometric attributes. They could instead be precomputed and stored within the persistence diagram in a preprocessing step, yielding an intrinsic peak memory closer to the original Wasserstein distance $\wasserstein{2}$.}
\end{minipage}
\end{table}

Regarding peak memory, we observe in \autoref{tab_memory} that, for the workstation configuration, the mean per-ensemble peak memory overhead of $\regionWasserstein{2}$ compared to $\wasserstein{2}$ is $3.49\times$.
Averaged over all ensembles, $\regionWasserstein{2}$ reaches roughly 12 GB of peak memory while $\wasserstein{2}$ reaches 6.5 GB (giving a ratio of mean peak memory values of $1.85\times$, while the mean per-ensemble memory overhead is more representative).
This is expected since our method needs access to the original data while the classical Wasserstein distance can work only on the topological representations.
The difference in peak memory mainly comes from loading the input data into RAM.
For the reference configuration, we observe a mean per-ensemble memory overhead of $3.09\times$, and averaged over all ensembles, $\regionWasserstein{2}$ reaches approximately 31.4 GB compared to 26.4 GB for $\wasserstein{2}$ (with a ratio of mean peak memory values of $1.19\times$).
The difference with the workstation configuration is mainly explained by the higher parallelism of the reference one: more cores allow more pairwise distances to be computed concurrently, which increases the amount of memory accumulated at peak usage.
Regarding $\liftingWasserstein{2}$ and $\volumeWasserstein{2}$, they only require additional coarse geometric attributes, such as feature coordinates or volumes, but the implementations used in our experiments load the full input data to compute these attributes, making their peak memory similar to that of $\regionWasserstein{2}$. In principle, these attributes could be precomputed and stored with the topological representations on disk, resulting in lower peak memory closer to that of the original Wasserstein distance.

We recall that the use of compression (\secCompression{} of the main manuscript) does not affect RAM usage but only disk storage, since the data must be decompressed in a preprocessing step to access the vertex values at runtime.
To conclude, we see that 
even on a workstation-class configuration, all methods remain relatively fast for the smallest ensembles, typically requiring only a few seconds and a few GB of peak memory. For the largest ensembles, runtimes can reach tens of minutes to a few hours depending on the configuration, while peak memory remains moderate, on the order of a few tens of GB.

}

\section{Regions compression}

In this section we briefly present the compression methods considered in this work, and how given the compression parameter $\compressionParameter$ (\secCompression{} of the main manuscript) these models need to be defined in order to provide a compressed size as close as possible to the target one.

Given $\compressionParameter$, we compute the desired number of parameters in the compressed size as $p = \lfloor \compressionParameter n \rfloor$ with $n$ the number of values in the input data to compress (we consider 32 bits float data).

\textbf{ZFP} \cite{zfp} is a lossy compression method that operates by dividing data into small blocks, transforming them into a fixed-point representation, and applying a customized encoding scheme that balances precision and compression efficiency. 
ZFP supports a fixed-rate mode allowing to bound the compressed size given an input parameter.
Specifically, we need to provide a rate $r$ defined as the number of bits per value that we allow.
Considering 32 bits float values, the total target bit budget is $32p$.
The entire dataset has $n$ values, therefore the average number of bits allowed per value is $r = \frac{32p}{n}$.

\textbf{Neural fields} \cite{neural_fields} are neural networks applied to compression, it consists \revisionTVCGminor{of} the use of an architecture whose size is lower than the input data, the network is then trained to predict the scalar value given the input coordinates and the architecture is itself the compressed representation.
Inspired by Lu et al. \cite{neural_fields} we use an architecture with a total of $l$ layers consisting of an input layer of $d$ neurons ($d$ being the dimension of the data to compress), $l - 2$ hidden layers of $k$ neurons each and an output layer of 1 neuron (to predict a scalar field).
The total number of parameters (weights and biases) is given by:

\begin{equation}
    \underbrace{d k + k}_{\text{input}} + 
    \underbrace{(l - 2) (k^2 + k)}_{\text{hidden}} + 
    \underbrace{k + 1}_{\text{output}},
\end{equation}

We consider that the total number of layers $l$ is an input parameter of the method, we now want to solve for $k$, the number of neurons in each hidden layer:

\begin{equation}
\begin{aligned}
    & d k + k + (l - 2) (k^2 + k) + k + 1 = p \\
    \Leftrightarrow \quad & d k + k + (l - 2) (k^2 + k) + k + 1 - p = 0 \\
    \Leftrightarrow \quad & k^2 (l - 2) + k (d + 1 + l - 2 + 1) + 1 - p = 0 \\
    \Leftrightarrow \quad & k^2 (l - 2) + k (d + l) + 1 - p = 0.
\end{aligned}
\end{equation}

We solve this using the quadratic formula, we first set \mbox{$a = l - 2$}, $b = d + l$ and $c = 1 - p$, we then compute \mbox{$\Delta = b^2 - 4ac$} and finally get $k = \text{round}\big(\frac{-b + \sqrt{\Delta}}{2a}\big)$.

In practice we use $l = 18$ with therefore 16 hidden layers, we also use residual connections for the hidden layers.
We found that the following parameters gave the best results averaged over all ensembles: a learning rate of 0.005, a batch size of 1024, a number of epochs of 1000, a learning rate scheduling multiplying by 0.9 the learning rate every 20 epochs and the leaky ReLU activation function.
We optimize the model using a NVIDIA GeForce RTX 4090 GPU with 24 GB of RAM.

\textbf{B-splines} \cite{b_splines} consist in the optimization of a grid of control points to best fit the input data.
Each point value in the input data is approximated by \revisionTVCGminor{a} weighted sum of the optimized control points where the weights correspond to the so-called \emph{basis functions}.
The grid of control points acts as the compressed representation.
We define the grid such that its aspect ratio is as close as possible to \revisionTVCGminor{that} of the input data.
It amounts to set the number of points in each dimension of the grid to be \revisionTVCGminor{proportional} to the extent of that dimension in the input data.
Specifically, let $e_d$ \revisionTVCGminor{be} the \revisionTVCGminor{extent} of the $d$\textsuperscript{th} dimension in the input data (i.e. the difference between the maximum and the minimum coordinate for this specific dimension over all the points).
Let $g_d$ \revisionTVCGminor{be} the number of control points in the $d$\textsuperscript{th} dimension in the grid.
In 2D we want that:

\begin{empheq}[left=\empheqlbrace]{align}
    \frac{g_0}{g_1} &= \frac{e_0}{e_1} \label{eq_splines2D_1} \\
    g_0g_1      &= p \label{eq_splines2D_2}
\end{empheq}

We can therefore write \autoref{eq_splines2D_1} as $g_0 = \frac{e_0}{e_1}g_1$, then replacing $g_0$ in \autoref{eq_splines2D_2} by this new expression gives:

\begin{equation}
\begin{aligned}
    & \frac{e_0}{e_1} g_1^2 = p \\
    \Leftrightarrow \quad & g_1^2 = p \frac{e_1}{e_0} \\
    \Leftrightarrow \quad & g_1 = \sqrt{p\frac{e_1}{e_0}}
\end{aligned}
\end{equation}

In practice, we therefore set the number of control points in the first dimension as $g_0 = \text{round}\big(\frac{e_0}{e_1}\sqrt{p\frac{e_1}{e_0}}\big)$ and as \mbox{$g_1 = \text{round}\big(\sqrt{p\frac{e_1}{e_0}}\big)$} for the second dimension.

In 3D we want that:

\begin{empheq}[left=\empheqlbrace]{align}
    \frac{g_0}{g_2} & = \frac{e_0}{e_2} \label{eq_splines3D_1} \\
    \frac{g_1}{g_2} & = \frac{e_1}{e_2} \label{eq_splines3D_2} \\
    g_0 g_1 g_2 & = p \label{eq_splines3D_3}
\end{empheq}

We rewrite \autoref{eq_splines3D_1} as $g_0 = \frac{e_0}{e_2}g_2$.
Then, we rewrite \autoref{eq_splines3D_2} as $g_1 = \frac{e_1}{e_2}g_2$.
Replacing these expressions in \autoref{eq_splines3D_3} gives:

\begin{equation}
\begin{aligned}
    & \frac{e_0}{e_2}\frac{e_1}{e_2}g_2^3 = p \\
    \Leftrightarrow \quad & g_2^3 = p \frac{e_2}{e_0} \frac{e_2}{e_1} \\
    \Leftrightarrow \quad & g_2 = \sqrt[\leftroot{-1}\uproot{2}\scriptstyle 3]{p\frac{e_2^2}{e_0 e_1}}
\end{aligned}
\end{equation}

In practice, we therefore set the number of control points in the first dimension as $g_0 = \text{round}\big(\frac{e_0}{e_2}\sqrt[\leftroot{-1}\uproot{2}\scriptstyle 3]{p\frac{e_2^2}{e_0 e_1}}\big)$, as \mbox{$g_1 = \text{round}\big(\frac{e_1}{e_2}\sqrt[\leftroot{-1}\uproot{2}\scriptstyle 3]{p\frac{e_2^2}{e_0 e_1}}\big)$} for the second dimension and finally as $g_2 = \text{round}\big(\sqrt[\leftroot{-1}\uproot{2}\scriptstyle 3]{p\frac{e_2^2}{e_0 e_1}}\big)$ for the third dimension.

\section{\revisionTVCGminor{Coarse} geometric properties definitions}

We define in this section the \revisionTVCGminor{coarse} geometric properties that were studied in previous works\revision{\mbox{\cite{soler_ldav18, lin_geometryAware23, mingzheTracking, SaikiaSW14_branch_decomposition_comparison}}} and consist of the baselines with which we compare our method to (in addition to the classical Wasserstein distances \cite{edelsbrunner09, pont_vis21}).
These two methods extend the classical ground metric (\equationGroundMetric{} of the main manuscript) to not only compare the birth and the death of two persistence pairs but also some \revisionTVCGminor{coarse} geometric properties about them.

First, the so-called \emph{geometrical lifting}\revision{\mbox{\cite{soler_ldav18, lin_geometryAware23, mingzheTracking}}} uses the distance between the coordinates in the domain of the extrema of the persistence pairs.
We will define it for 3D domains but its general definition is not restricted to this specific dimension.
Let two persistence pairs $p_i = (b_i, d_i)$ and $p_j = (b_j, d_j)$ and their 3D coordinates $(x_i, y_i, z_i)$ and $(x_j, y_j, z_j)$, the geometrical lifting ground metric is defined as:

\begin{equation}
\begin{aligned}
\pointMetric_{q}^\lifting(p_i, p_j) = & \Big(|b_j - b_i|^q + |d_j - d_i|^q \\
& + w_\lifting \big(|x_j - x_i|^q + |y_j - y_i|^q + |z_j - z_i|^q \big) \Big)^{1/q},
\end{aligned}
\end{equation}

\noindent
where $w_\lifting$ allows to weight the contribution of the coordinates distance in the final distance.
As suggested by Soler et al. \cite{soler_ldav18}, the coordinates can be normalized by the bounds of the input data.
We note $\liftingWasserstein{2}$ and $\liftingWassersteinTree$ the Wasserstein distance for respectively persistence diagrams and merge trees using $\pointMetric_{q}^\lifting(p_i, p_j)$ as a ground metric.

Second, we define the volume difference \cite{SaikiaSW14_branch_decomposition_comparison}.
Let two persistence pairs $p_i = (b_i, d_i)$ and $p_j = (b_j, d_j)$ and the volume, i.e. the number of vertices, of their regions in the domain $v_i$ and $v_j$, the volume difference ground metric is defined as:

\begin{equation}
\begin{aligned}
\pointMetric_{q}^\volume(p_i, p_j) = & \big(|b_j - b_i|^q + |d_j - d_i|^q + w_\volume |v_j - v_i|^q\big)^{1/q},
\end{aligned}
\end{equation}

\noindent
where $w_\volume$ allows to weight the contribution of the volume difference in the final distance.
As suggested by Saikia et al. \cite{SaikiaSW14_branch_decomposition_comparison}, the volume of a persistence pair can be normalized by the global volume of the input data.
We note $\volumeWasserstein{2}$ and $\volumeWassersteinTree$ the Wasserstein distance for respectively persistence diagrams and merge trees using $\pointMetric_{q}^\volume(p_i, p_j)$ as a ground metric.

\revision{
\section{Dimensionality Reduction Indicators}

To quantify how well 2D embedding preserves the original cluster structure, we use the \emph{Normalized Mutual Information
(NMI)} \cite{strehl2002cluster,vinh2010information} and the \emph{Adjusted Rand Index (ARI)} \cite{hubert1985comparing}, two widely-used, label-permutation–invariant measures for comparing a predicted partition~$\mathcal{C}$ with a ground-truth partition~$\mathcal{G}$.

Both metrics (i) are \emph{label-invariant}, (ii) account for chance agreement, and (iii) are recommended in comparative studies of clustering-quality indices~\cite{vinh2010information}.  
Together they offer complementary perspectives—information-theoretic fidelity (NMI) and pairwise correctness (ARI)—on how faithfully the 2D embedding preserves the high-dimensional clusters.

\smallskip
\noindent\textbf{Normalized Mutual Information (NMI).}
For two partitions $\mathcal{C}=\{\mathcal{C}_k\}_{k=1}^{K}$ and $\mathcal{G}=\{\mathcal{G}_\ell\}_{\ell=1}^{L}$ of the same set of $n$ points, let
\(p_k = \lvert\mathcal{C}_k\rvert/n\) and \(q_\ell = \lvert\mathcal{G}_\ell\rvert/n\) be the empirical cluster-label probabilities, and \(p_{k\ell} = \lvert\mathcal{C}_k\cap\mathcal{G}_\ell\rvert/n\) the joint probabilities.  
The \emph{Shannon entropy} of a partition is:
\[
  H(\mathcal{C}) \;=\; -\sum_{k=1}^{K} p_k \log p_k,
  \qquad
  H(\mathcal{G}) \;=\; -\sum_{\ell=1}^{L} q_\ell \log q_\ell.
\]
The \emph{mutual information} between the two partitions is:
\[
  I(\mathcal{C};\mathcal{G})
  \;=\;
  \sum_{k=1}^{K}\sum_{\ell=1}^{L}
  p_{k\ell}\,
  \log\frac{p_{k\ell}}{p_{k}\,q_{\ell}},
\]
which measures the reduction in uncertainty about $\mathcal{C}$ given $\mathcal{G}$ (and vice versa).
The \emph{Normalized Mutual Information} is the symmetric normalization:
\begin{equation}
  \mathrm{NMI}(\mathcal{C},\mathcal{G})
  \;=\;
  \frac{2\,I(\mathcal{C};\mathcal{G})}{H(\mathcal{C})+H(\mathcal{G})},
  \label{eq:nmi}
\end{equation}
taking values in $[0,1]$, with $1$ for perfect agreement and $0$ for statistical independence between the partitions.  
Because it derives from information theory, NMI is invariant to cluster-label permutations and rewards embeddings that preserve the full \emph{information content} of the original clustering.

\smallskip
\noindent\textbf{Adjusted Rand Index (ARI).}
Let $\mathcal{C}=\{\mathcal{C}_i\}_{i=1}^{K}$ and \mbox{$\mathcal{G}=\{\mathcal{G}_j\}_{j=1}^{L}$} be two partitions of $n$ points.
Let the $K\times L$ \emph{contingency table} $[n_{ij}]$, where:
\[
  n_{ij}\;=\;\lvert\mathcal{C}_i\cap\mathcal{G}_j\rvert
\]
counts how many points are simultaneously assigned to cluster $\mathcal{C}_i$ and to cluster $\mathcal{G}_j$.
The row sums $n_{i\cdot}=\sum_{j}n_{ij}$ and column sums $n_{\cdot j}=\sum_{i}n_{ij}$ are the cluster sizes in $\mathcal{C}$ and $\mathcal{G}$, respectively.

\medskip
\noindent
The \emph{chance agreement baseline} is defined as:
\[
  E \;=\;
  \frac{\displaystyle
        \sum_{i}\binom{n_{i\cdot}}{2}\;
        \sum_{j}\binom{n_{\cdot j}}{2}}
       {\displaystyle\binom{n}{2}},
\]
i.e. the expected number of agreeing pairs under a random matching that preserves the given row and column totals.
Then, the \emph{Adjusted Rand Index} is:
\begin{equation}
  \mathrm{ARI}(\mathcal{C},\mathcal{G})
  \;=\;
  \frac{\displaystyle
        \sum_{i,j}\binom{n_{ij}}{2}\;-\;E}
       {\displaystyle
        \tfrac12
        \Bigl[
          \sum_{i}\binom{n_{i\cdot}}{2}+
          \sum_{j}\binom{n_{\cdot j}}{2}
        \Bigr]\;-\;E}.
  \label{eq:ari}
\end{equation}
ARI equals $1$ for perfect agreement, $0$ for random agreement, and can
be negative when the partitions disagree more than would be expected by
chance.
}

\end{document}